\documentclass[prx,twocolumn,amsmath,amssymb,floatfix,footinbib,bibnotes,longbibliography]{revtex4-2}
\pdfoutput=1

\usepackage[colorlinks=true,citecolor=blue,linkcolor=magenta]{hyperref}
\usepackage{multirow}
\usepackage{amsmath}
\usepackage{bbm} 
\usepackage{graphicx}
\usepackage{dsfont}
\usepackage{changepage}
\usepackage{fancyhdr}
\usepackage{amsthm, amssymb}
 \usepackage{floatflt}
 \usepackage{float}
\usepackage{array}
\newcolumntype{P}[1]{>{\centering\arraybackslash}p{#1}}
\usepackage[usenames,dvipsnames]{color}
\usepackage[section]{placeins}
\usepackage[T1]{fontenc}
\usepackage[active]{srcltx}
\usepackage{color}
\usepackage{amssymb}
\usepackage{esint}
\usepackage[version=4]{mhchem}
\usepackage[normalem]{ulem}

\usepackage{mathtools}

\def\bm{\boldsymbol}

\newcommand {\apgt} {\ {\raise-.5ex\hbox{$\buildrel>\over\sim$}}\ }
\newcommand {\aplt} {\ {\raise-.5ex\hbox{$\buildrel<\over\sim$}}\ }

\newcommand{\veck}{\mathbf{k}}
\newcommand{\vecr}{\mathbf{r}}

\newcommand{\Z}{\mathbb{Z}}

\def\titlename{Fractional Wannier Orbitals and Tight-Binding Gauge Fields in Kitaev Honeycomb Superlattices with Flat Majorana Bands}

\def \authornames{K. B. Yogendra$^{1}$, G. Baskaran$^{2,3,4}$ and, Tanmoy Das$^{1}$}
\def \affiliations{$^1$Department of Physics, Indian Institute of Science, Bangalore 560012, India \\
\textit{$^2$ The Institute of Mathematical Sciences, CIT Campus, Chennai 600 113, India} \\
\textit{$^3$ Department of Physics, Indian Institute of Technology Madras, Chennai 600036, India}\\
\textit{$^4$ Perimeter Institute for Theoretical Physics, Waterloo, ON N2L 2Y5, Canada} \\}

\begin{document}

\title{\titlename}
\author{\authornames}
\affiliation{\affiliations}


\begin{abstract}
Fractional excitations hold immense promise for both fundamental physics and quantum technologies. However, constructing lattice models for their dynamics under gauge fields remains a formidable challenge due to inherent obstructions. Here, we introduce a novel and systematic framework for deriving low-energy lattice models of fractional orbitals coupled to tight-binding gauge fields. Departing from conventional geometric approaches, our method systematically eliminates the high-energy states via virtual hopping, thereby deriving the gauge potential and quantum metric through a superexchange-like mechanism. We demonstrate the framework by constructing Wannier orbitals for Majorana states and a tight-binding $Z_2$ gauge field across various flux crystalline phases in the Kitaev spin model on a honeycomb lattice. Our study reveals a striking phase transition between two non-trivial topological phases characterized by gapless flat-band with extensive degeneracy. Furthermore, we develop a gauge-invariant mean-field theory for interacting Majorana orbitals, leading to a correlation-induced fractional Chern state. Our work establishes a general framework for gauge-mediated tight-binding models and a gauge-invariant mean-field theory for interacting fractional orbitals that can be readily extended to $U(1)$, $SU(N)$ lattice gauge theories.
\end{abstract}

\maketitle

\section{Introduction}
The key to harnessing quantum materials for quantum technologies lies in engineering and controlling emergent excitations that obey unique statistics. \cite{Wilczek_2024,Ramon2020,Moore2017,Basov_2017,Tokura_2017,Milestones_2016,Sarma_2015} The sought-after excitations such as spinon, anyon, Majorana, and parafermion live on the enlarged states embedded in physical many-body Hilbert space of electrons or quantum spins. In the effective field theories for these emergent excitations, the influence of the remaining degrees of freedom is incorporated through a geometric term.\cite{wen_Book,Fradkin_Book,Sachdev_Book,Gay_Mackie_2009,Beenakker_2015RMP,Rahmani_2019,Zhang_1989,TARASOV_2013}. This couples the matter excitations with the emergent gauge fields to commence exotic statistics that enjoy constrained, protected, and slow dynamics of various characteristics.\cite{Chamon2005,Prem2017,Hart2021Logarithmic,yogendra2023emergent} 

Understanding the local wavefunction properties of emergent fractional particles is crucial for their detection and manipulation. Because computations of matrix elements, selection rules, and the linear/nonlinear response to \textit{local} probe fields\cite{Armitage_2019,NandKishore_2021,McGinley_2024}  rely critically on the symmetry and localization of the Wannier orbital wavefunction. Therefore, a systematic derivation of effective low-energy lattice models for fractional particles is essential.

Within a lattice, how are the local Wannier orbital states of fractional particles characterized when they undergo hopping under tight-binding (TB) gauge fields? Standard methods such as maximally localized Wannier orbitals (MLWOs),\cite{Vanderbilt_2012RMP,Marzari1997,Souza_2001} perturbation theory,\cite{Klein_1974,shavitt_1980,cohen_1977quantum,Anderson_1987RVB,Fazekas_1974RVB,Anderson_1987RVB,Schrieffer_1966,BRAVYI_2011}, renormalized Hamiltonian,\cite{Giuliani_Vignale_2005}, rotating-wave approximation, \cite{Ying_2007,Nesbet_1961}, Hubbard-Stratonovich transformation\cite{Stratonovich_1957,Hubbard_1959} produce effective low-energy models for conventional quasiparticles hopping in a lattice potential. In contrast, entangled or fractional particles traverse a lattice under a lattice gauge potential. Symmetry arguments and dualities are often employed to postulate such lattice-gauge theory coupled with particle-like excitations \cite{Kitaev2003, Kogut_RMP, Fradkin_Book, Wahl_2013, Jutho_2015}. Wegner realized that the high-temperature disorder phase of the Ising excitations is dual to a $\Z_2$ lattice gauge theory.\cite{Wegner_1971}. Kitaev introduced an exactly solvable low-energy spin model exhibiting a spin-liquid ground state, which directly translates to a $\Z_2$ lattice gauge theory.\cite{KITAEV2006,Baskaran2007,Tupitsyn_2010,Gu_2014} However, a systematic derivation of an effective $\Z_2$ gauge field-mediated TB model for the low-energy states for fractional particles that are embedded in a larger Hilbert space is lacking in the literature.

Conventional approaches to eliminating high-energy states typically involve projecting onto a low-energy subspace, but this method can introduce limitations such as Wannier obstructions to localization and constraints on correlations. In this work, we present a novel framework that circumvents these issues by applying the projection operator to the Hamiltonian rather than the states. We start with a Wannier orbital ansatz without direct projection, instead incorporating the projection operator and a variational geometric potential directly into the Hamiltonian.  This allows us to extract the tight-binding (TB) gauge field and quantum metric through a superexchange-like mechanism. Importantly, the method facilitates the imposition of a flux-preservation constraint, ensuring consistency between the global topology in the extended states and the uniform flux phase in the local orbital states of the fractional particles. Additionally, higher-order virtual hopping processes naturally give rise to quartic interactions between fractional orbitals, which we solve via a gauge-invariant mean-field theory.

We apply our framework to investigate Majorana orbitals in $\Z_2$ gauge fields in the Kitaev spin-1/2 model on the honeycomb lattice \cite{KITAEV2006}. The model, known for its exact solution, exhibits uniform $\Z_2$ fluxes in the ground state and Majorana fermions as excitations. With an applied magnetic field \cite{Nandini2019,Pollmann2018,LiangFu2018,MiangLu2018,Trebst2019,Valenti2019,Sorensen2021}, the Majorana excitations can proliferate, leading to flux crystallization \cite{Trebst_2012,Trebst_2014,Batista2019,Zhang_BAtista_2020} and their glassiness\cite{yogendra2023emergent}. In the flux crystalline phase, the model remains nearly exactly solvable, resulting in dispersive Majorana bands that have been extensively studied in the literature.\cite{Trebst_2012,Rudro2013,Trebst_2014,Nasu2017,Batista2019,Zhang_BAtista_2020,Vidal_2020,Nasu2021,Vojta2021,Koga2023,Vidal_2024} These Majorana bands exhibit transitions between gapless and gapped behavior depending on the flux configurations and also feature flat bands with integer Chern numbers. Our work advances beyond these studies by (1) Deriving an effective low-energy $\Z_2$ gauge theory via superexchange mechanism, and constructing Wannier Majorana orbitals that evade obstructions. (2) Investigating the quantum geometry, including the quantum metric and topology across the flat-Majorana dispersion phase. (3) Developing a gauge-invariant mean-field theory to describe geometry-induced interaction effects between fractional orbitals. (4) Proposing an analytical model for fractional Chern states.

The derived lattice model for Majorana orbitals with $\Z_2$ gauge fields reproduces the low-energy band structure.  To ensure topological consistency, we fix the gauge with a uniform flux per unit cell, guaranteeing an integer Chern number in the extended Majorana states. Notably, we find an interesting topological phase transition with a flat-band degeneracy at the critical point, distinct from conventional topological phase transitions, involving band gap closing at a Dirac point or nodal line degeneracy. The presence of flat bands allows for the emergence of fractional Chern bands for interacting Majorana fermions. To describe this phenomenon, we introduce a mean-field theory for the gauge-invariant Majorana density-wave state in the flat bands, providing an analytically tractable description of a fractional Chern insulator state.  This mean-field analysis offers valuable analytical insights into the general `trace/vortexibility' condition for fractional Chern states in flat bands.\cite{Ledwith2023}

\section{The Kitaev model with staggered fluxes}\label{Sec:Supercell}

\begin{figure}[ht] 
\centering
  \includegraphics[width=1\linewidth]{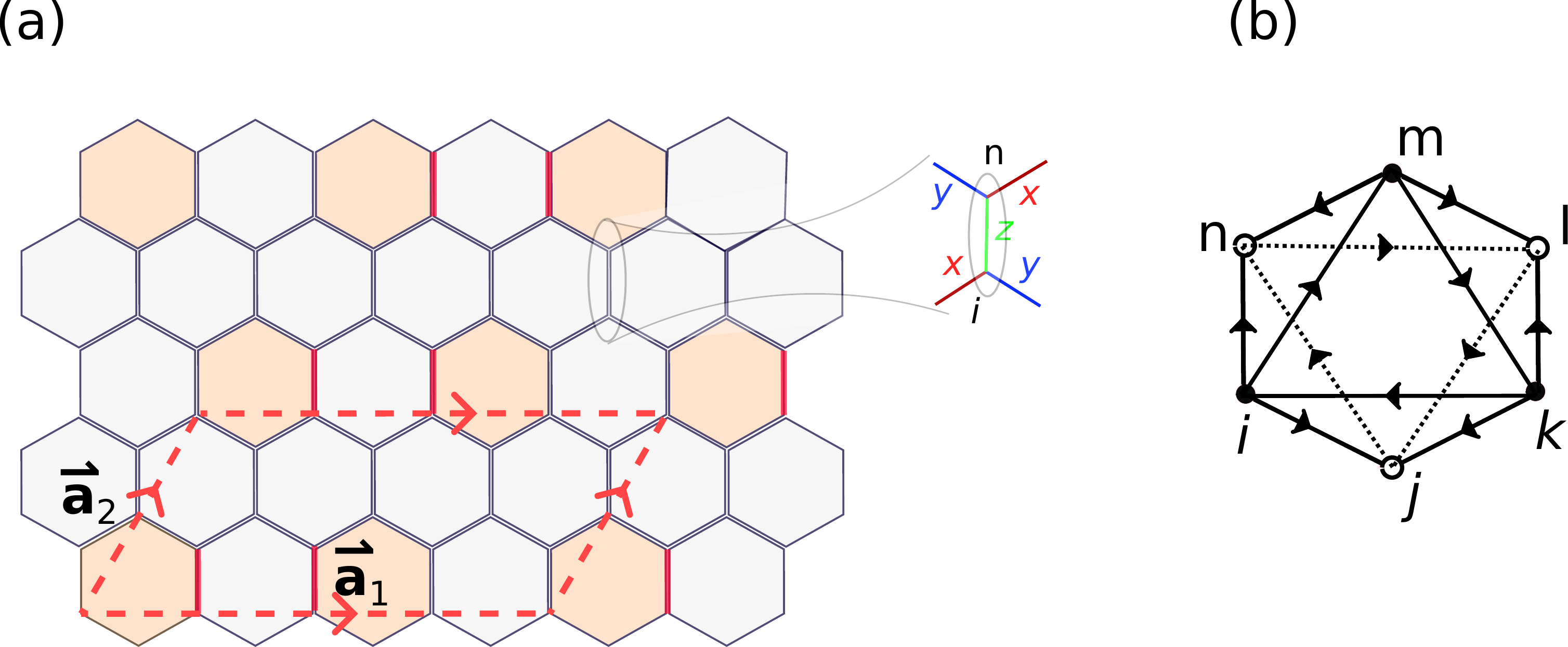}
\caption{$\Z_2$ flux crystal structure and the supercell formation in the Kitaev model. (a) We show a 2 $\times$ 2 supercell of honeycomb lattice containing a $\pi$ - flux  ($W_p = -1$) pair (orange plaquette) separated by two vertical bonds having $u_{ij}=-1$, while the rest of the plaquettes (white) have zero flux and bond have $u_{ij}=+1$.  ${\bf a}_1=(4a\sqrt{3},0)$ and ${\bf a}_2=a(\sqrt{3},3)$, where $a$ is the nearest neighbor distance in the honeycomb lattice. {\it Inset:} Three different exchange interactions $J$ along the three nearest neighbors at each site are highlighted in different colors.  (b) The chosen gauge convention for the $u$ operators in the Kitaev model is shown as $u_{ij} = +1$ if an arrow points from the site $i$ to $j$, or $-1$ otherwise. The same is shown for the next nearest neighbors by dashed lines where two $u$ operators for the intervening nearest neighbor bonds are multiplied. For example, the bond from the site, $i$ to $k$ term has $u_{ij} u_{jk}$ is $-1$. This convention is for the uniform flux sector, where each plaquette has zero flux.}
\label{fig:Lattice}
\end{figure}
 
The Kitaev model is a particular lattice model of the spin-$1/2$ fields ${\bf S}_i$ sitting at the $i^{\rm th}$ site on a honeycomb lattice and interacting with the nearest neighboring sites with bond-dependent exchange coupling $J$. A magnetic field of strength $h$ is applied along the [111] - direction, giving the corresponding model as \cite{KITAEV2006} as

\begin{equation}
    H = -J \sum_{\alpha=x,y,z, \langle ij \rangle_{\alpha}} S_i^{\alpha} S_j^{\alpha} - h \sum_{i,\alpha = x,y,z} S_{i}^{\alpha}.
\end{equation}
In the Majorana fermion representation $S_{i}^\alpha = ib_{i}^{\alpha}c_i$, the model reduces to a model of nearest-neighbor Majorana ($c_i$) hoppings mediated by a bond-dependent $\Z_2$ gauge field $u_{ij}={ib_{i}^{\alpha}}b_{j}^{\alpha}=\pm 1$. For a small magnetic field $h$, the lowest-order perturbation term produces a next nearest-neighbor hopping with the coupling constant $K \sim h^3/J^2$ (from the term $\sum_{\langle \langle ik \rangle \rangle} S_i^y S_j^z S_k^x$ ).\cite{KITAEV2006,Nasu2017,Fabian2019,Feng2020,Brenig2021,Heyl2021,Perkins2021,Kao2021168506} The model is expressed as 
\begin{equation}\label{eq:ham}
 H = iJ\sum_{\langle ij \rangle} u_{ij}c_i c_j + iK \sum_{\langle \langle ik \rangle \rangle} u_{ij}u_{jk} c_i c_k.
\end{equation}
The candidacy of the gauge field $u_{ij}$ demands it to be an antisymmetric tensor: $u_{ij} =- u_{ji}$, living on the bond between the $i$ and $j$ sites. And, the gauge field for the next-nearest neighbor $u_{ik}$ is a path-ordered product of two subsequent nearest-neighbor gauge fields  $u_{ik}=u_{ij}u_{jk}$, where $j$ is the intermediate site between $i$ and $k$. This is reflected in the second term in Eq.~\eqref{eq:ham}. A gauge choice of $u_{ij}$ is shown in Fig.~\ref{fig:Lattice}(b). A $\Z_2$ operator, defined on six consecutive links forming a loop on a plaquette $p$, is defined as $W_p= u_{ij}u_{jk}u_{kl}u_{lm}u_{mn}u_{ni}$. $W_p$ gives the $\Z_2$ flux monopole charge with $W_p=\pm 1$ for zero ($\pi$) flux. Another flux operator of importance is defined at the $r^{\rm th}$ site (called a vertex) as $X_r=u_{rs}u_{rt}u_{rq}$ where $t,s,q$ are the nearest neighbor site indices of the site $r$, which acts as a $\Z_2$ electric charge, which introduces quartic interaction between Majoranas (see Sec.~\ref{Sec:MF}).

The Hamiltonian's gauge redundancy manifests in gauge-dependent Majorana dispersions. However, the essential properties of these dispersions, such as the presence of gapless point (point degeneracy), flat band (extensive degeneracy), or topological features (band inversions), are gauge-invariant. The specific location of these gapless or band inversion points depends on the chosen gauge. In uniform flux configuration with all $W_p = +1$ and a gauge fixing of all $u=+1$ gives a graphene-like gapless Dirac node for $K=0$. $K \neq 0$ breaks the time-reversal symmetry, opening a band gap at the Dirac cone to topological Chern bands for Majorana fermions. This topological phase has Majorana zero modes (MZMs) at the boundary. Bound states of MZM with $\pi$-flux excitations can be created in bulk by thermal energy or vacancies. MZMs are topologically protected Ising anyons, which can be detected by thermal probes \cite{Reinhold2020} and by scanning tunneling microscopic techniques \cite{takashi2021}.

The focus of this work is to study different supercell formations of staggered $W_p$ fluxes and their impact on low-energy Majorana dispersions.\cite{Batista2019,Vidal_2020,Vojta2021,Nasu2021,Koga2023,Vidal_2024} Analogous to the $U(1)$ magnetic monopole, the creation of a $\Z_2$ monopole $W_p$ is topologically protected. A $\Z_2$ flux pair is defined by two $\pi$ fluxes separated by $d$ number of plaquettes with $u=-1$ in the intermediate links, as shown in Fig~\ref{fig:Lattice}(a). There is gauge redundancy in defining the supercell, and we fix the gauge for all considered supercells in the same way, as shown in the figure in Fig.~\ref{fig:Lattice}(b). This produces a $d\times d$ supercell of honeycomb lattice containing $2N=4d^2$ number of Majorana sublattices.  A  $\Z_2$ flux pair  for $d>1$ naturally breaks the $\mathsf{C}_6$ symmetry of the honeycomb lattice; however, the alignments of the $\Z_2$ flux pair along, say, ${\bf a}_1$ or ${\bf a}_2$ primitive lattice vectors, are gauge equivalent. 

We chose a Majorana spinor $\mathit{C}_I=\left(c_{1}~c_{2} ~...~c_{2N} \right)^T$ at the $I^{\rm th}$ supercell site. Then, the matrix-valued Hamiltonian in this spinor can be written from Eq.~\eqref{eq:ham} as
\begin{equation}\label{eq:ham2}
 H = i\sum_{I} C_I^{T}\mathcal{T}_{II} C_I + i \sum_{\langle IJ\rangle}C_I^{T} \mathcal{T}_{IJ} C_J .
\end{equation}
Here, the $\mathcal{T}_{IJ}$ is an (anti-symmetric) rank-2 tensor, with each element being a $2N\times 2N$ matrix. Their explicit forms are given in the Appendix~\ref{App:Supercell}. The basis vectors of the supercell are  ${\bf a}_1 = 2da(\sqrt{3},0)$, ${\bf a}_2=\frac{da}{2}(\sqrt{3},3)$, where $a$ is the nearest neighbor distance of the honeycomb primitive unit cell. The corresponding reciprocal vectors are $\mathbf{G}_1 = \frac{2\pi}{a}\left(\frac{\sqrt{3}}{6d},\frac{-1}{6d}\right)$, $\mathbf{G}_2= \frac{2\pi}{a}\left(0,\frac{2}{3d}\right)$. $\mathbf{a}_2$ vector of these supercells is half compared to the $\mathsf{C}_6$ symmetric honeycomb lattice; see Fig.~\ref{fig:Lattice}(a). Hence, the first Brillouin zone has two graphene-like BZs along the $\mathbf{G}_2$ vector. 

The (virtual) Majorana spinor state in the momentum space is  $C({\bf k}) = \frac{1}{\sqrt{L}}\sum_{I}e^{-i\veck.\mathbf{R}_I} C_I$, where ${\bf R}_{I\in \mathbb{Z}}=\sum_{i}I_i{\bf a}_i$ are the lattice sites of the supercell, and correspondingly, ${\bf k}$ is defined in the reciprocal space of ${\bf G}_{1,2}$. The corresponding Hamiltonian in the momentum space is obtained from Eq.~\ref{eq:ham2}:  $H=\frac{1}{\sqrt{N}}\sum_{{\bf k}\in{\rm BZ_+}}\mathit{C}^\dagger({\bf k})\mathcal{H}({\bf k})\mathit{C}({\bf k})$, where the matrix-elements of $\mathcal{H}$ are given in Appendix~\ref{App:Supercell}. The physical Majorana fermions $c_i =  c_i^\dagger$  turn into particle-hole symmetric virtual Majorana fermions in the ${\bf k}$-space $c^\dagger({\bf k}) = c({-\bf k})$, leading to ${\bf k\ge 0}$ being restricted to the positive quadrant in the first Brillouin zone (${\rm BZ}_+ $). The (anti-unitary) particle-hole symmetry $\mathsf{C}$ relates the Hamiltonian between different BZ quadrants as $\mathsf{C}\mathcal{H}({\bf k})\mathsf{C}^{-1}=-\mathcal{H}^T({\bf k})$. 
The final task is to diagonalize the $2N\times 2N$ particle-hole symmetric matrix $\mathcal{H}({\bf k})$. We denote the eigenvector states as $|n,\pm,{\bf k\rangle}$ corresponding to the eigenvalues of $\pm E_n({\bf k})$, where $n=1,2,...,N$. In this eigenbasis, the matter fields are the complex fermions (particles and holes), defined by the creation operators $|n,\pm,{\bf k\rangle}=\gamma^{\dagger}_{n,\pm}({\bf k})|0\rangle$, and related to the virtual Majoranas by a unitary transformation ($\Gamma$) as  $\gamma^\dagger_{n,\pm}({\bf k}) = \sum_{\alpha} \Gamma_{n,\pm,\alpha}({\bf k})c_{\alpha}({\bf k})$, for $ \Gamma_{n\pm,\alpha}({\bf k})\in \mathbb{C}$. The corresponding results are presented in Sec.~\ref{Sec:Results} for several representative $d\times d$ supercell configurations.

\section{Tight-binding Gauge-field model for Majorana orbitals}\label{Sec:TBModel}

Our task now is to obtain an effective TB model for a few low-energy eigenstates $|n,\pm,{\bf k}\rangle$. Since $|n,\pm,{\bf k}\rangle$ states are for complex fermions, we may treat the corresponding Wannier orbitals to be of the usual complex fermionic nature. However, owing to the underlying physics of Majorana Wannier orbitals hopping under lattice $\Z_2$ gauge field in real space, the TB model construction becomes non-trivial. In essence, we need to construct `Wannier' fields for both Majorana matter fields and the $\Z_2$ gauge fields while keeping all the symmetries and flux-preservation constraints intact.  

To avoid overloading with many new symbols, for the TB model, we adopt the same set of symbols, such as $\mathcal{H}$, $\mathcal{T}$, $c$, $\mathsf{C}$ and others used in the above Sec.~\ref{Sec:Supercell} with the same meanings. Should confusion arise, we explicitly mention the corresponding definition. 

We are interested in modeling the $P<N$ number of low-energy Majorana pair states $|p,{\pm},{\bf k}\rangle\equiv|{\bm p},{\bf k}\rangle$, where we combine the indices ${\bm p}\equiv(p,\pm)$ for $p=1,...,P$ with eigenvalues $\pm E_{p}({\bf k})$. These states are obtained from the full Hilbert space $|{\bf n}\equiv(n,\pm),{\bf k}\rangle$ by the projector $\mathcal{P}=\mathcal{P}_+ + \mathcal{P}_-=\sum_{p<N}\left(|{p,+}\rangle\langle {p,+}|+|{p,-}\rangle\langle {p,-}|\right)$, where ${\bf k}$ dependence in each term is kept implicit for simplicity in notation. $\mathcal{Q}=I-\mathcal{P}$ is the projection outside the low-energy states of our interest. Since $|{\bm p},{\bf k}\rangle$ states are incomplete,  their Fourier transformation to the Wannier orbitals in real space would not be useful. 

\begin{figure}[t] 
\centering
\includegraphics[width=0.9\linewidth]{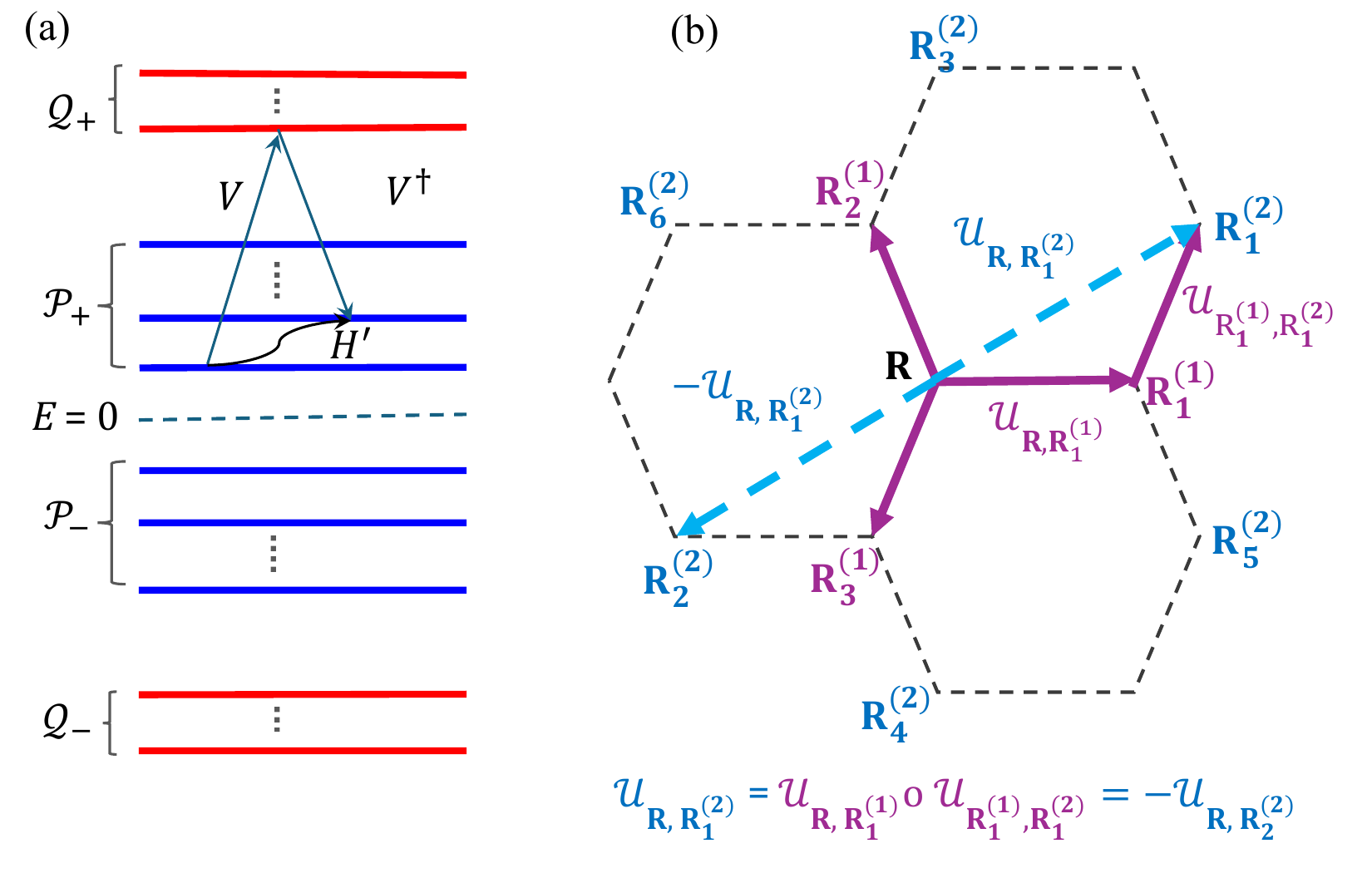}
\caption{(a) Schematic plots of the superexchange mechanism of gauge potential $H'$ in the low-energy spectrum (blue lines) due to virtual hopping to the eliminated high-energy states (red lines). Note that the virtual hopping potential V is a fitting parameter in the TB model. (b) The gauge fields in the effective theory in a honeycomb lattice. The sets of first and second nearest neighbor sites are denoted by ${\bf R}^{(1)}=\{{\bf R}^{(1)}_{1-3}\}$ and ${\bf R}^{(2)}=\{{\bf R}^{(2)}_{1-6}\}$. The gauge field for the second nearest neighbor across diagonally opposite directions must be opposite in sign to commence odd-parity hopping (superconducting term in Eq.~\eqref{eq:HM}, i.e., $\mathcal{U}_{{\bf R},{\bf R}^{(2)}_2}=-\mathcal{U}_{{\bf R},{\bf R}^{(2)}_1}$). }
\label{fig:Schematic}
\end{figure}

Our aim is to obtain complete, orthogonal states denoted by $|\tilde{{\bm p}},{\bf k}\rangle$ with corresponding eigenenergies $\tilde{E}_{p}({\bf k})\approx E_{p}({\bf k})$.  One typically defines a complex quantum geometric tensor from the $\mathcal{Q}$ projector and affix it with the $|{\bm p},{\bf k}\rangle$ states to obtain corresponding complete, orthogonal states (with a quantum metric) $|\tilde{\bm p},{\bf k}\rangle$.\cite{Vanderbilt_2012RMP,Vafek2018,Peotta_2015,Herzog_2022,Zou_Vishwanath_2018,Kruchkov_2022,Koshino_2018} Here, we devise an alternative bottom-up approach to construct a (variational) effective Hamiltonian $H_{\rm eff}$ with eigenstates $|\tilde{\bm p},{\bf k}\rangle$, and eigenenergies $\pm\tilde{E}_{\bf p}({\bf k})$. We introduce a superexchange interaction that produces tunneling between $|{\bm p},{\bf k}\rangle$ and $|{\bm p}',{\bf k}\rangle$ states with intermediate hopping to the $\mathcal{Q}$ states, see Fig.~\ref{fig:Schematic}. We call such a superexchange potential a `gauge' potential, through which we can define the $\Z_2$ gauge fields and topology in a lattice.

In what follows, we seek an effective Majorana Hamiltonian of the form $H_{\rm eff}=\mathcal{P}(H+H')\mathcal{P}$, $\forall {\bf k}$, where $H$ is the full supercell Hamiltonian, and  $H'$ is an unknown superexchange/`gauge' potential to be evaluated self-consistently. $H'$ gives off-diagonal terms in $H_{\rm eff}$ arising from the transitions between different Majorana states $|{\bm p}=(p,\pm),{\bf k}\rangle$ via intermediate hopping to the $\mathcal{Q}$ states. 
Since $H_{\rm eff}$ is an Hermitian matrix, its Hilbert space is complete. We denote the corresponding complete basis states by$|{\bm \alpha}=({\alpha},\pm),{\bf k}\rangle$ for $\alpha=1,...,P$.  In the basis of $|\alpha,+,{\bf k}\rangle\oplus|\alpha,-,{\bf k}\rangle$, we denote the matrix elements of $H_{\rm eff}({\bf k}) $  as
\begin{eqnarray}
%
\mathcal{H}_{\rm eff} ({\bf k}) 
&=& \left(\begin{array}{cc}
     {\Delta}_{R} & {\Delta}_I+ih_S  \\
    {\rm H.c.}  & - \Delta_R
\end{array}\right),
\label{eq:HM}
\end{eqnarray}
where ${\bf k}$ dependence on R.H.S. is kept implicit. $h_S({\bf k})=h({\bf k})+h^{T}(-{\bf k})$, and  $\Delta({\bf k})=\Delta_R({\bf k})+i\Delta_I({\bf k})$. Here, we restrict $\mathcal{H}_{\rm eff}$  to this particular form with the goal of obtaining Majorana orbitals, though a more general form could be considered for other types of excitations. The diagonal and off-diagonal $P\times P$ matrices of $\mathcal{H}_{\rm eff}$ are re-labeled in terms of $h$ and $\Delta$, so that the corresponding Hamiltonian in the complex fermion basis turns into a Bogolyubov-de-Gennes Hamiltonian with $h$ and $\Delta$ being their dispersion and pairing terms.\cite{KITAEV2006,Schindler_Neupert_2020} The block-off-diagonal term $({\Delta}_I+ih_S)_{\alpha,\alpha'}$ couples different Majoranas $|{\alpha},\pm,{\bf k}\rangle$ and $|{\alpha}',\mp,{\bf k}\rangle$, while the block-diagonal terms $\pm({\Delta}_{R})_{\alpha,{\alpha}'}$ give the dispersion for the same type of Majoranas $|{\alpha},\pm,{\bf k}\rangle$ and $|{\alpha}',\pm,{\bf k}\rangle$. $\mathcal{H}_{\rm eff} $ follows all the symmetries of the original Hamiltonian; in addition, the fermion-odd-parity in the pairing term $\Delta$ is also imposed.  Explicit expressions of $h$ and $\Delta$ in terms of $H$ and $H'$ are derived in  Appendix~\ref{Sec:AppBdG}.  

Our next task is to Fourier transform $H_{\rm eff}$ to real space by converting $h$ and $\Delta$ into $\mathbb{Z}_2$ gauge-field induced hoppings between Wannier Majorana orbitals in a lattice. We consider a lattice of $\mathcal{N}$ unit cells at positions ${\bf R}$.  The Fourier basis states are the Bloch phases at the ${\bf R}^{\rm th}$ cell as $z_{\bf R}({\bf k})$ = $\langle {\bf R}|{\bf k}\rangle$ = $e^{i{\bf k}\cdot{\bf R}}$. Then, we define the Majorana orbital states in real space as
\begin{eqnarray}\label{eq:FT}
 |{\bm \alpha},{\bf R}\rangle=\frac{1}{\sqrt{\mathcal{N}}}\sum_{{\bf k}>0}\bar{z}_{\bf R}({\bf k})|{\bm \alpha},{\bf k}\rangle.
 \end{eqnarray}
The physical Majorana operators are defined in real space as $|{\bm \alpha},{\bf R}\rangle=c_{{\bm \alpha},{\bf R}}|0\rangle$ where $c_{{\bm \alpha},{\bf R}}=c_{{\bm \alpha},{\bf R}}^{\dagger}$. The corresponding orthogonal Majorana wavefunctions at position ${\bf r}\in$ ${\bf R}$ unit cell are called the Bloch states $\psi_{{\bm \alpha},{\bf k}}({\bf r})=e^{i{\bf k}\cdot{\bf r}}u_{{\bm \alpha},{\bf k}}({\bf r})=\langle {\bf r}|{\bm \alpha},{\bf k}\rangle$, and Wannier states $w_{{\bm \alpha},{\bf R}}({\bf r})=\langle {\bf r}|{\bm \alpha},{\rm R}\rangle$. In the TB orbital case, the real space wavefunctions are fully localized to $w_{{\bm \alpha},{\bf R}}({\bf r})\sim\delta({\bf r}-{\bf R})$. In the Wannier orbital model $w_{{\bm \alpha},{\bf R}}$ is  (exponentially) maximally localized at $\langle{\bf r}\rangle_{\bm \alpha}=\int_{{\bf R}} d{\bf r}~{\bf r}|w_{\bm {\alpha},{\bf R}}({\bf r})|^2$, with its spread $\Delta {\bf r}_{\bm \alpha}=\langle {\bf r}^2\rangle_{\bm \alpha}-\langle {\bf r}\rangle_{\bm \alpha}^2$ also contained within the unit cell (see Appendix~\ref{App:Wannier}). The two particle-hole Majorana pairs may have different Wannier centers $\langle{\bf r}\rangle_{\alpha,+}\ne\langle {\bf r}\rangle_{\alpha,-}$, generally connected by a string $\Z_2$ vortex. However, their linear combination gives particle-hole complex fermion wavefunction which must be at the same position such that the $U(1)$ charge is conserved in each unit cell. 

It is convenient to represent the ${\bf R}$ positions in terms of sets of 1st, 2nd, 3rd, and higher nearest neighbors rather than primitive lattice vectors.  For the $n^{\rm th}$ nearest neighbor with $d_n$ number of sites (called coordination number), we define a $d_n$-dimensional vector as ${\bf Z}_{n}:=\big(z_{1}~...~z_{d_n}\big)^T$, $\forall {\bf k}$. We split the $\mathcal{N}$-dimensional vector of the Bloch phases as ${\bf Z}({\bf k})={\bf Z}_{1}({\bf k})\oplus{\bf Z}_{2}({\bf k})\oplus...$. \footnote{We make an approximation that the single-particle dispersion, many-body interaction, and superconducting order parameters are short-ranged, restricting to a few nearest neighbors only. (This truncation of the Fourier series to a polynomial of few sites gives a finite width of the single-particle states in both position and momentum space, and the number of nearest neighbors $\mathcal{N}$ to be considered is determined within a numerical procedure by fitting to the band structure at all ${\bf k}$-points. This yields the so-called compact localized orbitals for the flat band in the Wannierization procedure).}\footnote{Note that in our procedure, it is easy to implement the lattice (point-/space-) group symmetry by doing the invariant rotation on the Bloch phase spinor ${\bf Z}_{n}({\bf k})$.}

We now expand the dispersion relations in the Bloch basis ${\bf Z}({\bf k})$ to obtain the TB hopping tensor as $\mathcal{T}={\bf Z}({\bf k})\mathcal{H}_{\rm eff}({\bf k}){\bf Z}^{\dagger}({\bf k})$. $\mathcal{T}$ is a rank-2 tensor with component $\mathcal{T}_{{\bf R},{\bf R}'}$ corresponding to Majorana hoppings between ${\bf R},{\bf R}'\in \mathcal{N}$ sites. (Here, the symbol $\mathcal{T}$ is redefined for the Wannier states and not to be confused with those in the supercell in Eq.~\eqref{eq:ham2}.) Each component $\mathcal{T}_{{\bf R},{\bf R}'}$ is a $2P\times 2P$ matrix in the $2P$-dimensional Majorana basis present at the ${\bf R},{\bf R}'$ sites. We split the Hamiltonian into hoppings between different neighboring sites $n$, $n'$ as 
\begin{eqnarray}
\mathcal{H}_{\rm eff}({\bf k}) &=& 
\sum_{n,n'}{\bf Z}^{\dagger}_{n}({\bf k})\mathcal{T}_{nn'}{\bf Z}_{n'}({\bf k}) + {\rm h.c.} .
\label{eq:TBH_gauge}
\end{eqnarray}
$\mathcal{T}_{nn'}$ gives a set of Majorana hopping tensors between the $n$ and $n'$ neighbors: $\mathcal{T}_{n,n'}=\{\mathcal{T}_{{\bf R},{\bf R}'} | {\bf R}\in d_n, {\bf R}'=\in d_{n'}\}$. Due to the translational invariance, only the difference between $n$ and $n'$ is relevant. 

The intra-site hopping gives the onsite energy $\mathcal{T}_{nn}=i\mathcal{K}_0$ between $2P$-Majoranas with the particle-hole symmetric constraint ${\rm Tr}(\mathcal{T}_{nn}) = 0$. 

Now, we consider the first nearest neighbor term $\mathcal{T}_{{\bf R},{\bf R}'}\in \mathcal{T}_{n,n+1}$, where ${\bf R}'-{\bf R}\in d_1$ and ${\bf R}\ne {\bf R}'$. Using Taylor's expansion (assuming analyticity), we obtain the hopping tensor as 
\begin{equation}\label{eq:DefT}
\mathcal{T}_{{\bf R},{\bf R}'}=\left.i\mathcal{K}_1^{-1}\frac{\partial^2 \mathcal{H}_{\rm eff}}{\partial \bar{z}_{\bf R}\partial z_{\bf R'}}\right|_{z_{\bf R}=z_{\bf R'}=0}.
\end{equation}
$\mathcal{K}_1$ is, in general, orbital-dependent ($2P\times 2P$ non-singular matrices) as well as bond (i.e., ${\bf R},{\bf R}'$) dependent. (This expansion holds when $\mathcal{H}_{\rm eff}$ are polynomials in terms of $z_{\bf R}({\bf k})$, which holds for most band structures, except for non-compact flat bands \cite{Rhim_2019, Zhang_Jin_2020}.) $\mathcal{K}_1$ absorbs the energy dimension such that $\mathcal{T}_{{\bf R},{\bf R}'}$ becomes dimensionless, which is now to be defined in terms of gauge fields. The crux of the gauge theory is that $\mathcal{T}_{{\bf R},{\bf R}'}\ne \mathcal{T}_{{\bf R}',{\bf R}}$ in general. We separate the symmetric and anti-symmetric parts as 
\begin{eqnarray}\label{eq:GaugeTensors}
\mathcal{G}_{{\bf R},{\bf R}'}&=&\frac{1}{2}(\mathcal{T}_{{\bf R},{\bf R}'}+\mathcal{T}_{{\bf R}',{\bf R}}), \\
\mathcal{U}_{{\bf R},{\bf R}'}&=&-\frac{i}{2}(\mathcal{T}_{{\bf R},{\bf R}'}-\mathcal{T}_{{\bf R}',{\bf R}}).
\end{eqnarray}
Roughly speaking, $\mathcal{G}$ and $\mathcal{U}$ produce the amplitude and phase variation of the hopping term in $H_{\rm eff}$ between ${\bf R}$ and ${\bf R}'$. In the momentum space, these are precisely what the Fubini-Study metric ($\mathcal{G}_{\mu\nu}$) and the curvature  ($\mathcal{U}_{\mu\nu}$) terms constitute the symmetric and anti-symmetric components of the quantum geometric tensor.\cite{Peotta_2015,Herzog_2022,Liang_2024,Bouhon2023} Their expressions in terms of the projectors $\mathcal{P}({\bf k})$ are as follows
\begin{eqnarray}
\mathcal{G}_{\mu\nu}({\bf k})&=&\frac{1}{2}\mathcal{P}({\bf k})\{\partial_{\mu}\mathcal{P}({\bf k}),\partial_{\nu}\mathcal{P}({\bf k})\},
\label{eq:QMetric}\\
    \mathcal{U}_{\mu\nu}({\bf k})&=&-\frac{i}{2}\mathcal{P}({\bf k})\big[\partial_{\mu}\mathcal{P}({\bf k}),\partial_{\nu}\mathcal{P}({\bf k})\big],
    \label{eq:BCurvature}
\end{eqnarray}
where $\partial_{\mu}=\frac{\partial}{\partial k_{\mu}}$ with $\mu=1,2$ for ${\bf k}_{\mu}$ spanned along the reciprocal lattice vector ${\bf G}_{\mu}$. $\{\}$ and $[]$ are the anti-commutator and commutator. It is now obvious that $\mathcal{U}$ acts as parallel transport or Wilson line, which is $\Z_2$-valued in our case. In their present forms,  $\mathcal{G}$ and $\mathcal{U}$ are not gauge invariant in both real and momentum space formalism. Then, matter fields are attached at the two ends to commence gauge invariance. Otherwise, we take a trace over the matrix components and their product in a loop/plaquette in real/momentum space, giving topological invariants such as flux ($W$)/monopole in real space, Chern number ($C$) in momentum space, and similar quantum metric invariants as defined in Sec.~\ref{Sec:Chern}

The diagonal term of $\mathcal{G}$ tensor is zero as the ${\bf R}={\bf R}'$ terms are separated into the onsite energy matrix $\mathcal{K}_0$. The off-diagonal components $\mathcal{G}_{{\bf R},{\bf R}'}$ give the symmetric hopping matrix element between the orbitals localized at ${\bf R}$ and ${\bf R}'$ sites. Such symmetric hoppings are mediated by periodic lattice potential (e.g., potential due to nucleus in solid state systems) and depend on the symmetries of the two orbitals (e.g., it is present if the two orbitals have the same parity or absent if the parity of the two orbitals is opposite such as for the $s$ and $p$ orbitals\cite{DasWeyl}).  In our particular example below, we will seek a fully gauge-field mediated hopping between the two sites and set $\mathcal{G}=0$ in the effective theory.   

So far, we have kept the formalism general for any quasiparticles under $U(1)$ or $SU(N)$ gauge fields. From now on, we will impose the $\Z_2$ symmetry on the gauge fields for Majorana quasiparticles. We identify $\mathcal{U}_{{\bf R},{\bf R}'}$ as an anti-symmetric tensor that mediates tunneling between the Majoranas at the ${\bf R}$ and ${\bf R}'$ sites. For the gauge invariance of the theory for Majorana, the gauge fields must be $\mathbb{Z}_2$, which puts the constraints that $\mathcal{U}_{{\bf R},{\bf R}'}^2=\mathbb{I}$. So we interpret $\mathcal{U}_{{\bf R},{\bf R}'}$ as the non-Abelian  ($2P$-dimensional matrix-valued) $\mathbb{Z}_2$ Wilson line operator, which can be written as (path-ordered) exponentials of a (non-Abelian) gauge field $\mathcal{A}$. 

Next, we consider the second nearest neighbor term $\mathcal{T}_{{\bf R},{\bf R}''}\in \mathcal{T}_{n,n+2}$, where ${\bf R}''-{\bf R}\in d_2$ and ${\bf R}\ne {\bf R}''$. Proceeding similarly, we define the $\mathbb{Z}_2$ gauge field as the second nearest neighbor as $\mathcal{U}_{{\bf R},{\bf R}''}$. All gauge fields $\mathcal{U}$ are localized at the link/bond between the two sites. So we can smoothly deform the path to pass through a site ${\bf R}'$ corresponding to the 1st nearest neighbor to both ${\bf R}$ and ${\bf R}''$ sites, as shown in Fig.~\ref{fig:Schematic}. In other words, we can write $\mathcal{U}_{{\bf R},{\bf R}''}=\mathcal{U}_{{\bf R},{\bf R}'}\circ\mathcal{U}_{{\bf R}',{\bf R}''}$, where the composition operation $\circ$ reflects a matrix product for the tensor components. Therefore, for the $n^{\rm th}$-nearest neighbor gauge field, we have 
\begin{eqnarray}
\mathcal{U}_{{\bf R},{\bf R}^{(n)}} = \prod_{{\bf R}^{(m)}\in d_{n-1}}
\mathcal{U}_{{\bf R},{\bf R}^{(m)}}\circ\mathcal{U}_{{\bf R}^{(m)},{\bf R}^{(n)}},
\label{eq:nGF}
\end{eqnarray}
where ${\bf R}^{(m)}$ runs over the $n-1$ intermediate sites that minimize the distance between the ${\bf R}$ and ${\bf R}^{(n)}$ sites. Substituting these considerations in Eq.~\eqref{eq:TBH_gauge} we get
\begin{eqnarray}
\mathcal{H}_{\rm eff}({\bf k}) &=&i\mathcal{K}_0 + i\mathcal{K}_1\sum_{\{{\bf R},{\bf R}^{(1)}\}\in d_1}\mathcal{U}_{{\bf R},{\bf R}^{(1)}}~\bar{{z}}_{\bf R}({\bf k}){z}_{{\bf R}'}({\bf k})\nonumber\\
&&+  i\mathcal{K}_2\sum_{\{{\bf R},{\bf R}^{(2)}\}\in d_2}\mathcal{U}_{{\bf R},{\bf R}^{(2)}}~\bar{{z}}_{\bf R}({\bf k}){z}_{{\bf R}^{(2)}}({\bf k}) + ... \nonumber\\
\label{eq:HMK2}
\end{eqnarray}
The above Hamiltonian can be expressed in terms of physical Majorana orbitals $c_{{\bm \alpha},{\bf R}}$ in real space up to any number of nearest neighbor ($\mathcal{N}$) hoppings as
\begin{eqnarray}
    H_{\rm eff}
&=&i\sum_{n=0}^{\mathcal{N}}\mathcal{K}_n\sum_{\{\bf R,{\bf R}^{(n)}\}\in d_n}\sum_{{\bm \alpha},{\bm \alpha}'}\nonumber\\
&&\times\left(\mathcal{U}_{{\bf R},{\bf R}^{(n)}}\right)_{{\bm \alpha},{\bm \alpha}'}c_{{\bm \alpha},{\bf R}}c_{{\bm \alpha}',{\bf R}^{(n)}}.  
\label{eq:HMR}
\end{eqnarray}
Summation over $n$ corresponds to a different nearest neighbors. The gauge fields $\mathcal{U}_{{\bf R},{\bf R}^{(n)}}$ sit on the link between the ${\bf R},{\bf R}^{(n)}$ sites, and hence, there is no gauge field for the $n=0$ term, while $n=1$ term has one gauge field, $n=2$ has two gauge fields, and so on. Here, we have assumed the coupling constants $\mathcal{K}_n$ to be independent of the orbital and bond-independent and only depending on the $n^{\rm th}$ nearest neighbor distance. This is a reasonable assumption as at the $n^{\rm th}$ nearest neighbor site, only one type of orbital is placed.

We note that the derivation of $H_{\rm eff}$ in Eq.~\eqref{eq:HMR} is quite general and can be extended for other fractional orbitals by relaxing the $\Z_2$ constraint on the gauge field and the particle-hole symmetry on the matter fields. Moreover, the evaluation of the parameters $\mathcal{T}_{{\bf R},{\bf R}'}$ in Eq.~\eqref{eq:DefT} and gauge fields in Eqs.~\eqref{eq:GaugeTensors} does not necessitate the knowledge of the full spectrum of the original theory in Eq.\eqref{eq:ham2}. They can either be treated as fitting parameters if the excitation spectrum is known or variational parameters within a many-body wavefunction ansatz, though the latter approach is not explored in this work.

\subsection{Gauge fixing and topological invariants}
An important property of the gauge theory is the gauge constraint, which restricts gauge redundancy to the physical states. Although the gauge operators $\mathcal{U}_{{\bf R},{\bf R}^{(n)}}$ are gauge-dependent, the flux is a gauge-invariant physical operator. Therefore, the total flux in the supercell must be preserved in both the effective model and the supercell model. 

The total flux in a supercell is defined as $W_S=\prod_{p\in S}W_p$, where $S$ is the supercell index containing $S$ number of original unit cells. The $W_S$, written in terms of the effective $\mathbb{Z}_2$ gauge field, is 
\begin{equation}\label{eq:W}
W_S={\rm Tr}\left(\prod_{{\bf R},{\bf R^{(n)}}\in S}\mathcal{U}_{{\bf R},{\bf R}^{(n)}}\right).
\end{equation}
In the effective theory, we can define a similar invariant for the symmetric tensor $\mathcal{G}$ as $G_S={\rm Tr}\left(\prod_{{\bf R},{\bf R^{(n)}}\in S}\mathcal{G}_{{\bf R},{\bf R}^{(n)}}\right)$. 

Their counterparts in the momentum space are called the quantum metric invariant and the Chern number as defined to be \cite{Peotta_2015,Herzog_2022,Liang_2024,Bouhon2023}:
\begin{eqnarray}
G &=& \frac{1}{(2\pi)^2}\int_{\rm BZ} \sqrt{{\rm det}(\eta)} \eta_{\mu\nu} d k_{\mu}dk_{\nu}{\rm Tr}\mathcal{G}_{\mu\nu}({\bf k}),
\label{eq:QMI}\\
C &=& \frac{1}{2\pi}\int_{\rm BZ} d k_{1}dk_{2}{\rm Tr}\mathcal{U}_{12}({\bf k}),
\label{eq:Chern}
\end{eqnarray}
where $\eta_{\mu\nu}=\hat{{\bf G}}_{\mu}\cdot\hat{{\bf G}}_{\nu}$ is a symmetric tensor that measures the curvature of the torus geometry and depends on the lattice under consideration. $W_S$ and $G_S$ can describe both local and global topological properties in real space, depending on the size of the 2-loop $S$. In contrast, $C$ and $G$ in the momentum space only capture the global topology on the torus. Do they correspond to the same topological invariant? Indeed, this is the case. Because the flux crystal in the original lattice is taken into account in the formation of the supercell, the total flux is uniform among all the supercells. This is reflected in the effective theory, as well, in that  each band corresponds to uniform flux values $W_S=\pm 1$, $\forall S$, corresponding to Chern numbers $C=\pm 1$ for the two respective bands. 

Both $\mathcal{G}_{\mu\nu}({\bf k})$ and  $\mathcal{U}_{\mu\nu}({\bf k})$ can also be calculated directly from the effective Hamiltonian $\mathcal{H}_{\rm eff}({\bf k})$. The corresponding formulas appear similar by replacing $\mathcal{P}({\bf k})$ with $\mathcal{H}_{\rm eff}({\bf k})$. For the Chern number, the formula coincides with the Kubo formula for Hall conductivity, while the same formula for $G$ has no analog in any previous analysis. We compute them in Sec.~\eqref{Sec:Chern} .

We consider $\mathcal{K}_n$ to be the TB hopping parameters related to the gauge potential $H'$ in a self-consistent way to fit the low-energy band structure under an additional constraint of flux preservation. The values of the $\mathcal{K}_n$ parameters are the same as those obtained on the Wannier orbital basis, as shown below.

\subsection{Examples of two Majorana bands in a honeycomb lattice}\label{Sec:Example}

As an example, appropriate for the Kitaev model of present interest, we consider a honeycomb lattice with one ($P=1$) pair of Majorana bands, see Fig.~\ref{fig:Schematic}(b). Here we have $d_1=3$ first nearest neighbors ${\bf R}^{(1)}-{\bf R}=\delta{\bf R}^{(1)}=\{\frac{1}{2}(1,\pm\sqrt{3}),~(-1,0)\}$ and $d_2=6$ second nearest neighbors ${\bf R}^{(2)}-{\bf R}=\delta{\bf R}^{(2)}=\{\pm \frac{1}{2}(3,\sqrt{3}),~\pm \frac{1}{2}(3,-\sqrt{3}),~\pm(0,-\sqrt{3})\}$, and so on. 

Since only one type of Majorana orbital is positioned at each site, we can split the position and orbital indices from the gauge field as $\mathcal{U}_{{\bf R},{\bf R}'}=u_{{\bf R},{\bf R}'}{\bm \sigma}$. Here $u_{{\bf R},{\bf R}'}=\pm 1$, and $u_{{\bf R},{\bf R}'}=-u_{{\bf R}',{\bf R}}$ are the $\mathbb{Z}_2$ gauge fields for two orbitals positions at ${\bf R}$, and ${\bf R}'$ sites, and ${\bm \sigma}$ are the Pauli matrices in two  ($a=\pm$) Majorana basis. In this bipartite lattice, the same (different) Majorana orbitals are positioned at the first (second) nearest neighbor sites. Hence, the first nearest neighbor gauge field is off-diagonal: $\mathcal{U}_{{\bf R},{\bf R}^{(1)}}=u_{{\bf R},{\bf R}^{(1)}}\sigma^x$. The second nearest neighbor is diagonal $\mathcal{U}_{{\bf R},{\bf R}^{(2)}}= u_{{\bf R},{\bf R}^{(1)}}u_{{\bf R}^{(1)},{\bf R}^{(2)}}\sigma^z$, where ${\bf R}^{(1)}$ is the 1st nearest neighbor that connects ${\bf R}$ and ${\bf R}^{(2)}$ sites in the shortest distance. In all the diagonal terms, $a=\pm$ orbitals must have opposite gauge fields for the Hamiltonian to be particle-hole symmetric, and hence, we have $\sigma^z$ here.   

Taking into account the above properties, we have the TB Majorana orbital Hamiltonian (up to the second nearest neighbors): 
\begin{eqnarray}
H_{\rm eff}
&=&K_0\sum_{{\bf R}}\sum_{a=\pm}c_{a,{\bf R}}\sigma^z_{aa'}c_{a',{\bf R}}\nonumber\\
&&+ iK_1\sum_{{\bf R},{\bf R}^{(1)}_i\in {d_1}} u_{{\bf R},{\bf R}^{(1)}_i}\sum_{a=\pm}c_{a,{\bf R}}\sigma^x_{aa'}c_{a',{\bf R}^{(1)}_i}\nonumber\\
&&
+iK_2\sum_{{\bf R},{\bf R}^{(2)}_j\in {d_2}}u_{{\bf R},{\bf R}^{(1)}_l}u_{{\bf R}^{(1)}_l,{\bf R}^{(2)}_j}\sum_{a=\pm}c_{a,{\bf R}}\sigma^z_{aa'}c_{a',{\bf R}^{(2)}_j}.\nonumber\\
\label{eq:HMhoneycom}
\end{eqnarray}
$u_{{\bf R},{\bf R}^{(1)}}=\pm 1$ can take any value in a link, provided the total flux in a unit cell is conserved to the value in the full Hamiltonian. $\mathcal{K}_i=K_i\mathbb{I}$ are set to be orbital-independent coupling constants for simplicity in notation in this example, however, in the fitting procedure in Sec.~\ref{Sec:TBfitting} they are considered orbital dependent. Going to the momentum space, we obtain the diagonal and off-diagonal terms as
\begin{eqnarray}
&&\Delta_I({\bf k})+ih_S({\bf k})=i{K}_1\sum_{{\bf R},{\bf R}^{(1)}_i\in {d_1}}u_{{\bf R},{\bf R}^{(1)}_i}e^{i{\bf k}\cdot\delta{\bf R}^{(1)}_i},\nonumber\\ 
&&\Delta_{R}({\bf k})={K}_0+i{K}_2\sum_{{\bf R},{\bf R}^{(2)}_j\in {d_2}}u_{{\bf R},{\bf R}^{(1)}_l}u_{{\bf R}^{(1)}_l,{\bf R}^{(2)}_j}e^{\delta{\bf R}^{(2)}_j}.\nonumber\\
\end{eqnarray}
It is interesting to notice here that the imaginary and real parts of the superconducting (complex-fermion) pairing gaps arise from the first and second nearest neighbor Majorana hoppings, respectively. 

Note that $h_S$, $\Delta_{R,I}$ are real. We set  ${K}_i$ to be real and ${K}_0=0$. This makes $\Delta_I=-2{K}_1\sum_{i=1}^3\sin{({\bf k}\cdot\delta{\bf R}_i^{(1)}})$, $h_S=-2{K}_1\sum_{i=1}^3\cos{({\bf k}\cdot\delta{\bf R}_i^{(1)}})$. $\Delta_R$ arises from the  second next-nearest neighbor, which gives $-2{K}_2\sum_{j=1,3,5}\sin{({\bf k}\cdot\delta{\bf R}^{(2)}_j})$. This reduces the gauge choices for nearest neighbors to be $u_{{\bf R},{\bf R}^{(2)}_1}=-u_{{\bf R},{\bf R}^{(2)}_2}$, $u_{{\bf R},{\bf R}^{(2)}_3}=-u_{{\bf R},{\bf R}^{(2)}_4}$, and $u_{{\bf R},{\bf R}^{(2)}_5}=-u_{{\bf R},{\bf R}^{(2)}_6}$. This affects the gauge choices for the nearest neighbors and also the flux-modulation-induced supercell constructions, shown in Fig.~\ref{fig:Lattice}. Both $\Delta_{R, I}$ are odd under spatial parity and are consistent with odd-fermion parity for the fermionic odd-parity for complex fermion pairing for the same spin states. This gives the well-known $p+ip$ pairing state for the corresponding complex fermion state.

\section{Results}\label{Sec:Results}

\subsection{Majorana band structure of the full supercell Hamiltonian}

\begin{figure*}[ht] 
\centering
\includegraphics[width=1\linewidth]{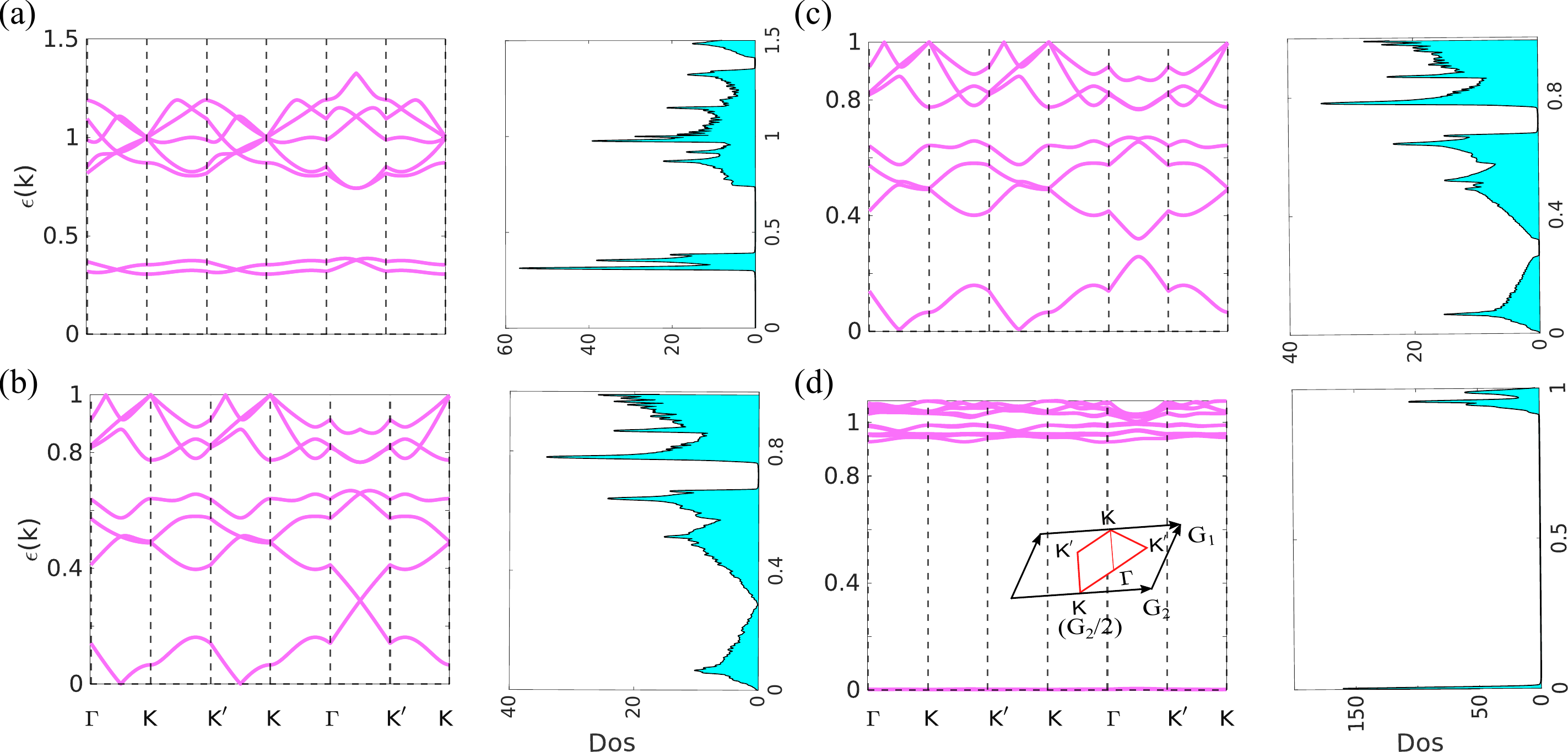}
\caption{The dispersions of the Majorana fermions for two representative flux configurations. (a) The dispersions for the $3\times 3$ for $K/J=0$, which shows a gapped dispersion. (b$-$d) The Majorana dispersions for $4\times 4$ supercell are shown for $K/J=0.0$ in (b), $K/J=0.01$ in (c), and $K/J=0.175$ in (d). $K/J>0$ opens a band gap across the zero energy. The adjacent right-hand panel shows the density of states for all three cases. The ${\bf k}-$  directions are given in the inset.The dispersions are shown only for $\epsilon(k) > 0$; redundant particle-hole symmetric $\epsilon(k) < 0$ are excluded. }
\label{fig:Flat}
\end{figure*}

We consider here several representative superlattices of dimension $d\times d$ containing a single $\Z_2$ flux pair of length $d$, i.e., the number of $u=-1$ gauge fields flipped between the two $\Z_2$ fluxes, while $u=+1$ in the rest of the bonds in the supercell. This makes the supercell Hamiltonian dimension to be $2N=4d^2$. 
A typical superlattice for $d=2$ is shown in Fig.~\ref{fig:Lattice}(a). It turns out the band structure properties are characteristically similar for all $3d\times 3d$ supercells, which differ from the characteristically similar band structure for other supercells. Therefore, we present the numerical results for two representative values of $d=3$, $4$ in Fig.~\ref{fig:Flat} by diagonalizing the supercell Hamiltonian given in Eq.~\eqref{eq:ham2}. We remind the reader that although the band dispersion depends on the gauge choice, but different gauge choices give equivalent dispersion along different momentum directions. Moreover, the salient properties such as gapless (degeneracy), gap, flat bands, Chern number, and quantum metric indices are gauge invariant. We show the gauge choice and the orientation of the $\Z_2$ flux pair for one example case of $d=4$ in Fig.~\ref{fig:wann}. 

Interestingly, we find that only for the  $d=3$ (and its integer multiples) flux configuration, the Majorana dispersions are gapped even for $K=0$, and render nearly flat-band, see Fig.~\ref{fig:Flat}(a). The reason for the gapped behavior is the broken sublattice symmetry that protects the degeneracy at energy $E=0$, although the particle-hole symmetry remains intact. 

For other flux configurations, the Majorana bands show a gapless feature at energy $E=0$ at the high-symmetric momenta for $K=0$, a representative result of which is shown in  Fig.~\ref{fig:Flat}(b). The low-energy particle-hole symmetric bands have linear dispersions around the gapless point, as in graphene, and also show linearly dispersing gap-closing points with high-energy bands. All these gapless points acquire mass term for $K\ne 0$ value, \ref{fig:Flat}(c).  The gap to the higher energy bands is larger than that at $E=0$. We denote the gap at $E=0$ by $\Delta$, while the bandwidth of the corresponding two low-energy bands is denoted by $\delta$. The ratio $\delta/\Delta$, called the flatness ratio, measures the flatness of the low-energy bands, with $\delta/\Delta\rightarrow 0$ corresponding to complete flatness (i.e., all ${\bf k}$ points are degenerate), while $\delta/\Delta\rightarrow \infty$ corresponds to point degeneracy. $K$ controls the band gap $\Delta(K)$,  while the larger the length ($d$) of the $\Z_2 $ flux pair, the smaller is $\delta$, and typically $\delta$ scales allegorically as $\delta\sim 1/d^2$. 

Increasing $d$ while holding all other parameters constant elevates the sublattice dimension. In other words, it expands the dimension of the local Hilbert space $C_I$. This, in turn, enhances the level-repulsion from the eliminated high-energy bands to the target low-energy bands. This repulsion is captured by the quantum metric $\mathcal{G}$ in the wavefunction description or by the superexchange or gauge potential ($H'$) within our effective theory, see Fig.~\ref{fig:Flat}(c). 

As $K$ increases, we observe a fascinating topological phase transition, depicted in Figs.~\ref{fig:gap4by4}. Initially, the gap scales as $\Delta\sim K$ for $K\rightarrow 0$, before it reaches a maximum around $K\approx 0.06J$. This is an interesting point where the band gap varies minimally with $K$. 
With a further increase of $K$, $\Delta$ reduces and eventually vanishes entirely around $K\approx  0.175 J$. Notably, the bandwidth ($\delta$) also vanishes at this critical point, suggesting the formation of a completely flat band where both bands become degenerate across all ${\bf k}$-points. This results in an extensive degeneracy in the Hamiltonian. It is noteworthy that on either side of this flat band degeneracy, the system exhibits well-defined Chern bands with  $C=\pm 1$. \footnote{The higher-energy bands also carry finite Chern numbers, resulting in a total Chern number of $C=\pm 2$ for the filled and empty bands, consistent with Ref.~\cite{Vidal_2020}. However, for the construction of the effective lattice model, we focus exclusively on the two low-energy bands with $C=\pm 1$. }  This observation suggests a unique type of topological phase transition characterized by the emergence of an extensive band degeneracy, which is different from the quintessential Dirac cone degeneracy at other topological phase transitions; see Sec.~\ref{Sec:Chern}. 

\subsection{Effective model and Majorana Wannier centers}\label{Sec:TBfitting}

\begin{figure}[ht] 
\centering
\includegraphics[width=1\linewidth]{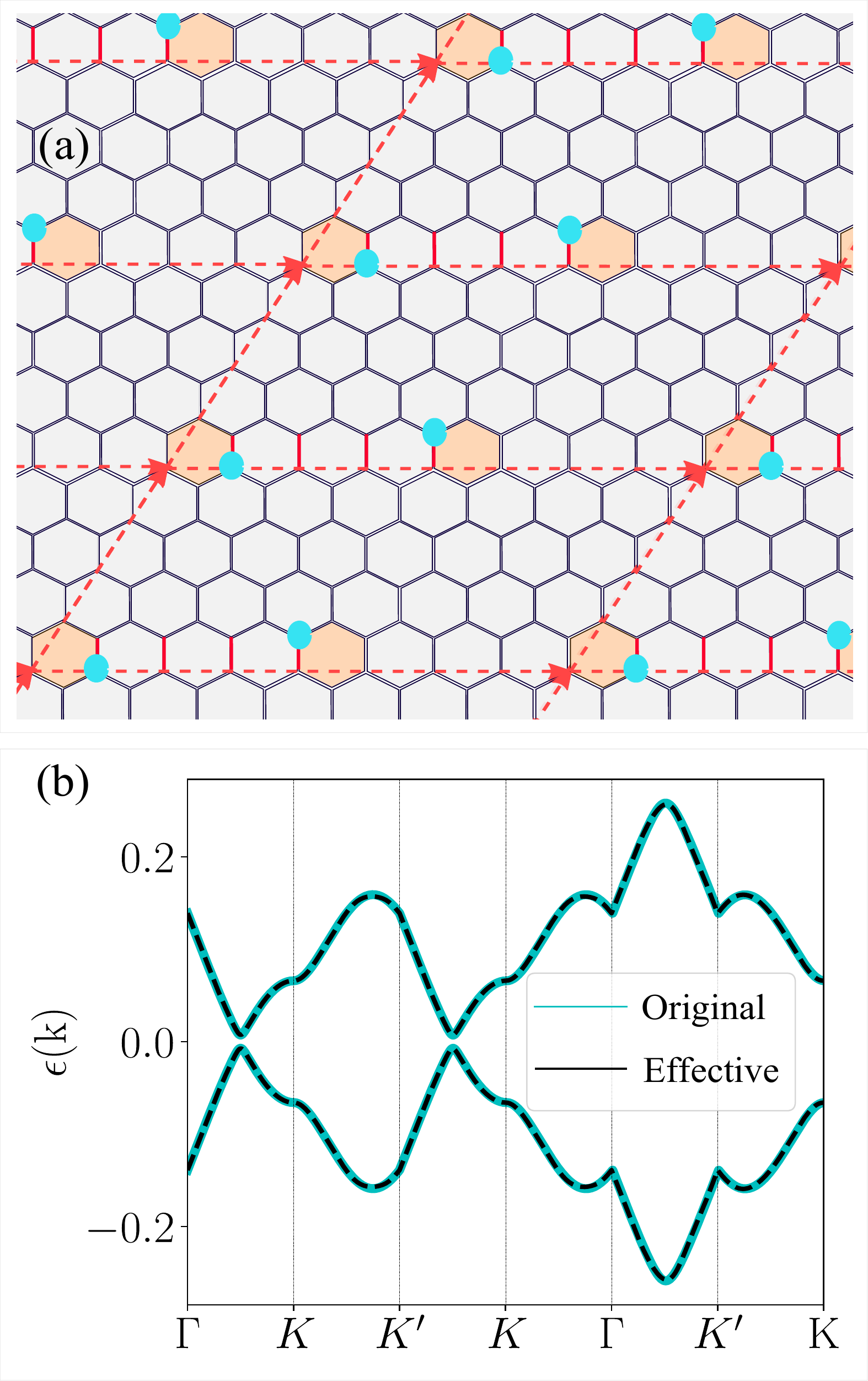}
\caption{(a) The locations of the Wannier centers (cyan color)  are shown to be adjacent to the $\pi$ flux plaquettes for the $4 \times 4$ supercell. (bottom)  (b) The band structures from the original band structure and the effective Hamiltonian are shown for $K/J =0.01$ on $4 \times 4$ configuration.}
\label{fig:wann}
\end{figure}

We proceed by constructing an effective Hamiltonian $H_{\rm eff}$ to capture the behavior of the two low-energy bands ($\alpha=1$, and $a=\pm$) and the localization of the corresponding Wannier orbitals $|\pm,{\bf R}\rangle$. Here, we focus on the supercell results for the $d=4$ case for $K\ne 0$, i.e., a gapped system. We note that stabilizing a $4 \times 4$ flux configuration typically requires third nearest neighbor hoppings,\cite{Batista2019,Zhang_BAtista_2020} while still preserving the essential physics including Majorana band structure, flat bands, topology, and fractional Chern states, though these features may appear at different values of $K$. The construction of the effective band mirrors the example provided in Sec.~\ref{Sec:Example}, utilizing the same sets of nearest neighbors, except the ${\bf G}_{1,2}$ are different here. We expand the Hamiltonian up to several nearest neighbors, incorporating the coupling constants, and the specific values of the TB parameters are given in Appendix~\ref{App:Parameters}. The fitting yields a near-perfect fit of the energy dispersions  $\pm E({\bf k})$ to the original supercell results, see Fig.~\ref{fig:wann}(b). For the fitting procedure, we use the Wannier90 code \cite{Wannier90_MOSTOFI}. A key advantage of using the Wannier90 code is its ability to provide the real-space projection of the $w_{\pm,{\bf R}}({\bf r})$, and their corresponding spread functions $\Delta{\bf r}_{\pm}$. Due to the non-zero Chern numbers of these flat bands, identifying their Wannier centers presents a challenge due to global gauge obstruction. In the effective theory, this gauge obstruction is evaded by choosing uniform flux $W_S$ at all unit cells.  The flux condition deduced in the effective Hamiltonian in Eq.~\eqref{eq:HMhoneycom} gives constraints on the fitting parameters. This method is discussed in Appendix~\ref{App:Wannier}

As anticipated, the Majorana Wannier orbitals $w_{\pm,{\bf R}}({\bf r})$ are localized at the $\Z_2$ flux sites, as shown in Fig.~\ref{fig:wann}(a). In fact, an isolated Majorana orbital with a $\Z_2$ flux makes a composite Majorana (zero-energy) Wannier orbital,\cite{KITAEV2006,Trebst_2012,Nasu_2023} and their periodic arrangements with finite nearest-neighbor tight-binding hopping strength produce dispersive bands. Within the original supercell, these two states were linked by the string operator that connects the $\Z_2$ flux pair. However, in effective theory, $w_{\pm,{\bf R}}({\bf r})$ represents the two basis states of a unit cell. Since the effective gauge fields reside on the links connecting lattice sites, there is no gauge field directly coupling the two Majorana orbitals within a unit cell. Their coupling is not parametrized by the anti-symmetric onsite interaction term $i\mathcal{K}_0=K_0\sigma^z+K'_0\sigma^y$, where $\sigma^{\mu}$ matrices are defined in the $a=\pm$ Wannier orbital basis, as in the example case given in Sec.~\ref{Sec:Example}. Here, $K_0$ captures the onsite energy difference of the two orbitals, while $K_0'$ describes the intra-unit cell coupling between them. The remaining terms in Eq.~\eqref{eq:HMhoneycom} remain the same. 

\subsection{Chern number and Quantum metric}\label{Sec:Chern}

\begin{figure}[ht] 
\centering
\includegraphics[height=1.0\linewidth,width=0.75\linewidth]{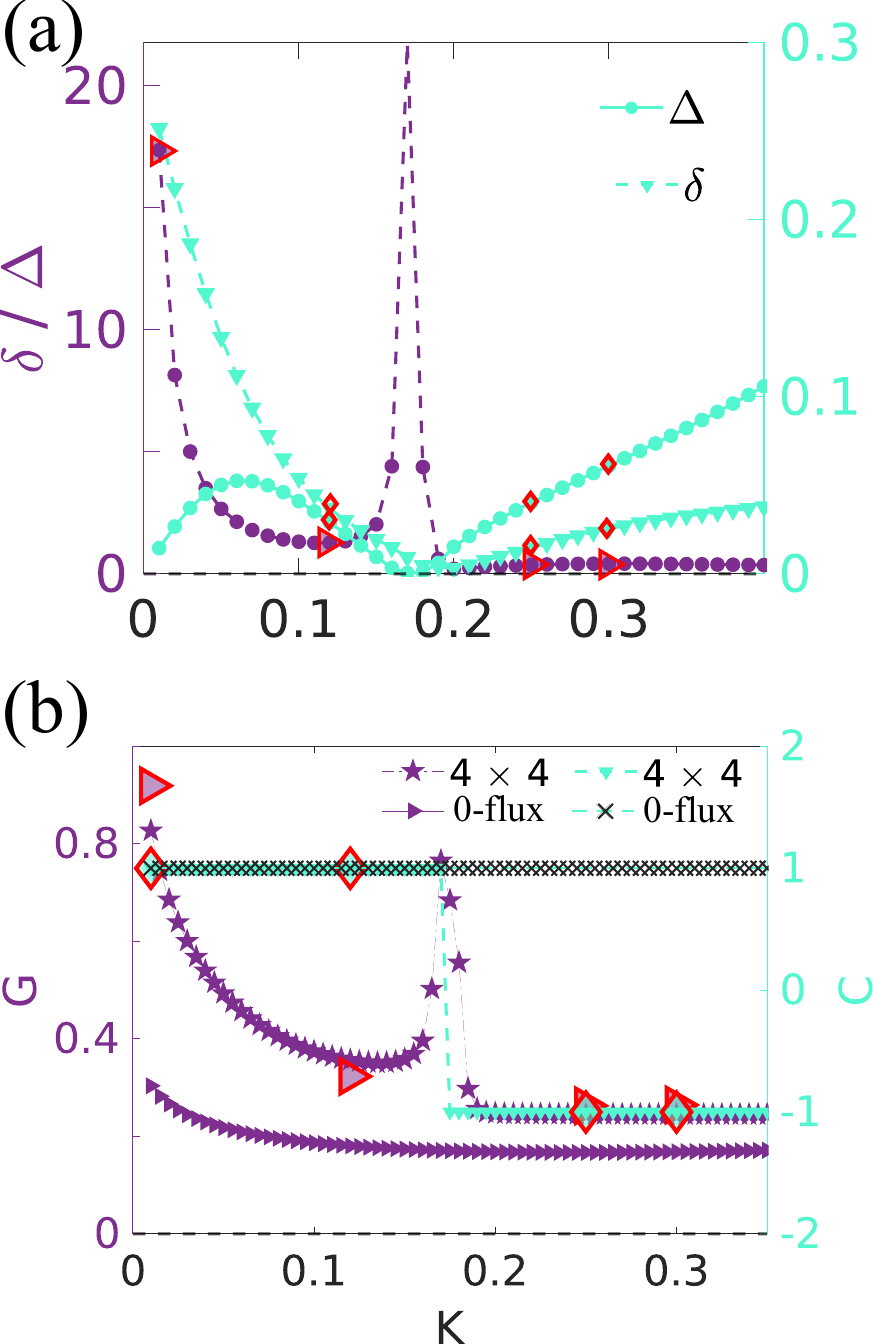}
\caption{(a) Band gap ($\Delta$), bandwidth ($\delta$), and the flatness ratio ($\delta/\Delta$) are plotted as a function of $K$ from original model in Eq.~\eqref{eq:ham2} (verified values from the effective model in Eq.~\ref{eq:HMhoneycom} are highlighted in red) for the $4\times 4$ supercell. (b) Computed values of the quantum metric invariant ($G$) and Chern number ($C$) are plotted as a function of $K$ and compared with zero-flux configuration. Two topological phase transitions are located here: at K=0, we have a topological phase transition from trivial to non-trivial phase with Dirac cone degeneracy, and another one at $K \thickapprox 0.175J$ between two non-trivial topologies with a flat-band degeneracy whose dispersions are shown in Fig.~\ref{fig:Flat}(d).}
\label{fig:gap4by4}
\end{figure}

This work suggests that the $\Z_2$ flux quantization condition within the supercell leads to similar properties as in the integer $U(1)$ flux quantization condition in a magnetic Brillouin zone in the TKNN theory for $U(1)$ quantum Hall insulators.\cite{Thouless_1982} Both mechanisms lead to a finite Chern number for each Majorana band, a characteristic that persists in both the full supercell Hamiltonian and the resulting effective theory. 

Alternatively, we can interpret this behavior by parametrizing the eliminated states either as a geometry term in the wave function or as gauge fields within the Hamiltonian. A trivial topological space would correspond to a product state between the low-energy states and the eliminated states. Conversely, a non-trivial topology signifies entangled states between them. In the case of flat band geometry encountered here, a non-trivial topology necessarily arises.

To ensure consistency, we compute the Chern number for both cases. In the supercell case, we employ the projector $\mathcal{P}({\bf k})$ to compute the Chern number ($C$) using Eq.~\eqref{eq:Chern}. Similarly, for the effective Hamiltonian, $C$ is obtained using it projector $\tilde{\mathcal{P}}({\bf k})=|\tilde{{\bf p}},{\bf k}\rangle\langle \tilde{{\bf p}},{\bf k}|$, where $|\tilde{{\bf p}},{\bf k}\rangle$ are the eigenstates of $\mathcal{H}_{\rm eff}({\bf k})$. We present the results for the $d=4$ case and compare them with the zero-flux ($d=0$) scenario. Our calculations consistently reveal that the $K\ne 0$ case exhibits a Chern number of  $C=+1$ for both $d=0$ and $d=4$ cases (for the $-E({\bf k})$ band). However, in the $d=4$ supercell, a sharp transition from $C=+1$ to $-1$ occurs at the critical point 
 $K=0.175J$, where the band gap ($\Delta$) closes and reopens.  The underlying physics governing the Chern number transition obtained from the effective Hamiltonian $\mathcal{H}_{\rm eff}({\bf k})$ is analogous in which the uniform flux sector changes from $W_s=+1$ to $W_s=-1$. The transition is different from the gap closing and reopening at a single Dirac point; the involvement of the flat band at this transition point to a novel topological phase transition. 

The influence of flat band physics and the geometry effect introduced by the eliminated high-energy states are effectively captured by the quantum metric ($\mathcal{G}({\bf k})$) term in Eq.~\eqref{eq:QMetric}. The corresponding invariant ($G$) is defined in Eq.~\eqref{eq:QMI} and is plotted in Fig.~\eqref{fig:gap4by4}(b) as a function of $K$. \footnote{The symmetric tensor $\eta_{\mu\nu}$ for the hexagonal lattice becomes 
$\eta_{\mu\nu} = \hat{\mathbf{G}_{\mu}}.\hat{\mathbf{G}_{\nu}} = \begin{pmatrix}
1 & -1/2 \\
-1/2 & 1 
\end{pmatrix}. $} As expected, the $G$ value for the zero-flux configuration exhibits no distinguishing features, reinforcing the notion that $G$ captures a distinct topological invariant arising from the projector, separate from the Chern number.  In the supercell case, however, $G$ displays an additional singularity at the gap-closing point of $K=0.175J$. Interestingly, both $G$ and the flatness ratio, $\delta/\Delta$, exhibit similar behavior. This suggests that the singularity in $G$ is sensitive to the characteristics of the flat band, particularly the presence of extensive band degeneracy. 

It's important to distinguish between the phase transition properties of interacting and non-interacting systems. In interacting theories, a second-order phase transition is characterized by the appearance of gapless collective modes and singular correlation functions.  In contrast, in this non-interacting theory, an extensive degeneracy emerges at the flat bands. Here, the flat bands exhibit maximal entanglement with the eliminated high-energy bands, and consequently,  we expect this unique phase transition feature to be reflected in the topological entanglement entropy.\cite{TEE_Kitaev,Levin_TEE,Nehra_2020,Kuno_2020}

A non-zero Chern number $C$ signifies an obstacle in smoothly changing the wavefunction's phase (${\rm arg}(w_{a,{\bf R}})$) throughout the material.\cite{Christian2007,Monaco2018}  In contrast, a non-zero quantum metric $G$ directly affects how "spread out"( $\Delta {\bf r}_{a}$) the wavefunction is\cite{Peotta2015,Torma2022}. More generally, $\mathcal{G}$ puts constraints on how different parts of the wavefunction are correlated, and $\Delta {\bf r}_{a}$ is a type of correlation function. In the effective Hamiltonian $\mathcal{H}_{\rm eff}({\bf k})$, the winding number of the wavefunction is determined by the complex phase of the off-diagonal term $\Delta_I({\bf k})+ih_s({\bf k})$.  $\Delta_R({\bf k})$ acts like a Dirac mass term, which gives the inverse correlation length of the wavefunction, essentially defining its spread. At discrete Dirac points, all these terms simultaneously vanish at a single ${\bf k}-$point, while for the degenerate flat bands, they vanish at all ${\bf k}$-points. In our effective theory, $\Delta_{I,R}$, and $h_S$ are assumed to be polynomials of Bloch phases $z_{\bf R}({\bf k})$, which are a set of linearly independent basis functions. Consequently,  a flat band arises when all the coefficients in these polynomials, i.e., the TB parameters $\mathcal{K}_i$, become zero.

\subsection{Gauge invariant Mean-field theory for Fractional Chern insulator}\label{Sec:MF}

\begin{figure*}[ht] 
\centering
\includegraphics[height=0.475\linewidth,width=0.8450\linewidth]{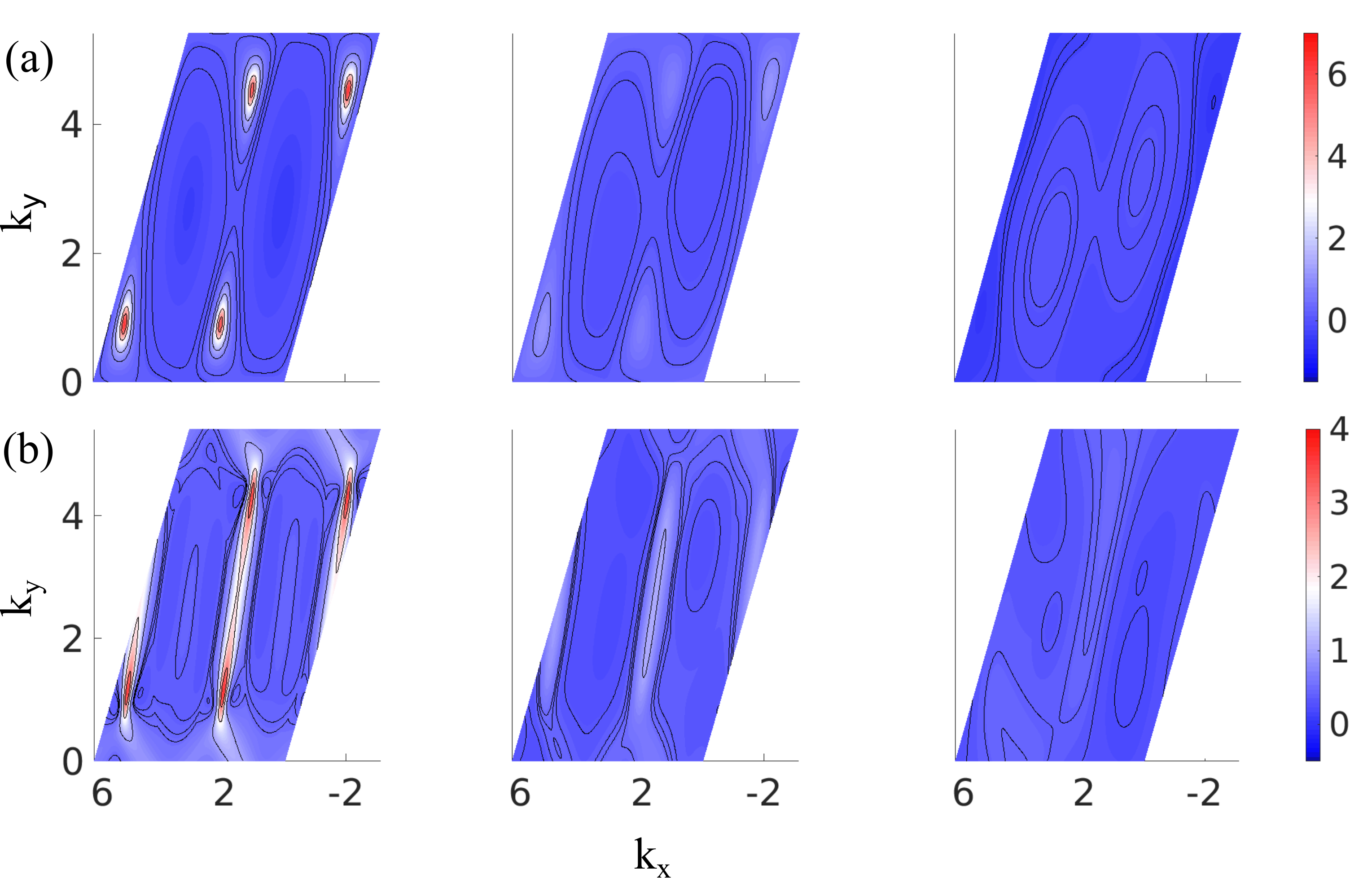}
\caption{We plot (a) the Berry curvature $G(\bf k)$ and (b) the difference $G(\bf k)-U(\bf k)$ from the trace condition (ii) in the first Brillouin zone. The plots in each column correspond to different $K$ values of 0.035, 0.12, and 0.22, from left to right.}
\label{fig:BCurvTDiff}
\end{figure*}

The interplay between $\mathcal{U}$ and $\mathcal{G}$ creates a promising platform for realizing fractional Chern insulating states through interactions \cite{Neupert_2011,Tang_2011,Sun_2011}. The current understanding of fractional Chern insulating state primarily relies on numerical results. \cite{FCI_Neupert_2015,Neupert_2011,Tang_2011,Sun_2011,Regnault_2011,FCI_Liu_2012,PARAMESWARAN_2013,FCI_Laumann_2015,FCI_Cooper_2015,FCI_Behrmann_2016,FCI_Thomale_2017,FCI_Zhao_2024} In this section, we propose a mean-field theory that predicts the emergence of a fractional Chern number in Majorana bands.  

We introduce a mean-field theory to split the Majorana flat bands by forming a density wave order state. A density wave state effectively folds the BZ into a reduced BZ, with the ordering vector ${\bf Q}$ defining the new reciprocal lattice vectors. The original Chern bands transform into a main band and folded (or shadow) bands within the reduced BZ. These bands share partial occupation density. This process leads to a fascinating consequence: a single, split Chern band becomes partially filled with a finite interacting gap separating it from another partially filled Chern band. \cite{Tang_2011,Neupert_2011,Sun_2011,Gupta_2017}

Here, we begin by verifying if the bands fulfill the essential criteria for an ideal 'vortexable' band, a prerequisite for realizing a fractional Chern phase. \cite{PARAMESWARAN_2013,Roy_2014,Simon_2020,Ledwith2023} Interestingly, the low-energy Majorana bands in the supercell configurations fulfill those conditions: i) uniform-in-${\bf k}$ Berry curvature $U({\bf k})={\rm Tr}\mathcal{U}_{\mu\nu}(\mathbf{k})$ and ii) the trace condition $G({\bf k})=\frac{1}{2}\sum_{\mu}{\rm Tr}\mathcal{G}_{\mu \mu}({\bf k}) \simeq U({\bf k}),$ $\forall \mathbf{k}$-point. Another quantity that measures how good is a flat band is called the flatness ratio ($\delta / \Delta$), where $\delta$ is the bandwidth of each flat band, and $\Delta$ is the band gap between the two flat bands under consideration. When $\delta/\Delta$ is less, as shown in Fig.~\ref{fig:gap4by4}(a) for the $4 \times 4$ supercell, the conditions for ideal Chern bands are satisfied more accurately. Fig.~\ref{fig:BCurvTDiff}(a) shows $U({\bf k})$ and Fig.~\ref{fig:BCurvTDiff} (b) gives the difference $G({\bf k})-U({\bf k})$ in ii) at three different $K$ values, 0.035, 0.12, and 0.22 from left to right. $\delta/\Delta$ decreases with increasing $K$, and $U({\bf k})$ becomes more uniform. The same is true for the trace condition in (ii), i.e., the difference $G({\bf k})-U({\bf k})$ is much less with increasing $K$.

Having established that the Chern Majorana bands in the supercell settings are prone to fractionalization, we now include an interaction term within the effective Hamiltonian $H_{\rm eff}$ in Eq.~\eqref{eq:HMR}, or more specifically in Eq.~\eqref{eq:HMhoneycom} for the case of  $2\times 2$ Honeycomb lattice. Interestingly, a $\Z_2$ `electric field' operator introduced below Eq.~\eqref{eq:ham} mediates a quartic Majorana interaction in the honeycomb lattice with three nearest neighbors.\cite{Kitaev2003,Zhang_BAtista_2020} The corresponding operator for the Wannier orbitals, in general, reads as $X_{\bf R}=\prod_{\{{\bf R}^{(1)}\}\in d_1}\mathcal{U}_{{\bf R},{\bf R}^{(1)}}$, where it is reminded that ${\bf R}^{(1)}$ is a set of first nearest neighbor sites containing $ d_{1}$ elements with respect to the ${\bf R}$ site. $X_{\bf R}$ is a gauge-dependent operator, and it couples to all the Majoranas sitting at ${\bf R}$ and ${\bf R}^{(1)}$, which then becomes gauge invariant. In the original Kitaev model with the small magnetic field, it was shown that such a term arises in the same third-order perturbation term as in Eq.~\eqref{eq:ham} and has the same coupling constant of $iK$. Here in the effective model, such a term can arise from the multiple virtual hoppings to the eliminated states and can be obtained by higher-order polynomial fittings along the line done for the quadratic term in Eq.~\eqref{eq:DefT}. We consider a general coupling term of $i\mathcal{K}_1'$ and write the interaction term on a honeycomb lattice and $a=\pm$ particle-hole Majorana pair in a single $\alpha=1$ orbital  state as
\begin{eqnarray}
    H_{\rm int} = -i\mathcal{K}_1' \sum_{{\bf R},a,a'=\pm} iu_{{\bf R},{\bf R}^{(1)}_1}u_{{\bf R},{\bf R}^{(1)}_2}u_{{\bf R},{\bf R}^{(1)}_3}\nonumber\\
    \times c_{a,{\bf R}}c_{a',{\bf R}^{(1)}_1}c_{a',{\bf R}^{(1)}_2}c_{a',{\bf R}^{(1)}_3}.
    \label{eq:Hint}
\end{eqnarray}
Consistent with the Elitzur’s theorem \cite{Elitzur_1975}, we can write down two gauge invariant mean-field order parameters, defined as follows:
\begin{eqnarray}
\Omega_{1} &=& \frac{i\mathcal{K}_1'}{3}\sum_{i=1}^3\big\langle u_{{\bf R},{\bf R}^{(1)}_i}c_{a,{\bf R}}c_{a',{\bf R}^{(i)}_i}\big\rangle,
\label{eq:MFOP1}\\
\Omega_{2} &=& -\frac{i\mathcal{K}_1'}{3}\sum_{i\ne j =1}^3\big\langle u_{{\bf R}^{(1)}_i,{\bf R}^{(1)}_j}c_{a',{\bf R}^{(1)}_i}c_{a',{\bf R}^{(1)}_j}\big\rangle.
\label{eq:MFOP2}
\end{eqnarray}
Here, the expectation value is self-consistently evaluated within the mean-field ground state. We have also assumed the order parameters $\Omega_{1,2}$ to be independent of ${\bf R}$ (uniform phase), and hence, it is dropped or from the L.H.S. of the above equation.
Both order parameters involve two Majorana operators, and the gauge field links them so that the order parameter remains gauge invariant. The resultant mean-field interaction (gauge-invariant) Hamiltonian becomes 
\begin{eqnarray}
    H_{\rm int}^{\rm (MF)} &=&  ~~~~~ i\Omega_{2}\sum_{{\bf R},i,a'}u_{{\bf R},{\bf R}^{(1)}_i}c_{a,{\bf R}}c_{a',{\bf R}^{(1)}_i}.\nonumber\\
   && 
  + i\Omega_{1}\sum_{{\bf R},i\ne j,a\ne a'}u_{{\bf R}^{(1)}_i,{\bf R}^{(1)}_j}c_{a',{\bf R}^{(1)}_i}c_{a',{\bf R}^{(1)}_j},
   \label{eq:HintMF}
\end{eqnarray}
In the second term in Eq.\eqref{eq:MFOP2} and the first term in Eq.~\eqref{eq:HintMF}, we have implemented the relation: $u_{{\bf R}^{(1)}_i,{\bf R}^{(1)}_j}=-u_{{\bf R},{\bf R}^{(1)}_i}u_{{\bf R},{\bf R}^{(1)}_j}$. In the Fourier space, we assume $\Omega_{1,2}$ breaks the translational symmetry to a staggered density wave state at a fixed wavevector ${\bf Q}$ such that $\Omega_1({\bf Q})=\sum_{{\bf R}_i-{\bf R}^{(1)}_j}\Omega_1 e^{i{\bf Q}\cdot({\bf R}_i-{\bf R}^{(1)}_j)}$, and $\Omega_2({\bf Q})=\sum_{{\bf R}-{\bf R}^{(1)}_j}\Omega_2 e^{i{\bf Q}\cdot({\bf R}-{\bf R}^{(1)}_j)}$. While the above gauge-invariant mean-field theory admits various generalizations, we focus on a simpler case here to illustrate the emergence of partially filled Chern bands within this framework.

Adding $H_{\rm int}^{\rm (MF)}$ to the effective Hamiltonian $H_{\rm eff}$ in Eq.~\eqref{eq:HMhoneycom}, we can express the matrix form of the Hamiltonian in the spinor $(c_{+}({\bf k})~ c_{-}({\bf k})~ c_{+}({\bf k}+{\bf Q})~ c_{-}({\bf k}+{\bf Q})^T$ for the $\alpha=1$ band as
\begin{eqnarray}
\mathcal{H}_{\rm MF} ({\bf k})
&=& \left(\begin{array}{cc}
     {\mathcal{H}_{\rm eff}(\mathbf{k})} & i{\Omega(\mathbf{Q})}  \\
    {\rm H.c.}  &  {\mathcal{H}_{\rm eff}(\mathbf{k}+\mathbf{Q})}
\end{array}\right), \label{eq:HMF}\\
\Omega(\mathbf{Q}) 
&=& \left( \begin{array}{cc}
        \Omega_1({\bf Q}) & \Omega_2({\bf Q}) \\
        \Omega_2^*({\bf Q}) & \Omega_1^*({\bf Q})
        \end{array}
        \right). \nonumber
\label{eq:OP}
\end{eqnarray}
The above form of $\Omega({\bf Q})$ maintains the particle-hole symmetry of the Hamiltonian. The eigenvectors of the mean-field Hamiltonian in Eq.~\eqref{eq:HMF} is used in Eqs.\eqref{eq:MFOP1},\eqref{eq:MFOP2} to calculate these order -parameter self-consistently.

We test out results for a simpler commensurate density wave order for $\mathbf{Q}= \mathbf{G}_2/2$, and $\Omega_1 = 0.01, \Omega_2=0.015$ with $K=0.12$. The corresponding energy eigenvalues of $\mathcal{H}_{\rm MF} ({\bf k})$ split the particle-hole symmetric eigenvalues $\pm E_{\bf k}$ into four bands with a finite gap between all of them. The Chern number of each band now corresponds to $C_n=\sum_{{\bf k}\in {\rm RBZ}}\sum_{a}\mathcal{V}_{na}^{\dagger}({\bf k})\mathcal{V}_{na}({\bf k}){\rm Tr }_a\mathcal{U}_{12}({\bf k})$, where $\mathcal{V}$ corresponds to the eigenvectors of $\mathcal{H}_{\rm MF}$, and  ${\rm Tr }_a$ corresponds to the trace operation with the $a=\pm$ eigenvector on the Berry curvature given in Eq.~\eqref{eq:BCurvature}. RBZ corresponds to the reduced BZ in the density wave state. The obtained values of the Chern number for all four bands are  $C= 0.129, 0.505, -0.873$, and $-0.4$. 

\section{Summary and outlook}
Our work has two major advancements: (i) We construct a lattice model for generalized fractional orbitals hopping under tailored gauge potentials, and (ii) We apply this model to Majorana fermions in the flux-crystalline phase of the Kitaev model, exploring interaction effects and fractional Chern states via a mean-field approach.

In the first part of the model development, we have kept the theory general for any fractional quasiparticles. This approach is extendable to other fractional quasiparticles and represents a significant step forward in addressing long-standing challenges like Wannier obstructions. The key idea lies in starting with a well-defined Wannier orbital ansatz $-$ a topologically trivial product state $-$ and then sculpting the desired topology and quantum metric through an effective Hamiltonian that incorporates a projector and a variational geometric potential. Secondly, we derive the tight-binding gauge fields from the quadratic terms in the Taylor expansion (Eqs.~\eqref{eq:TBH_gauge},\eqref{eq:DefT} ) and the interaction terms from the quartic terms (Eq.~\eqref{eq:Hint}). This elegantly transfers the constraints imposed by the projection operator on Wannier orbital correlations to the correlations encoded within the emergent geometric potential. This makes the formalism general and applicable to generic fractional states. 

We applied our framework to Majorana orbitals in the flux-crystalline phase of the Kitaev model, focusing on their interaction effects and fractional Chern states through a mean-field method. While the full dispersion of Majorana bands in various flux superlattices of the Kitaev model has been extensively studied \cite{Nasu2021,Koga2023,Batista2019,Vojta2021,Vidal_2024,Trebst_2014}, a crucial missing piece has been a proper tight-binding description, including the quantum metric and the emergence of fractional Chern states.  Our work fills this gap with several key advancements: (a)  We constructed a TB model for Majorana orbitals by introducing a geometric potential $-$ arising from virtual hopping to eliminate high-energy states $-$ and systematically derived both quadratic and quartic interactions mediated by this potential. We also elucidated the conditions under which these potentials manifest as a $\Z_2$ gauge field and quantum metric, ensuring flux preservation and consistency with the original supercell Hamiltonian.  (b) We then analyzed the evolution of the Berry curvature and quantum metric of the effective theory as flat bands emerge with increasing nearest-neighbor hopping. Strikingly, we uncovered a novel critical point where the quantum metric diverges, signaling a phase transition between $C=\pm 1$ phases within the same band. This hallmark of flat bands with extensive degeneracy suggests, even at the non-interacting level, a state of maximal entanglement between low- and high-energy states. (c) Finally, building on the existence of flat bands with a divergent quantum metric, we developed a mean-field theory to describe a gauge-invariant Majorana density wave order. The resulting split Chern bands enable partial filling with a gapped spectrum relative to other Majorana bands. This mean-field theory provides an analytically tractable framework for understanding fractional Chern insulator states in these fascinating systems. 

The proposals for attaining control over the creation/annihilation of $\Z_2$ fluxes are reviewed first. Theoretically, approximate studies predict the existence of flux superlattices in the Kitaev model with further nearest neighbor hoppings \cite{Trebst_2014} and also with the applied magnetic field \cite{Vojta2021}. An unbiased DMRG study of the Kitaev model with external magnetic fields provided evidence for the emergence of various flux superlattices.\cite{yogendra2023emergent} In general, the Heisenberg model on some frustrated lattices can host various flux crystals predicted by projective symmetry group analysis, giving rise to diverse spin liquid phases. \cite{wen_psg_2002,Wang_Vishwanath_2006,song_vishwanath_2020} Experimentally, creating and stabilizing these flux superlattices is challenging. In recent times, it has been theoretically proposed in Ref.~\onlinecite{Motome2021} that the local modulation of exchange interactions by introducing Dzyaloshinskii-Moriya interactions flips the sign of local bond interactions. That produces a $\Z_2$ flux pair.  Further desired configurations are obtained by creating/annihilating the sequence of pairs in neighboring plaquettes. In general, superconducting quantum interference device (SQUID) microscopy is helpful in experimentally visualizing these vortex networks.\cite{Kalisky_2022VisualSNet} 

This work centers on constructing a gauge-field mediated TB model for fractional particles. Since fractional/entangled excitations do not exist by themselves, their combinations must produce electronic states. Alternatively, one can view these fractional particles as residing within a medium of gauge fields, either confined or deconfined.  Our focus is twofold: understanding the origin of these gauge fields and establishing a systematic framework for their parameterization within a TB model.  The emergence of such gauge fields comes from the projection operation used to eliminate high-energy states. This operation effectively imposes constraints, leading to restricted dynamics and correlation functions pertaining to fractional particles. These constrained dynamics can give rise to more phenomena such as a distinct type of quantum glass,\cite{yogendra2023emergent} or deconfined critically,\cite{Senthil_Deconf_2004,Senthil_Deconf_Review} or extensive degeneracy in the formation of flat bands, leading to novel topological critically in the noninteracting theory as observed here. 

Our approach deviates from conventional methods by introducing a gauge potential directly within the Hamiltonian through a superexchange mechanism. This mechanism gives rise to gauge-mediated tight-binding (TB) hoppings arising from the anti-symmetric part of the superexchange potential. Notably, an additional constraint can be readily incorporated to ensure flux preservation and topology without worrying about the maximal localization of the Wannier orbitals.  The detailed constructions are provided in Sec.~\ref{Sec:TBModel} applies to a general $SU(N)$ gauge field, $\mathcal{U}$, and its corresponding fractional particle. In this  specific work, we have inserted the $\Z_2$ valuedness of the $\mathcal{U}$ operators towards the end and in the particle-hole symmetry of $\mathcal{H}_{\rm eff}$, which ensures real Majorana states in real space. Therefore, it will be rather straightforward to generalize the TB theory to the $SU(N)$ gauge field coupled to other fractional particles. These effective models of fractional particles are useful in constructing more precise spectroscopic probes for their detection.

Our analysis reveals an interacting critical point within the theory as a function of the second nearest-neighbor hopping strength ($K$). At this critical point, the quantum metric diverges, and the Chern number exhibits a transition between +1 and -1. This signifies a potential singularity in the entanglement entropy spectrum, where the entanglement is maximal between the low-energy and the eliminated high-energy states. Quantifying this entanglement spectrum in terms of gauge-mediated TB parameters remains an intriguing challenge for future investigations. 

Finally, leveraging the insights from the effective Hamiltonian, we propose a mean-field theory for a gauge-invariant Majorana density wave order. In conventional gauge theories, the mean-field order parameter is subject to an additional constraint arising from the requirement of gauge invariance. This constraint often presents significant challenges within the geometric framework, leading to a reliance on numerical methods for studying fractional Chern insulator states. Our effective Hamiltonian, however, allows us to derive a self-consistent mean-field theory for Majorana fermions. This method paves the way for future investigations into more exotic interaction effects within both $\Z_2$ and $SU(N)$ gauge theories. 

\section*{Acknowledgments}
K.B.Y. thanks Partha Sarathi Rana for his help with the Wannier90 code. The work is supported by research funding from the S.E.R.B. Department of Science and Technology, India, under Core Research Grant (CRG) Grant No. CRG/2022/00341m and under I.R.H.P.A Grant No. IPA/2020/000034. We also acknowledge the computational facility at S.E.R.C. Param Pravega under NSM grant No. DST/NSM/R\&D HPC Applications/2021/39. 
 
\appendix
\section{Matrix Elements of the Supercell Hamiltonian}\label{App:Supercell}

Here, we explicitly give the matrix elements of the $2N\times 2N$ matrices $\mathcal{H}_{IJ}$ as 
\begin{widetext}
\begin{eqnarray}
    \mathcal{T}_{II}=\left(\begin{array}{ccccccc}
         0 &Ju_{1,2} & Ku_{1,2}u_{2,3} & 0 &  \dots &0 &0 \\
         -Ju_{1,2} & 0 & Ju_{2,3} & Ku_{2,3}u_{3,4} & \dots &0 &0\\
         -Ku_{1,2}u_{2,3} & -Ju_{2,3} & 0 & Ju_{3,4} & \dots &0 &0\\
         0 & -Ku_{2,3}u_{3,4} & -Ju_{3,4} & 0 &  \dots  &0&0\\
         \vdots & \vdots & \vdots & \vdots & \ddots &0&Ju_{2N-1,2N} \\
         0 & 0 & 0 & 0& 0&-Ju_{2N-1,2N} &0
    \end{array}\right),  
    \end{eqnarray}
    and
    \begin{eqnarray}
    \mathcal{T}_{IJ}=\left(\begin{array}{ccccc}
      0  &  0   &    0  & \dots &0 \\
         \vdots & \vdots & \vdots & \ddots &\vdots\\
         0  &  0   &    0  & \dots &0 \\
         Ku_{2N-1,2N}u_{2N,1} & 0 & 0& \dots &0 \\
         Ju_{2N,1} & Ku_{2N,1}u_{12} & 0&\dots&0   
    \end{array}\right).
\end{eqnarray}
The matrix elements in the momentum  space become
\begin{eqnarray}
    \mathcal{H}({\bf k})=i\mathcal{T}_{II} + \left(\begin{array}{ccccccc}
         0 & 0 & 0&\dots &  0& \mathcal{K}_{1,2N-1}({\bf k}) &\mathcal{J}_{1,2N}({\bf k}) \\
         0  &  0   & 0&   \dots &0& 0&  \mathcal{K}_{2,2N}({\bf k})\\
         0  &  0   &   0& \dots &0& 0&  0\\
         \vdots  &  \vdots&\vdots   &  \ddots&  \vdots &\vdots&  \vdots\\
         0  &  0   &  0 & \dots &0& 0&  0\\
         \mathcal{K}^*_{1,2N-1}({\bf k}) & 0& 0&\dots & 0& 0&0 \\
         \mathcal{J}^*_{1,2N}({\bf k}) & \mathcal{K}^*_{2,2N}({\bf k}) & 0&\dots&0  &0 &0
    \end{array}\right),
\end{eqnarray}
where $\mathcal{K}_{1,2N-1}({\bf k})=iKu_{2N-1,2N}u_{2N,1}e^{-i{\bf k}\cdot({{\bf R}_{1}}-{\bf R}_{2N-1})}$, $\mathcal{J}_{1,2N}({\bf k})=iJu_{2N,1}e^{i{\bf k}\cdot({{\bf R}_{1}}-{\bf R}_{2N})}$,  $\mathcal{K}_{2,2N}({\bf k})=iKu_{2N,1}u_{1,2}e^{-i{\bf k}\cdot({{\bf R}_{2}}-{\bf R}_{2N})}$.
\end{widetext}

\section{Matrix-Elements of the Tight-Binding Model}\label{Sec:AppBdG}

To ensure the Majorana operators are well defined, and the corresponding complex fermion operators recover the $U(1)$ gauge fields, we transform the Hamiltonian on a complex fermion basis. For each $\alpha-$orbital, the two Majorana orbitals pair, $a=\pm$ constitute a complex fermion particle-hole pair states at different momenta  as
\begin{equation}
|\alpha,a,{\bf k}\rangle=+e^{i\phi_{\alpha,a}} |\alpha,-,-{\bf k}\rangle_F+e^{-i\phi_{\alpha,a}} |\alpha,+,{\bf k}\rangle_F.
\label{eq:TBMFDef}
\end{equation}
where $\phi_{\alpha,+}=0$ and $\phi_{\alpha,-}=\pi/2$. $|\alpha,\pm,{\bf k}\rangle_{F}$ are the complex fermionic hole and particle excitation states defined as $|\alpha,+,{\bf k}\rangle_F= f_{\alpha,{\bf k}}|G\rangle$ and  $|\alpha,-,{\bf k}\rangle_F = f^{\dagger}_{\alpha,{\bf k}}|G\rangle$ with $f_{\alpha,{\bf k}}$, and $f_{\alpha,{\bf k}}^{\dagger}$ corresponding annihilation and creation operators of complex fermions from some grand canonical ensemble state $|G\rangle$. It is easy to see that the corresponding Majorana and complex fermion operators are local in real space as 
$|\alpha,a,{\bf R}\rangle=e^{i\phi_{\alpha,a}} |\alpha,-,{\bf R}\rangle_c+e^{-i\phi_{\alpha,a}} |\alpha,+,{\bf R}\rangle_c$. Note that $|\alpha,a,{\bf R}\rangle$ are the physical Majorana states in real space, whereas complex fermions correspond to physical states in both real and momentum spaces. There is an inherent gauge obstruction between the two Majorana orbitals by a phase difference of $\phi_{\alpha}=\phi_{\alpha,+}-\phi_{\alpha,-}$. We have kept this phase difference to orbital independent, but it can be generalized to be orbital dependent, which may commence interesting properties.

We construct a $2P$ dimensional Majorana spinor as $|\alpha,+,{\bf k}\rangle\oplus|\alpha,-,{\bf k}\rangle$ and  complex fermionic particle-hole symmetric Nambu spinor $|\alpha,+,{\bf k}\rangle_F\oplus|\alpha,-,-{\bf k}\rangle_F$.The transformation between them is defined by the unitary operator: 
\begin{eqnarray}
    \mathcal{S}=\frac{1}{2}\left(\begin{array}{cc}
     1 & i \\
     1 & -i 
\end{array}\right)\otimes I_{P\times P}. 
\end{eqnarray}
The Majorana Hamiltonian given in the main text is written generally as
\begin{eqnarray}
H_{\rm eff}=h_A \otimes\mathcal{I}_{2\times 2} +
 \left(\begin{array}{cc}
     {\Delta}_R & {\Delta}_I+ih_S \\
     {\rm H.c} &  - {\Delta}_R
\end{array}\right),
\label{eq:HM}
\end{eqnarray}
where ${\bf k}$ dependence in all variables is kept implicit. Here $h_{S/A}({\bf k})=h({\bf k})\pm h^T({-\bf k})$, and $\Delta({\bf k})=\Delta_R({\bf k})+i\Delta_I({\bf k})$.  Then, by transforming this Hamiltonian to the complex fermion basis gives
\begin{equation}
    H_{\rm eff}^{\rm (F)}=\frac{1}{4}\mathcal{S}H_{\rm eff}({\bf k})\mathcal{S}^{\dagger}= \left(\begin{array}{cc}
     h({\bf k}) & \Delta({\bf k}) \\
     \Delta^{\dagger}({\bf k}) & -h^T(-{\bf k})
\end{array}\right).
\label{eq:HMC}
\end{equation}
This is a typical Bogoluybov-de-Gennes Hamiltonian in the particle-hole basis, where $h_{\bf k}$ is a $P\times P$ Hamiltonian for complex-fermions hopping, and $\Delta({\bf k})$ is the $P\times P$ matrix consisting of superconducting pairings of the complex fermions.  

The explicit form of the matrix elements of eq.~\eqref{eq:HM} can be written as
\begin{widetext}
\begin{eqnarray}
(\Delta_R)_{\alpha,\alpha'}({\bf k})&=& \sum_{p,p'=1}^{P} \langle \alpha,+,{\bf k} |p,+,{\bf k}\rangle \langle p,+,{\bf k}|(H+H')| p',+,{\bf k}\rangle \langle p',+,{\bf k}|\alpha',+,{\bf k}\rangle,
\label{eq:ME1}\\
(\Delta_I+iH_S)_{\alpha,\alpha'}({\bf k})&=& \sum_{p,p'=1}^{P} \langle \alpha,+,{\bf k} |p,+,{\bf k}\rangle \langle p,+,{\bf k}|(H+H')| p',-,{\bf k}\rangle \langle p',-,{\bf k}|\alpha',-,{\bf k}\rangle,
\label{eq:ME2}
\end{eqnarray}
\end{widetext}
Due to particle-hole symmetry, $|p,+,{\bf k}\rangle$ states give $|\alpha,+,{\bf k}\rangle$ Majorana orbital, while  $|p,-,{\bf k}\rangle$ states give $|\alpha,-,{\bf k}\rangle$ orbital, respectively. We define $(U_{\pm})_{\alpha,p}=\langle \alpha,\pm,{\bf k} |p,\pm,{\bf k}\rangle $ a $P\times P$ overlap matrix which consists of the probability amplitudes of the particle-hole symmetric eigenstates of the full Hamiltonian $|p,{\pm},{\bf k}\rangle$ to the effective Majorana orbital states. Note that $U$ is not a unitary operator as $|p,\pm,{\bf k}\rangle$ states are incomplete. We denote the matric elements of $H$ is $(\mathcal{D}_{\pm})_{p,p'}({\bf k})=\langle p,\pm,{\bf k}|H|p',\pm,{\bf k}\rangle=\pm E_p({\bf k})\delta_{p,p'}$. The matrix elements $H'$ can be written in second-order perturbation theory with respect to some gauge interaction/superexchange potential $V$ that makes the transition from the $\mathcal{P}$ to the $\mathcal{Q}$ states as
\begin{widetext}
\begin{eqnarray}
(\mathcal{H'}_{aa'})_{p,p'} &=& \langle p,a,{\bf k}|H'|p',a',{\bf k}\rangle=\frac{1}{2}\sum_{q,q'\in \mathcal{Q}} (V_a)_{pq}(V_{a'})_{q'p'}\left[\frac{{\rm sgn}(a)}{E_{p}-E_q}-\frac{{\rm sgn}(a')}{E_{q'}-E_{p'}}\right] + ....
\label{eq:H'}
\end{eqnarray}
\end{widetext}
Here $q\in \mathcal{Q}_{\pm}=\mathcal{I}-\mathcal{P}_{\pm}$ states are the particle-hole pairs outside the subspace of our interest. $ (V_{\pm})_{pq}=\langle p,\pm,{\bf k}|V|q,\pm,{\bf k}\rangle$ is the tunneling amplitude between the two eigenstates, our fitting parameters.    
$|p,{\pm},{\bf k}\rangle $ are the eigenstates of $H$ and hence are particle-hole symmetric. Interestingly,  $H'$ is not particle hole-symmetric in the $|p,{\pm},{\bf k}\rangle $ as it allows for transition between the particle-hole symmetric states in Eq.~\eqref{eq:ME2}, but its matrix elements must be particle-hole symmetric in the $|\alpha,\pm,{\bf k}\rangle$ states, by construction. If $\mathsf{C}$ is the particle-hole symmetric operator in the $|\alpha,\pm,{\bf k}\rangle$ basis, defined as $|\alpha,+,{\bf k}\rangle=\mathsf{C}|\alpha,-,{\bf k}\rangle$, the matrix elements transform as: $\mathsf{C}\mathcal{D}_+({\bf k})\mathsf{C}^{-1}=-\mathcal{D}_-({\bf k})$, $\mathsf{C}\mathcal{H}'_{++}({\bf k})\mathsf{C}^{-1}=-\mathcal{H}'_{--}({\bf k})$, and $\mathsf{C}\mathcal{H}'_{+-}({\bf k})\mathsf{C}^{-1}=-\mathcal{H}'_{-+}({\bf k})$.  Substituting them in Eqs.~\eqref{eq:ME1},\eqref{eq:ME2} we get $\Delta_R= -U_+(\mathcal{D}_++\mathcal{H'}_{++})U^{-1}_+=- U_-(\mathcal{D}_-+\mathcal{H'}_{--})U^{-1}_-$, and $(\Delta_I+i\mathcal{H}_S)= U_+\mathcal{H'}_{+-}U^{-1}_-=- U_- \mathcal{H'}_{-+}U^{-1}_+$, $\forall {\bf k}$.

\section{Numerical fitting procedure}\label{App:Wannier}

\subsection{Gauge Obstruction and spread function of Wannier Majorana states}\label{App:Spread}

Here, we address the issue of the gauge obstruction for Wannier orbitals of the electrons and how they transcend into the Wannier orbitals of fractional particles. Fixing a smooth global gauge for all Wannier orbitals of electrons within a unit cell can be hindered by several factors. Below, we discuss several such cases and their corresponding remedies. 

(a) Topological Insulators: In topological insulators, band inversion between two Wannier orbitals obstructs a global momentum-space gauge. Here, the Wannier orbitals differ by a well-defined local gauge connection, reflecting the non-trivial topology. For example, while a Chern number of 1 requires a band inversion within the Brillouin zone (BZ), its specific location can be shifted without affecting the overall topology (movable gauge obstruction). As discussed in \cite{TI_Kohmoto_1985,TI_Thonhauser_2006,Soluyanov2011,TI_Rui_2011,TI_Qi_2011,TI_Gunawardana_2024,TI_Xie_2024,wang2024_Landau,Vanderbilt_2012RMP}, this can be addressed by defining an appropriate gauge-fixing matrix.

(b)  Gapless Points: If a gapless point arises from a symmetry-protected degeneracy between two bands, it may not be readily movable (unless it is a gauge theory). \cite{GP_Strinati_1978,Rhim_2019,GP_Bergman_2008} This can be tackled by expressing the Wannier orbital states as superpositions within the degenerate manifold and carefully handling the singular point in the expansion coefficient (unitary matrix).

(c) Flat Bands.: Constructing Wannier orbitals becomes challenging for specific flat bands, particularly when they exhibit degeneracy with another band. In such cases, a complete set of localized compact Wannier orbitals may not be achievable. Instead, a combination of compact localized states and extended states might be necessary to form a complete basis set.\cite{Vafek2018,Herzog2022,Peotta_2015,FB_Flach_2014,FB_Dubail_2015,FB_Morales_2016,FB_Maimaiti_2017,FB_Read_2017,FB_Zhang_2020}

(d) Atomically Obstructed Insulators: This recently discovered class of (trivial or fragile) topological insulators presents a unique challenge. \cite{AOI_Bradlyn_2017,AOI_Po_2017,AOI_Schnidler_2021,AOI_Chen_2023} Here, each of the multiple Wannier orbitals must individually possess a sufficiently small spread function ($\Delta {\bf r}$) such that their combined spread stays confined within a unit cell.  

Can the aforementioned challenges be entirely overcome using Wannier orbitals for fractional particles within a gauge theory framework? The fractional particles of interest here arise from the superposition states of the original complex matter fermions. The fractional particles exhibit a physical separation in real space due to emergent local gauge fields and/or topology.  Their physical separation is linked by the gauge fields such as $W_p$, $X_v$, and $\mathcal{U}_{ij}$. For example, in the present case, the two Majorana orbitals are pinned at the two $\pi-$ flux pairs that are separated by a distance $d$. Therefore, in analogy with the atomically obstructed orbitals, the spread function $(\Delta {\bf r})$ associated with each Majorana Wannier orbital must be less than $d/2$ if there exists a finite trivial gap between the two Majorana bands. For the gapless case, $\Delta {\bf r}\sim d/2$, whereas in a topologically non-trivial case,  $\Delta {\bf r}>d/2$ such that the two Majorana Wannier orbitals overlap within the unit cell, and an intra-unit-cell gauge field between the two orbitals contains a winding or knot to produce the topological invariant. 

The gauge obstruction is incorporated within the eigenstates of the full Hamiltonian $|{\bf p},{\bf k}\rangle$ before fractionalizing them in the orbital states. This is done as in the standard method outlined in Ref.~\onlinecite{Marzari1997}. The procedure has two steps. First, we allow a unitary transformation $B({\bf k})$ to the eigenstates as $|\tilde{\bf p},{\bf k}\rangle=\sum_{p}B_{\tilde{\bf p},{\bf p}}({\bf k})|{\bf p},{\bf k}\rangle$ - which incorporates the singular gauge that needs to be added/subtracted from the global gauge. Next, we perform a smooth gauge fixing on the rotated states $|\tilde{\bf p},{\bf k}\rangle$ between the two nearest momenta differ by the grid size of $\delta{\bf k}=2\pi/L$, where $L$ is the sample length.  It turns out the spread function $\Delta {\bf r}=\sum_{{\tilde{\bf p}}}\langle {\bf r}^2\rangle_{\tilde{\bf p}}-\langle {\bf r}\rangle_{\tilde{\bf p}}^2$ is related to the overall matrix 
\begin{eqnarray}
M_{\tilde{\bf p},\tilde{\bf p}'}({\bf k},{\bf k}+\delta{\bf k})&=&\frac{1}{\sqrt{2N}}\sum_{i} \sum_{{\bf p},{\bf p}'}B^{\dagger}_{{\bf p},\tilde{\bf p}}({\bf k})\Gamma^{\dagger}_{i,{\bf p}}({\bf k}) \bar{z}_{{\bf t}_{i}}(\delta{\bf k})\nonumber\\
&&
\times \Gamma_{{\bf p}',i}({\bf k}+\delta{\bf k})B_{\tilde{\bf p}',{\bf p}'}({\bf k}+\delta{\bf k}),
\end{eqnarray}
where ${\bf t}_{i}$ are the positions of the original $2N$ Majorana sublattices in the full Hamiltonian within the ${\bf R}$ supercell. $\Gamma$ is the unitary matrix consisting of the eigenvector of the full Hamiltonian defined in Sec.~\ref{Sec:Supercell}. ${\bf R}_{\bf \alpha}$ is the position of the $\alpha$-sublattice within the supercell, and we sum over all ${\bf R}$ in the entire lattice. In the main text, we work with the $|\tilde{\bf p},{\bf k}\rangle$ after the gauge fixing, which we continue to denote by $|{\bf p},{\bf k}\rangle$ for simplicity in notation.

\subsection{Completeness of the Wannier Majorana states}\label{App:Complete}

In the above description, the $|\alpha,\pm,{\bf k}\rangle$ are defined to be the orthonormal complete Wannier states for the effective $2P\times 2P$ Hamiltonian, while  $|p,\pm,{\bf k}\rangle$ are the low-energy eigenstates of our interests of the full Hamiltonian which orthonormal but not complete. Our numerical procedure follows two steps. First, we construct $|\alpha,\pm,{\bf k}\rangle$ states iteratively and then use $\mathcal{K}_n$ as fitting parameters to find the corresponding energies $\pm E_{p}({\bf k})$ subject to the flux conservation constraint. The procedure followed is the same as \cite{Vanderbilt_2012RMP} and implemented in the Wannier90 package.

We assume $|\alpha,\pm,{\bf k}\rangle_t$ as some trial non-orthogogonal complete Wannier states related to the  $|p,\pm,{\bf k}\rangle$ states by an overall matrix $(U^{(t)}_{\pm})_{p,\alpha}({\bf k})=\langle p,\pm,{\bf k}|\alpha,\pm,{\bf k}\rangle_t$. Note that $U^{(t)}_{\pm}$ are not the same as the desired overlap matrix $U_{\pm}$ defined below Eq.~\eqref{eq:ME2} and we want to find a relation between them. To orthonormalize $|\alpha,\pm,{\bf k} \rangle_t$ we define their overlap matrix 
\begin{eqnarray}
    &(S_{\pm})_{\alpha\alpha'}&({\bf k}) \nonumber\\
    &=& \sum_{pp'}(U^{(t)\dagger}_{\pm})_{\alpha,p}({\bf k})(U^{(t)}_{\pm})_{p',\alpha'}({\bf k}) \langle p,\pm,{\bf k}|p',\pm,{\bf k}\rangle \nonumber \\
    &=& (U^{(t)\dagger}_{\pm}U^{(t)}_{\pm})_{\alpha\alpha'} \nonumber
\end{eqnarray}
Then the orthonormal Wannier states are defined as $|\alpha,\pm,{\bf k}\rangle=\sum_{\alpha'}(S_{\pm}^{-1/2})_{\alpha\alpha'}({\bf k})|\alpha',\pm,{\bf k}\rangle$. Then it is easy to show that the overlap matrix is defined as $U_{\pm}=U_{\pm}^{(t)}S_{\pm}^{-1/2}$, $\forall {\bf k}$. For the method to work, i.e., the Wannier orbitals to be smooth in the momentum space, the overall matrix $U^{(t)}_{\pm}$  must be non-singular. This is often not the case for topological insulators, atomically obstructed insulators, or flat bands with singular compact orbitals. For removable singularity, the procedure works well as described in \cite{TI_Kohmoto_1985,Rhim_2019,Soluyanov2011,Vafek2018}. Note that we do not need separate trial functions for the $\pm$ states as they are related by the particle-hole symmetry $\mathcal{C}$. 

\subsection{Choosing the trial wavefunction}\label{App:Trial}

\begin{figure}[t] 
\centering
\includegraphics[width=1\linewidth]{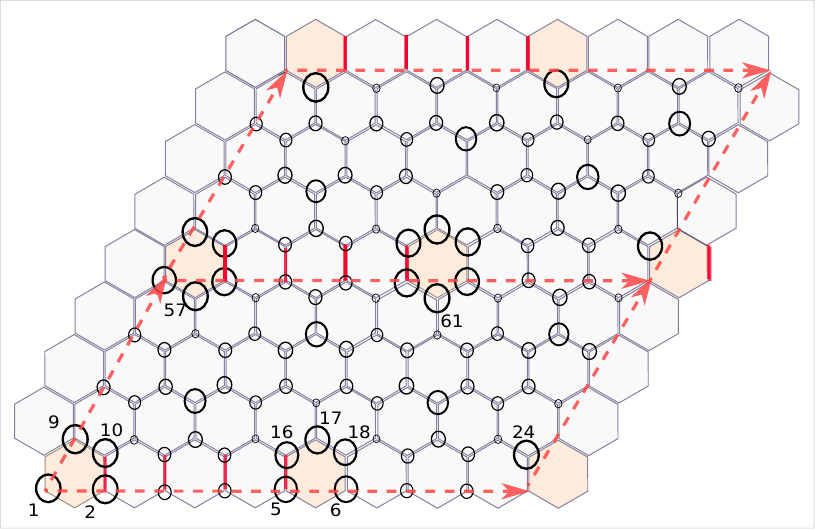}
\caption{We plot the orbital weight of the two lowest energy $(p=1)$ particle-hole states $\sum_{\veck}\chi_{p,i}\left(\veck\right)$ at different sites $i$, defined in Eq.~\ref{eq:coef}. The results are shown for the $4 \times 4-$ flux configuration. The size of the open circles denotes the orbital weight strength. We notice that the weight is largest around the $\pi-$ fluxes.}
\label{fig:Coef}
\end{figure}

How do we efficiently guess the trial Wannier states $|\alpha,\pm,{\bf k}\rangle_t$? We follow the procedure outlined in \onlinecite{Soluyanov2011} for complex fermions and make necessary modifications. Because we have the eigenvalues and eigenvectors of the full Hamiltonian $H$ in the supercell, we study first where our interested eigenvectors are localized in the supercell. This gives hints on the location of the Wannier centers for the trial states. 

The trial functions are considered as follows: following the procedure mentioned in Ref.~\onlinecite{Soluyanov2011}, we can expect the sites close to the $\pi-$fluxes to be the most probable regions for Majorana wave functions. We confirm this by plotting the probability amplitude of each Majorana sublattice of the Full supercell Hamiltonian: 
\begin{equation}\label{eq:coef}
    \chi_{p,i}\left({\bf k}\right)= \sum_{a=\pm} |\Gamma_{p,a,i}({\bf k})|^2.
\end{equation}
This signifies the occupation probability of the targeted bands for each sublattice index, $\alpha$ for a $\veck-$point. The highest probability indeed coincides at sites close to the $\pi-$ fluxes, as shown in Fig.~\ref{fig:Coef}. Based on this, we construct the trial functions for the ${\bm \alpha}$-Wannier Majorana orbital
\begin{equation}
    |{\alpha},\pm,{\bf R}\rangle_t = \frac{1}{\sqrt{2N}}\sum_{i}^{2N} \delta({\bf  r}-\mathbf{R}-\mathbf{t}_{i})|\vecr \rangle,
\end{equation}
where ${\bf t}_{i}$ are the positions of the original Majorana sublattices in the full Hamiltonian within the ${\bf R}$ supercell. For example, for the $4 \times 4-$ configuration shown in Fig.~\ref{fig:Coef}, the sublattice indices for the trial function with $(\alpha,+)\in$ \{$1,9,10$\}; and $(\alpha,-) \in$ \{$5,16,17,18$\}. There is no unique definition for these trial functions. The good choice is the functions that give $10 - 20 \%$ change in the spread function, from the initial spread function to the final spreads after the minimization procedure. In the $\veck-$ space, we obtained $|\alpha,\pm,{\bf k}\rangle_t$  by the Fourier transformation given in Eq.~\eqref{eq:FT}. With these trial functions, the Wannier centers are shown in Fig.~\ref{fig:wann}(a) for $4 \times 4$- configurations. In Fig.~\ref{fig:wann}(b), we plot the dispersions of Majorana fermions from Wannierised orbitals, and it fits ED results well with distances, $\mathbf{R} \leq |9 \mathbf{a}_1 + 9 \mathbf{a}_2|$.
\begin{widetext}
\subsection{Values of tight-binding parameters}\label{App:Parameters}
\begin{eqnarray}
\mathcal{T}_{{\bf R},{\bf R}}&=&i\left(
    \begin{array}{cc}
         0 &  -0.036\\
         0.036 & 0 
    \end{array}\right), \nonumber\\
    \mathcal{T}_{{\bf R},{\bf R}^{(1)}=(1,0)}&=&i\left(
    \begin{array}{cc}
         0.000 &  0.005 \\
        -0.023 & -0.000 
    \end{array}\right)=\mathcal{T}^{\dagger}_{{\bf R},-{\bf R}^{(1)}}, \nonumber\\
    \mathcal{T}_{{\bf R},{\bf R}^{(2)}=(0,1)}&=&i\left(
    \begin{array}{cc}
         0.028 &  -0.05\\
        -0.05 & -0.028 
    \end{array}\right)=\mathcal{T}^{\dagger}_{{\bf R},-{\bf R}^{(2)}}, \nonumber\\
    \mathcal{T}_{{\bf R},{\bf R}^{(3)}=(1,1)}&=&i\left(
    \begin{array}{cc}
         -0.004 &  0.005 \\
        - 0.008 & -0.004 
    \end{array}\right)=\mathcal{T}^{\dagger}_{{\bf R},-{\bf R}^{(3)}}, \nonumber\\
   \mathcal{T}_{{\bf R},{\bf R}^{(4)}=(1,-1)}&=&i\left(
    \begin{array}{cc}
         0.018 &  -0.008 \\
        - 0.038 & -0.018 
    \end{array}\right)=\mathcal{T}^{\dagger}_{{\bf R},-{\bf R}^{(4)}}, \nonumber\\
    \mathcal{T}_{{\bf R},{\bf R}^{(5)}=(2,0)}&=&i\left(
    \begin{array}{cc}
         0.025 &  -0.002 \\
        0.012 & -0.25
    \end{array}\right)=\mathcal{T}^{\dagger}_{{\bf R},-{\bf R}^{(5)}}, \nonumber\\
    \mathcal{T}_{{\bf R},{\bf R}^{(6)}=(0,2)}&=&i\left(
    \begin{array}{cc}
         0.005 &  0.006\\
        -0.029 & 0.005 
    \end{array}\right)=\mathcal{T}^{\dagger}_{{\bf R},-{\bf R}^{(6)}}, \nonumber\\
\mathcal{T}_{{\bf R},{\bf R}^{(7)}=(1,2)}&=&i\left(
    \begin{array}{cc}
         0.003 &  -0.002 \\
        0.011 & -0.003
    \end{array}\right)=-\mathcal{T}_{{\bf R},-{\bf R}^{(7)}}, \nonumber\\
    \mathcal{T}_{{\bf R},{\bf R}^{(8)}=(1,-2)}&=&i\left(
    \begin{array}{cc}
         -0.011 &  -0.004 \\
        -0.014 & 0.011
    \end{array}\right)=-\mathcal{T}_{{\bf R},-{\bf R}^{(8)}}, \nonumber\\
    \mathcal{T}_{{\bf R},{\bf R}^{(9)}=(2,-1)}&=&i\left(
    \begin{array}{cc}
         -0.006 &  0.003 \\
        -0.026 &   0.006
    \end{array}\right)=\mathcal{T}^{\dagger}_{{\bf R},-{\bf R}^{(9)}}, \nonumber\\
    \mathcal{T}_{{\bf R},{\bf R}^{(10)}=(2,1)}&=&i\left(
    \begin{array}{cc}
         0.001 &  -0.001 \\
        0.001 & -0.001
    \end{array}\right)=\mathcal{T}^{\dagger}_{{\bf R},-{\bf R}^{(10)}}, \nonumber\\
    \mathcal{T}_{{\bf R},{\bf R}^{(11)}=(2,2)}&=&i\left(
    \begin{array}{cc}
         -0.017 &  0.001 \\
        -0.004 &  0.017
    \end{array}\right)=\mathcal{T}^{\dagger}_{{\bf R},-{\bf R}^{(11)}}, \nonumber\\
       \mathcal{T}_{{\bf R},{\bf R}^{(12)}=(2,-2)}&=&i\left(
    \begin{array}{cc}
         0.003 &  0.001 \\
         0.025 &  -0.003
    \end{array}\right)=\mathcal{T}^{\dagger}_{{\bf R},-{\bf R}^{(12)}}, \nonumber\\
        \mathcal{T}_{{\bf R},{\bf R}^{(13)}=(2,3)}&=&i\left(
    \begin{array}{cc}
         0.00 &  0.000 \\
         0.002 &  0.00
    \end{array}\right)=\mathcal{T}^{\dagger}_{{\bf R},-{\bf R}^{(13)}}, \nonumber\\
\mathcal{T}_{{\bf R},{\bf R}^{(14)}=(2,-3)}&=&i\left(
    \begin{array}{cc}
         0.003 &  -0.003 \\
         -0.010 &  -0.003
    \end{array}\right)=\mathcal{T}^{\dagger}_{{\bf R},-{\bf R}^{(14)}}, \nonumber\\
    \mathcal{T}_{{\bf R},{\bf R}^{(15)}=(1,3)}&=&i\left(
    \begin{array}{cc}
         0.000 &  0.00 \\
         -0.004 &  0.00
    \end{array}\right)=\mathcal{T}^{\dagger}_{{\bf R},-{\bf R}^{(15)}}, \nonumber\\
    \mathcal{T}_{{\bf R},{\bf R}^{(16)}=(1,-3)}&=&i\left(
    \begin{array}{cc}
         -0.002 &  0.005 \\
         0.013 &  0.002
    \end{array}\right)=\mathcal{T}^{\dagger}_{{\bf R},-{\bf R}^{(16)}}, \nonumber\\
     \mathcal{T}_{{\bf R},{\bf R}^{(17)}=(3,2)}&=&i\left(
    \begin{array}{cc}
         0.00 &  0.000 \\
         0.002 &  0.00
    \end{array}\right)=\mathcal{T}^{\dagger}_{{\bf R},-{\bf R}^{(17)}}, \nonumber\\
    \mathcal{T}_{{\bf R},{\bf R}^{(18)}=(3,-2)}&=&i\left(
    \begin{array}{cc}
         0.00 &  0.000 \\
         -0.006 &  0.00
    \end{array}\right)=\mathcal{T}^{\dagger}_{{\bf R},-{\bf R}^{(18)}}, \nonumber\\
    \mathcal{T}_{{\bf R},{\bf R}^{(19)}=(3,3)}&=&i\left(
    \begin{array}{cc}
         0.000 &  0.000 \\
         -0.001 &  0.000
    \end{array}\right)=\mathcal{T}^{\dagger}_{{\bf R},-{\bf R}^{(19)}}, \nonumber\\
    \mathcal{T}_{{\bf R},{\bf R}^{(20)}=(3,-3)}&=&i\left(
    \begin{array}{cc}
         -0.002 &  0.001 \\
         -0.002 &  0.003
    \end{array}\right)=\mathcal{T}^{\dagger}_{{\bf R},-{\bf R}^{(20)}}, \nonumber\\
      \end{eqnarray}
The lattice vectors are given above in terms of the Miller indices and ${\bf R}=(0,0)$. The rest of the parameters are nearly zero and hence ignored in this table. The $2\times 2$ matrices for each tensor component are in the two particle-hole Majorana basis of $a=\pm$. 
\end{widetext}
\bibliography{main.bib}

\begin{thebibliography}{144}%
\makeatletter
\providecommand \@ifxundefined [1]{%
 \@ifx{#1\undefined}
}%
\providecommand \@ifnum [1]{%
 \ifnum #1\expandafter \@firstoftwo
 \else \expandafter \@secondoftwo
 \fi
}%
\providecommand \@ifx [1]{%
 \ifx #1\expandafter \@firstoftwo
 \else \expandafter \@secondoftwo
 \fi
}%
\providecommand \natexlab [1]{#1}%
\providecommand \enquote  [1]{``#1''}%
\providecommand \bibnamefont  [1]{#1}%
\providecommand \bibfnamefont [1]{#1}%
\providecommand \citenamefont [1]{#1}%
\providecommand \href@noop [0]{\@secondoftwo}%
\providecommand \href [0]{\begingroup \@sanitize@url \@href}%
\providecommand \@href[1]{\@@startlink{#1}\@@href}%
\providecommand \@@href[1]{\endgroup#1\@@endlink}%
\providecommand \@sanitize@url [0]{\catcode `\\12\catcode `\$12\catcode
  `\&12\catcode `\#12\catcode `\^12\catcode `\_12\catcode `\%12\relax}%
\providecommand \@@startlink[1]{}%
\providecommand \@@endlink[0]{}%
\providecommand \url  [0]{\begingroup\@sanitize@url \@url }%
\providecommand \@url [1]{\endgroup\@href {#1}{\urlprefix }}%
\providecommand \urlprefix  [0]{URL }%
\providecommand \Eprint [0]{\href }%
\providecommand \doibase [0]{https://doi.org/}%
\providecommand \selectlanguage [0]{\@gobble}%
\providecommand \bibinfo  [0]{\@secondoftwo}%
\providecommand \bibfield  [0]{\@secondoftwo}%
\providecommand \translation [1]{[#1]}%
\providecommand \BibitemOpen [0]{}%
\providecommand \bibitemStop [0]{}%
\providecommand \bibitemNoStop [0]{.\EOS\space}%
\providecommand \EOS [0]{\spacefactor3000\relax}%
\providecommand \BibitemShut  [1]{\csname bibitem#1\endcsname}%
\let\auto@bib@innerbib\@empty
\bibitem [{\citenamefont {Greiter}\ and\ \citenamefont
  {Wilczek}(2024)}]{Wilczek_2024}%
  \BibitemOpen
  \bibfield  {author} {\bibinfo {author} {\bibfnamefont {M.}~\bibnamefont
  {Greiter}}\ and\ \bibinfo {author} {\bibfnamefont {F.}~\bibnamefont
  {Wilczek}},\ }\bibfield  {title} {\bibinfo {title} {Fractional statistics},\
  }\href
  {https://doi.org/https://doi.org/10.1146/annurev-conmatphys-040423-014045}
  {\bibfield  {journal} {\bibinfo  {journal} {Annual Review of Condensed Matter
  Physics}\ }\textbf {\bibinfo {volume} {15}},\ \bibinfo {pages} {131}
  (\bibinfo {year} {2024})}\BibitemShut {NoStop}%
\bibitem [{\citenamefont {Aguado}\ and\ \citenamefont
  {Kouwenhoven}(2020)}]{Ramon2020}%
  \BibitemOpen
  \bibfield  {author} {\bibinfo {author} {\bibfnamefont {R.}~\bibnamefont
  {Aguado}}\ and\ \bibinfo {author} {\bibfnamefont {L.~P.}\ \bibnamefont
  {Kouwenhoven}},\ }\bibfield  {title} {\bibinfo {title} {{Majorana qubits for
  topological quantum computing}},\ }\href {https://doi.org/10.1063/PT.3.4499}
  {\bibfield  {journal} {\bibinfo  {journal} {Physics Today}\ }\textbf
  {\bibinfo {volume} {73}},\ \bibinfo {pages} {44} (\bibinfo {year}
  {2020})}\BibitemShut {NoStop}%
\bibitem [{\citenamefont {{Keimer}}\ and\ \citenamefont
  {{Moore}}(2017)}]{Moore2017}%
  \BibitemOpen
  \bibfield  {author} {\bibinfo {author} {\bibfnamefont {B.}~\bibnamefont
  {{Keimer}}}\ and\ \bibinfo {author} {\bibfnamefont {J.~E.}\ \bibnamefont
  {{Moore}}},\ }\bibfield  {title} {\bibinfo {title} {{The physics of quantum
  materials}},\ }\href {https://doi.org/10.1038/nphys4302} {\bibfield
  {journal} {\bibinfo  {journal} {Nature Physics}\ }\textbf {\bibinfo {volume}
  {13}},\ \bibinfo {pages} {1045} (\bibinfo {year} {2017})}\BibitemShut
  {NoStop}%
\bibitem [{\citenamefont {{Basov}}\ \emph {et~al.}(2017)\citenamefont
  {{Basov}}, \citenamefont {{Averitt}},\ and\ \citenamefont
  {{Hsieh}}}]{Basov_2017}%
  \BibitemOpen
  \bibfield  {author} {\bibinfo {author} {\bibfnamefont {D.~N.}\ \bibnamefont
  {{Basov}}}, \bibinfo {author} {\bibfnamefont {R.~D.}\ \bibnamefont
  {{Averitt}}},\ and\ \bibinfo {author} {\bibfnamefont {D.}~\bibnamefont
  {{Hsieh}}},\ }\bibfield  {title} {\bibinfo {title} {{Towards properties on
  demand in quantum materials}},\ }\href {https://doi.org/10.1038/nmat5017}
  {\bibfield  {journal} {\bibinfo  {journal} {Nature Materials}\ }\textbf
  {\bibinfo {volume} {16}},\ \bibinfo {pages} {1077} (\bibinfo {year}
  {2017})}\BibitemShut {NoStop}%
\bibitem [{\citenamefont {{Tokura}}\ \emph {et~al.}(2017)\citenamefont
  {{Tokura}}, \citenamefont {{Kawasaki}},\ and\ \citenamefont
  {{Nagaosa}}}]{Tokura_2017}%
  \BibitemOpen
  \bibfield  {author} {\bibinfo {author} {\bibfnamefont {Y.}~\bibnamefont
  {{Tokura}}}, \bibinfo {author} {\bibfnamefont {M.}~\bibnamefont
  {{Kawasaki}}},\ and\ \bibinfo {author} {\bibfnamefont {N.}~\bibnamefont
  {{Nagaosa}}},\ }\bibfield  {title} {\bibinfo {title} {{Emergent functions of
  quantum materials}},\ }\href {https://doi.org/10.1038/nphys4274} {\bibfield
  {journal} {\bibinfo  {journal} {Nature Physics}\ }\textbf {\bibinfo {volume}
  {13}},\ \bibinfo {pages} {1056} (\bibinfo {year} {2017})}\BibitemShut
  {NoStop}%
\bibitem [{\citenamefont {Aasen}\ \emph {et~al.}(2016)\citenamefont {Aasen},
  \citenamefont {Hell}, \citenamefont {Mishmash}, \citenamefont {Higginbotham},
  \citenamefont {Danon}, \citenamefont {Leijnse}, \citenamefont {Jespersen},
  \citenamefont {Folk}, \citenamefont {Marcus}, \citenamefont {Flensberg},\
  and\ \citenamefont {Alicea}}]{Milestones_2016}%
  \BibitemOpen
  \bibfield  {author} {\bibinfo {author} {\bibfnamefont {D.}~\bibnamefont
  {Aasen}}, \bibinfo {author} {\bibfnamefont {M.}~\bibnamefont {Hell}},
  \bibinfo {author} {\bibfnamefont {R.~V.}\ \bibnamefont {Mishmash}}, \bibinfo
  {author} {\bibfnamefont {A.}~\bibnamefont {Higginbotham}}, \bibinfo {author}
  {\bibfnamefont {J.}~\bibnamefont {Danon}}, \bibinfo {author} {\bibfnamefont
  {M.}~\bibnamefont {Leijnse}}, \bibinfo {author} {\bibfnamefont {T.~S.}\
  \bibnamefont {Jespersen}}, \bibinfo {author} {\bibfnamefont {J.~A.}\
  \bibnamefont {Folk}}, \bibinfo {author} {\bibfnamefont {C.~M.}\ \bibnamefont
  {Marcus}}, \bibinfo {author} {\bibfnamefont {K.}~\bibnamefont {Flensberg}},\
  and\ \bibinfo {author} {\bibfnamefont {J.}~\bibnamefont {Alicea}},\
  }\bibfield  {title} {\bibinfo {title} {Milestones toward majorana-based
  quantum computing},\ }\href {https://doi.org/10.1103/PhysRevX.6.031016}
  {\bibfield  {journal} {\bibinfo  {journal} {Phys. Rev. X}\ }\textbf {\bibinfo
  {volume} {6}},\ \bibinfo {pages} {031016} (\bibinfo {year}
  {2016})}\BibitemShut {NoStop}%
\bibitem [{\citenamefont {Sarma}\ \emph {et~al.}(2015)\citenamefont {Sarma},
  \citenamefont {Freedman},\ and\ \citenamefont {Nayak}}]{Sarma_2015}%
  \BibitemOpen
  \bibfield  {author} {\bibinfo {author} {\bibfnamefont {S.~D.}\ \bibnamefont
  {Sarma}}, \bibinfo {author} {\bibfnamefont {M.}~\bibnamefont {Freedman}},\
  and\ \bibinfo {author} {\bibfnamefont {C.}~\bibnamefont {Nayak}},\ }\bibfield
   {title} {\bibinfo {title} {Majorana zero modes and topological quantum
  computation},\ }\bibfield  {journal} {\bibinfo  {journal} {npj Quantum
  Information}\ }\textbf {\bibinfo {volume} {1}},\ \href
  {https://doi.org/10.1038/npjqi.2015.1} {10.1038/npjqi.2015.1} (\bibinfo
  {year} {2015})\BibitemShut {NoStop}%
\bibitem [{\citenamefont {Wen}(2007)}]{wen_Book}%
  \BibitemOpen
  \bibfield  {author} {\bibinfo {author} {\bibfnamefont {X.-G.}\ \bibnamefont
  {Wen}},\ }\href {https://doi.org/10.1093/acprof:oso/9780199227259.001.0001}
  {\emph {\bibinfo {title} {{Quantum Field Theory of Many-Body Systems: From
  the Origin of Sound to an Origin of Light and Electrons}}}}\ (\bibinfo
  {publisher} {Oxford University Press},\ \bibinfo {year} {2007})\BibitemShut
  {NoStop}%
\bibitem [{\citenamefont {Fradkin}(2013)}]{Fradkin_Book}%
  \BibitemOpen
  \bibfield  {author} {\bibinfo {author} {\bibfnamefont {E.}~\bibnamefont
  {Fradkin}},\ }\href@noop {} {\emph {\bibinfo {title} {Field Theories of
  Condensed Matter Physics}}},\ \bibinfo {edition} {2nd}\ ed.\ (\bibinfo
  {publisher} {Cambridge University Press},\ \bibinfo {year}
  {2013})\BibitemShut {NoStop}%
\bibitem [{\citenamefont {Sachdev}(2011)}]{Sachdev_Book}%
  \BibitemOpen
  \bibfield  {author} {\bibinfo {author} {\bibfnamefont {S.}~\bibnamefont
  {Sachdev}},\ }\href@noop {} {\emph {\bibinfo {title} {Quantum Phase
  Transitions}}},\ \bibinfo {edition} {2nd}\ ed.\ (\bibinfo  {publisher}
  {Cambridge University Press},\ \bibinfo {year} {2011})\BibitemShut {NoStop}%
\bibitem [{\citenamefont {Gay}\ and\ \citenamefont
  {Mackie}(2009)}]{Gay_Mackie_2009}%
  \BibitemOpen
  \bibinfo {editor} {\bibfnamefont {S.}~\bibnamefont {Gay}}\ and\ \bibinfo
  {editor} {\bibfnamefont {I.}~\bibnamefont {Mackie}},\ eds.,\ \href@noop {}
  {\emph {\bibinfo {title} {Semantic Techniques in Quantum Computation}}}\
  (\bibinfo  {publisher} {Cambridge University Press},\ \bibinfo {year}
  {2009})\BibitemShut {NoStop}%
\bibitem [{\citenamefont {Beenakker}(2015)}]{Beenakker_2015RMP}%
  \BibitemOpen
  \bibfield  {author} {\bibinfo {author} {\bibfnamefont {C.~W.~J.}\
  \bibnamefont {Beenakker}},\ }\bibfield  {title} {\bibinfo {title}
  {Random-matrix theory of majorana fermions and topological superconductors},\
  }\href {https://doi.org/10.1103/RevModPhys.87.1037} {\bibfield  {journal}
  {\bibinfo  {journal} {Rev. Mod. Phys.}\ }\textbf {\bibinfo {volume} {87}},\
  \bibinfo {pages} {1037} (\bibinfo {year} {2015})}\BibitemShut {NoStop}%
\bibitem [{\citenamefont {Rahmani}\ and\ \citenamefont
  {Franz}(2019)}]{Rahmani_2019}%
  \BibitemOpen
  \bibfield  {author} {\bibinfo {author} {\bibfnamefont {A.}~\bibnamefont
  {Rahmani}}\ and\ \bibinfo {author} {\bibfnamefont {M.}~\bibnamefont
  {Franz}},\ }\bibfield  {title} {\bibinfo {title} {Interacting majorana
  fermions},\ }\href {https://doi.org/10.1088/1361-6633/ab28ef} {\bibfield
  {journal} {\bibinfo  {journal} {Reports on Progress in Physics}\ }\textbf
  {\bibinfo {volume} {82}},\ \bibinfo {pages} {084501} (\bibinfo {year}
  {2019})}\BibitemShut {NoStop}%
\bibitem [{\citenamefont {Zhang}\ \emph {et~al.}(1989)\citenamefont {Zhang},
  \citenamefont {Hansson},\ and\ \citenamefont {Kivelson}}]{Zhang_1989}%
  \BibitemOpen
  \bibfield  {author} {\bibinfo {author} {\bibfnamefont {S.~C.}\ \bibnamefont
  {Zhang}}, \bibinfo {author} {\bibfnamefont {T.~H.}\ \bibnamefont {Hansson}},\
  and\ \bibinfo {author} {\bibfnamefont {S.}~\bibnamefont {Kivelson}},\
  }\bibfield  {title} {\bibinfo {title} {Effective-field-theory model for the
  fractional quantum hall effect},\ }\href
  {https://doi.org/10.1103/PhysRevLett.62.82} {\bibfield  {journal} {\bibinfo
  {journal} {Phys. Rev. Lett.}\ }\textbf {\bibinfo {volume} {62}},\ \bibinfo
  {pages} {82} (\bibinfo {year} {1989})}\BibitemShut {NoStop}%
\bibitem [{\citenamefont {TARASOV}(2013)}]{TARASOV_2013}%
  \BibitemOpen
  \bibfield  {author} {\bibinfo {author} {\bibfnamefont {V.~E.}\ \bibnamefont
  {TARASOV}},\ }\bibfield  {title} {\bibinfo {title} {Review of some promising
  fractional physical models},\ }\href
  {https://doi.org/10.1142/s0217979213300053} {\bibfield  {journal} {\bibinfo
  {journal} {International Journal of Modern Physics B}\ }\textbf {\bibinfo
  {volume} {27}},\ \bibinfo {pages} {1330005} (\bibinfo {year}
  {2013})}\BibitemShut {NoStop}%
\bibitem [{\citenamefont {Chamon}(2005)}]{Chamon2005}%
  \BibitemOpen
  \bibfield  {author} {\bibinfo {author} {\bibfnamefont {C.}~\bibnamefont
  {Chamon}},\ }\bibfield  {title} {\bibinfo {title} {Quantum glassiness in
  strongly correlated clean systems: An example of topological
  overprotection},\ }\href {https://doi.org/10.1103/PhysRevLett.94.040402}
  {\bibfield  {journal} {\bibinfo  {journal} {Phys. Rev. Lett.}\ }\textbf
  {\bibinfo {volume} {94}},\ \bibinfo {pages} {040402} (\bibinfo {year}
  {2005})}\BibitemShut {NoStop}%
\bibitem [{\citenamefont {Prem}\ \emph {et~al.}(2017)\citenamefont {Prem},
  \citenamefont {Haah},\ and\ \citenamefont {Nandkishore}}]{Prem2017}%
  \BibitemOpen
  \bibfield  {author} {\bibinfo {author} {\bibfnamefont {A.}~\bibnamefont
  {Prem}}, \bibinfo {author} {\bibfnamefont {J.}~\bibnamefont {Haah}},\ and\
  \bibinfo {author} {\bibfnamefont {R.}~\bibnamefont {Nandkishore}},\
  }\bibfield  {title} {\bibinfo {title} {Glassy quantum dynamics in translation
  invariant fracton models},\ }\href
  {https://doi.org/10.1103/PhysRevB.95.155133} {\bibfield  {journal} {\bibinfo
  {journal} {Phys. Rev. B}\ }\textbf {\bibinfo {volume} {95}},\ \bibinfo
  {pages} {155133} (\bibinfo {year} {2017})}\BibitemShut {NoStop}%
\bibitem [{\citenamefont {Hart}\ \emph {et~al.}(2021)\citenamefont {Hart},
  \citenamefont {Gopalakrishnan},\ and\ \citenamefont
  {Castelnovo}}]{Hart2021Logarithmic}%
  \BibitemOpen
  \bibfield  {author} {\bibinfo {author} {\bibfnamefont {O.}~\bibnamefont
  {Hart}}, \bibinfo {author} {\bibfnamefont {S.}~\bibnamefont
  {Gopalakrishnan}},\ and\ \bibinfo {author} {\bibfnamefont {C.}~\bibnamefont
  {Castelnovo}},\ }\bibfield  {title} {\bibinfo {title} {Logarithmic
  entanglement growth from disorder-free localization in the two-leg compass
  ladder},\ }\href {https://doi.org/10.1103/PhysRevLett.126.227202} {\bibfield
  {journal} {\bibinfo  {journal} {Phys. Rev. Lett.}\ }\textbf {\bibinfo
  {volume} {126}},\ \bibinfo {pages} {227202} (\bibinfo {year}
  {2021})}\BibitemShut {NoStop}%
\bibitem [{\citenamefont {Yogendra}\ \emph {et~al.}(2023)\citenamefont
  {Yogendra}, \citenamefont {Das},\ and\ \citenamefont
  {Baskaran}}]{yogendra2023emergent}%
  \BibitemOpen
  \bibfield  {author} {\bibinfo {author} {\bibfnamefont {K.~B.}\ \bibnamefont
  {Yogendra}}, \bibinfo {author} {\bibfnamefont {T.}~\bibnamefont {Das}},\ and\
  \bibinfo {author} {\bibfnamefont {G.}~\bibnamefont {Baskaran}},\ }\bibfield
  {title} {\bibinfo {title} {Emergent glassiness in the disorder-free kitaev
  model: Density matrix renormalization group study on a one-dimensional ladder
  setting},\ }\href {https://doi.org/10.1103/PhysRevB.108.165118} {\bibfield
  {journal} {\bibinfo  {journal} {Phys. Rev. B}\ }\textbf {\bibinfo {volume}
  {108}},\ \bibinfo {pages} {165118} (\bibinfo {year} {2023})}\BibitemShut
  {NoStop}%
\bibitem [{\citenamefont {Wan}\ and\ \citenamefont
  {Armitage}(2019)}]{Armitage_2019}%
  \BibitemOpen
  \bibfield  {author} {\bibinfo {author} {\bibfnamefont {Y.}~\bibnamefont
  {Wan}}\ and\ \bibinfo {author} {\bibfnamefont {N.~P.}\ \bibnamefont
  {Armitage}},\ }\bibfield  {title} {\bibinfo {title} {Resolving continua of
  fractional excitations by spinon echo in thz 2d coherent spectroscopy},\
  }\href {https://doi.org/10.1103/PhysRevLett.122.257401} {\bibfield  {journal}
  {\bibinfo  {journal} {Phys. Rev. Lett.}\ }\textbf {\bibinfo {volume} {122}},\
  \bibinfo {pages} {257401} (\bibinfo {year} {2019})}\BibitemShut {NoStop}%
\bibitem [{\citenamefont {Nandkishore}\ \emph {et~al.}(2021)\citenamefont
  {Nandkishore}, \citenamefont {Choi},\ and\ \citenamefont
  {Kim}}]{NandKishore_2021}%
  \BibitemOpen
  \bibfield  {author} {\bibinfo {author} {\bibfnamefont {R.~M.}\ \bibnamefont
  {Nandkishore}}, \bibinfo {author} {\bibfnamefont {W.}~\bibnamefont {Choi}},\
  and\ \bibinfo {author} {\bibfnamefont {Y.~B.}\ \bibnamefont {Kim}},\
  }\bibfield  {title} {\bibinfo {title} {Spectroscopic fingerprints of gapped
  quantum spin liquids, both conventional and fractonic},\ }\href
  {https://doi.org/10.1103/PhysRevResearch.3.013254} {\bibfield  {journal}
  {\bibinfo  {journal} {Phys. Rev. Res.}\ }\textbf {\bibinfo {volume} {3}},\
  \bibinfo {pages} {013254} (\bibinfo {year} {2021})}\BibitemShut {NoStop}%
\bibitem [{\citenamefont {McGinley}\ \emph {et~al.}(2024)\citenamefont
  {McGinley}, \citenamefont {Fava},\ and\ \citenamefont
  {Parameswaran}}]{McGinley_2024}%
  \BibitemOpen
  \bibfield  {author} {\bibinfo {author} {\bibfnamefont {M.}~\bibnamefont
  {McGinley}}, \bibinfo {author} {\bibfnamefont {M.}~\bibnamefont {Fava}},\
  and\ \bibinfo {author} {\bibfnamefont {S.~A.}\ \bibnamefont {Parameswaran}},\
  }\bibfield  {title} {\bibinfo {title} {Signatures of fractional statistics in
  nonlinear pump-probe spectroscopy},\ }\href
  {https://doi.org/10.1103/PhysRevLett.132.066702} {\bibfield  {journal}
  {\bibinfo  {journal} {Phys. Rev. Lett.}\ }\textbf {\bibinfo {volume} {132}},\
  \bibinfo {pages} {066702} (\bibinfo {year} {2024})}\BibitemShut {NoStop}%
\bibitem [{\citenamefont {Marzari}\ \emph {et~al.}(2012)\citenamefont
  {Marzari}, \citenamefont {Mostofi}, \citenamefont {Yates}, \citenamefont
  {Souza},\ and\ \citenamefont {Vanderbilt}}]{Vanderbilt_2012RMP}%
  \BibitemOpen
  \bibfield  {author} {\bibinfo {author} {\bibfnamefont {N.}~\bibnamefont
  {Marzari}}, \bibinfo {author} {\bibfnamefont {A.~A.}\ \bibnamefont
  {Mostofi}}, \bibinfo {author} {\bibfnamefont {J.~R.}\ \bibnamefont {Yates}},
  \bibinfo {author} {\bibfnamefont {I.}~\bibnamefont {Souza}},\ and\ \bibinfo
  {author} {\bibfnamefont {D.}~\bibnamefont {Vanderbilt}},\ }\bibfield  {title}
  {\bibinfo {title} {Maximally localized wannier functions: Theory and
  applications},\ }\href {https://doi.org/10.1103/RevModPhys.84.1419}
  {\bibfield  {journal} {\bibinfo  {journal} {Rev. Mod. Phys.}\ }\textbf
  {\bibinfo {volume} {84}},\ \bibinfo {pages} {1419} (\bibinfo {year}
  {2012})}\BibitemShut {NoStop}%
\bibitem [{\citenamefont {Marzari}\ and\ \citenamefont
  {Vanderbilt}(1997)}]{Marzari1997}%
  \BibitemOpen
  \bibfield  {author} {\bibinfo {author} {\bibfnamefont {N.}~\bibnamefont
  {Marzari}}\ and\ \bibinfo {author} {\bibfnamefont {D.}~\bibnamefont
  {Vanderbilt}},\ }\bibfield  {title} {\bibinfo {title} {Maximally localized
  generalized wannier functions for composite energy bands},\ }\href
  {https://doi.org/10.1103/PhysRevB.56.12847} {\bibfield  {journal} {\bibinfo
  {journal} {Phys. Rev. B}\ }\textbf {\bibinfo {volume} {56}},\ \bibinfo
  {pages} {12847} (\bibinfo {year} {1997})}\BibitemShut {NoStop}%
\bibitem [{\citenamefont {Souza}\ \emph {et~al.}(2001)\citenamefont {Souza},
  \citenamefont {Marzari},\ and\ \citenamefont {Vanderbilt}}]{Souza_2001}%
  \BibitemOpen
  \bibfield  {author} {\bibinfo {author} {\bibfnamefont {I.}~\bibnamefont
  {Souza}}, \bibinfo {author} {\bibfnamefont {N.}~\bibnamefont {Marzari}},\
  and\ \bibinfo {author} {\bibfnamefont {D.}~\bibnamefont {Vanderbilt}},\
  }\bibfield  {title} {\bibinfo {title} {Maximally localized wannier functions
  for entangled energy bands},\ }\href
  {https://doi.org/10.1103/PhysRevB.65.035109} {\bibfield  {journal} {\bibinfo
  {journal} {Phys. Rev. B}\ }\textbf {\bibinfo {volume} {65}},\ \bibinfo
  {pages} {035109} (\bibinfo {year} {2001})}\BibitemShut {NoStop}%
\bibitem [{\citenamefont {Klein}(1974)}]{Klein_1974}%
  \BibitemOpen
  \bibfield  {author} {\bibinfo {author} {\bibfnamefont {D.~J.}\ \bibnamefont
  {Klein}},\ }\bibfield  {title} {\bibinfo {title} {{Degenerate perturbation
  theory}},\ }\href {https://doi.org/10.1063/1.1682018} {\bibfield  {journal}
  {\bibinfo  {journal} {The Journal of Chemical Physics}\ }\textbf {\bibinfo
  {volume} {61}},\ \bibinfo {pages} {786} (\bibinfo {year} {1974})}\BibitemShut
  {NoStop}%
\bibitem [{\citenamefont {Shavitt}\ and\ \citenamefont
  {Redmon}(1980)}]{shavitt_1980}%
  \BibitemOpen
  \bibfield  {author} {\bibinfo {author} {\bibfnamefont {I.}~\bibnamefont
  {Shavitt}}\ and\ \bibinfo {author} {\bibfnamefont {L.~T.}\ \bibnamefont
  {Redmon}},\ }\bibfield  {title} {\bibinfo {title} {{Quasidegenerate
  perturbation theories. A canonical van Vleck formalism and its relationship
  to other approaches}},\ }\href {https://doi.org/10.1063/1.440050} {\bibfield
  {journal} {\bibinfo  {journal} {The Journal of Chemical Physics}\ }\textbf
  {\bibinfo {volume} {73}},\ \bibinfo {pages} {5711} (\bibinfo {year}
  {1980})}\BibitemShut {NoStop}%
\bibitem [{\citenamefont {Cohen-Tannoudji}\ \emph {et~al.}(1977)\citenamefont
  {Cohen-Tannoudji}, \citenamefont {Diu},\ and\ \citenamefont
  {Lalo{\"e}}}]{cohen_1977quantum}%
  \BibitemOpen
  \bibfield  {author} {\bibinfo {author} {\bibfnamefont {C.}~\bibnamefont
  {Cohen-Tannoudji}}, \bibinfo {author} {\bibfnamefont {B.}~\bibnamefont
  {Diu}},\ and\ \bibinfo {author} {\bibfnamefont {F.}~\bibnamefont
  {Lalo{\"e}}},\ }\href {https://books.google.co.in/books?id=iHcpAQAAMAAJ}
  {\emph {\bibinfo {title} {Quantum Mechanics}}},\ \bibinfo {series} {A Wiley -
  Interscience publication}\ No.\ \bibinfo {number} {v. 1}\ (\bibinfo
  {publisher} {Wiley},\ \bibinfo {year} {1977})\BibitemShut {NoStop}%
\bibitem [{\citenamefont {Anderson}\ \emph {et~al.}(1987)\citenamefont
  {Anderson}, \citenamefont {Baskaran}, \citenamefont {Zou},\ and\
  \citenamefont {Hsu}}]{Anderson_1987RVB}%
  \BibitemOpen
  \bibfield  {author} {\bibinfo {author} {\bibfnamefont {P.~W.}\ \bibnamefont
  {Anderson}}, \bibinfo {author} {\bibfnamefont {G.}~\bibnamefont {Baskaran}},
  \bibinfo {author} {\bibfnamefont {Z.}~\bibnamefont {Zou}},\ and\ \bibinfo
  {author} {\bibfnamefont {T.}~\bibnamefont {Hsu}},\ }\bibfield  {title}
  {\bibinfo {title} {Resonating--valence-bond theory of phase transitions and
  superconductivity in ${\mathrm{la}}_{2}$${\mathrm{cuo}}_{4}$-based
  compounds},\ }\href {https://doi.org/10.1103/PhysRevLett.58.2790} {\bibfield
  {journal} {\bibinfo  {journal} {Phys. Rev. Lett.}\ }\textbf {\bibinfo
  {volume} {58}},\ \bibinfo {pages} {2790} (\bibinfo {year}
  {1987})}\BibitemShut {NoStop}%
\bibitem [{\citenamefont {Fazekas}\ and\ \citenamefont
  {Anderson}(1974)}]{Fazekas_1974RVB}%
  \BibitemOpen
  \bibfield  {author} {\bibinfo {author} {\bibfnamefont {P.}~\bibnamefont
  {Fazekas}}\ and\ \bibinfo {author} {\bibfnamefont {P.~W.}\ \bibnamefont
  {Anderson}},\ }\bibfield  {title} {\bibinfo {title} {On the ground state
  properties of the anisotropic triangular antiferromagnet},\ }\href
  {https://doi.org/10.1080/14786439808206568} {\bibfield  {journal} {\bibinfo
  {journal} {Philosophical Magazine}\ }\textbf {\bibinfo {volume} {30}},\
  \bibinfo {pages} {423} (\bibinfo {year} {1974})}\BibitemShut {NoStop}%
\bibitem [{\citenamefont {Schrieffer}\ and\ \citenamefont
  {Wolff}(1966)}]{Schrieffer_1966}%
  \BibitemOpen
  \bibfield  {author} {\bibinfo {author} {\bibfnamefont {J.~R.}\ \bibnamefont
  {Schrieffer}}\ and\ \bibinfo {author} {\bibfnamefont {P.~A.}\ \bibnamefont
  {Wolff}},\ }\bibfield  {title} {\bibinfo {title} {Relation between the
  anderson and kondo hamiltonians},\ }\href
  {https://doi.org/10.1103/PhysRev.149.491} {\bibfield  {journal} {\bibinfo
  {journal} {Phys. Rev.}\ }\textbf {\bibinfo {volume} {149}},\ \bibinfo {pages}
  {491} (\bibinfo {year} {1966})}\BibitemShut {NoStop}%
\bibitem [{\citenamefont {Bravyi}\ \emph {et~al.}(2011)\citenamefont {Bravyi},
  \citenamefont {DiVincenzo},\ and\ \citenamefont {Loss}}]{BRAVYI_2011}%
  \BibitemOpen
  \bibfield  {author} {\bibinfo {author} {\bibfnamefont {S.}~\bibnamefont
  {Bravyi}}, \bibinfo {author} {\bibfnamefont {D.~P.}\ \bibnamefont
  {DiVincenzo}},\ and\ \bibinfo {author} {\bibfnamefont {D.}~\bibnamefont
  {Loss}},\ }\bibfield  {title} {\bibinfo {title} {Schrieffer–wolff
  transformation for quantum many-body systems},\ }\href
  {https://doi.org/https://doi.org/10.1016/j.aop.2011.06.004} {\bibfield
  {journal} {\bibinfo  {journal} {Annals of Physics}\ }\textbf {\bibinfo
  {volume} {326}},\ \bibinfo {pages} {2793} (\bibinfo {year}
  {2011})}\BibitemShut {NoStop}%
\bibitem [{\citenamefont {Giuliani}\ and\ \citenamefont
  {Vignale}(2005)}]{Giuliani_Vignale_2005}%
  \BibitemOpen
  \bibfield  {author} {\bibinfo {author} {\bibfnamefont {G.}~\bibnamefont
  {Giuliani}}\ and\ \bibinfo {author} {\bibfnamefont {G.}~\bibnamefont
  {Vignale}},\ }\href@noop {} {\emph {\bibinfo {title} {Quantum Theory of the
  Electron Liquid, Chapter 8.6}}}\ (\bibinfo  {publisher} {Cambridge University
  Press},\ \bibinfo {year} {2005})\BibitemShut {NoStop}%
\bibitem [{\citenamefont {Wu}\ and\ \citenamefont {Yang}(2007)}]{Ying_2007}%
  \BibitemOpen
  \bibfield  {author} {\bibinfo {author} {\bibfnamefont {Y.}~\bibnamefont
  {Wu}}\ and\ \bibinfo {author} {\bibfnamefont {X.}~\bibnamefont {Yang}},\
  }\bibfield  {title} {\bibinfo {title} {Strong-coupling theory of periodically
  driven two-level systems},\ }\href
  {https://doi.org/10.1103/PhysRevLett.98.013601} {\bibfield  {journal}
  {\bibinfo  {journal} {Phys. Rev. Lett.}\ }\textbf {\bibinfo {volume} {98}},\
  \bibinfo {pages} {013601} (\bibinfo {year} {2007})}\BibitemShut {NoStop}%
\bibitem [{\citenamefont {Nesbet}(1961)}]{Nesbet_1961}%
  \BibitemOpen
  \bibfield  {author} {\bibinfo {author} {\bibfnamefont {R.~K.}\ \bibnamefont
  {Nesbet}},\ }\bibfield  {title} {\bibinfo {title} {Approximate methods in the
  quantum theory of many-fermion systems},\ }\href
  {https://doi.org/10.1103/RevModPhys.33.28} {\bibfield  {journal} {\bibinfo
  {journal} {Rev. Mod. Phys.}\ }\textbf {\bibinfo {volume} {33}},\ \bibinfo
  {pages} {28} (\bibinfo {year} {1961})}\BibitemShut {NoStop}%
\bibitem [{\citenamefont {{Stratonovich}}(1957)}]{Stratonovich_1957}%
  \BibitemOpen
  \bibfield  {author} {\bibinfo {author} {\bibfnamefont {R.~L.}\ \bibnamefont
  {{Stratonovich}}},\ }\bibfield  {title} {\bibinfo {title} {{On a Method of
  Calculating Quantum Distribution Functions}},\ }\href@noop {} {\bibfield
  {journal} {\bibinfo  {journal} {Soviet Physics Doklady}\ }\textbf {\bibinfo
  {volume} {2}},\ \bibinfo {pages} {416} (\bibinfo {year} {1957})}\BibitemShut
  {NoStop}%
\bibitem [{\citenamefont {Hubbard}(1959)}]{Hubbard_1959}%
  \BibitemOpen
  \bibfield  {author} {\bibinfo {author} {\bibfnamefont {J.}~\bibnamefont
  {Hubbard}},\ }\bibfield  {title} {\bibinfo {title} {Calculation of partition
  functions},\ }\href {https://doi.org/10.1103/PhysRevLett.3.77} {\bibfield
  {journal} {\bibinfo  {journal} {Phys. Rev. Lett.}\ }\textbf {\bibinfo
  {volume} {3}},\ \bibinfo {pages} {77} (\bibinfo {year} {1959})}\BibitemShut
  {NoStop}%
\bibitem [{\citenamefont {Kitaev}(2003)}]{Kitaev2003}%
  \BibitemOpen
  \bibfield  {author} {\bibinfo {author} {\bibfnamefont {A.}~\bibnamefont
  {Kitaev}},\ }\bibfield  {title} {\bibinfo {title} {Fault-tolerant quantum
  computation by anyons},\ }\href
  {https://doi.org/10.1016/s0003-4916(02)00018-0} {\bibfield  {journal}
  {\bibinfo  {journal} {Annals of Physics}\ }\textbf {\bibinfo {volume}
  {303}},\ \bibinfo {pages} {2} (\bibinfo {year} {2003})}\BibitemShut {NoStop}%
\bibitem [{\citenamefont {Kogut}(1979)}]{Kogut_RMP}%
  \BibitemOpen
  \bibfield  {author} {\bibinfo {author} {\bibfnamefont {J.~B.}\ \bibnamefont
  {Kogut}},\ }\bibfield  {title} {\bibinfo {title} {An introduction to lattice
  gauge theory and spin systems},\ }\href
  {https://doi.org/10.1103/RevModPhys.51.659} {\bibfield  {journal} {\bibinfo
  {journal} {Rev. Mod. Phys.}\ }\textbf {\bibinfo {volume} {51}},\ \bibinfo
  {pages} {659} (\bibinfo {year} {1979})}\BibitemShut {NoStop}%
\bibitem [{\citenamefont {Wahl}\ \emph {et~al.}(2013)\citenamefont {Wahl},
  \citenamefont {Tu}, \citenamefont {Schuch},\ and\ \citenamefont
  {Cirac}}]{Wahl_2013}%
  \BibitemOpen
  \bibfield  {author} {\bibinfo {author} {\bibfnamefont {T.~B.}\ \bibnamefont
  {Wahl}}, \bibinfo {author} {\bibfnamefont {H.-H.}\ \bibnamefont {Tu}},
  \bibinfo {author} {\bibfnamefont {N.}~\bibnamefont {Schuch}},\ and\ \bibinfo
  {author} {\bibfnamefont {J.~I.}\ \bibnamefont {Cirac}},\ }\bibfield  {title}
  {\bibinfo {title} {Projected entangled-pair states can describe chiral
  topological states},\ }\href {https://doi.org/10.1103/PhysRevLett.111.236805}
  {\bibfield  {journal} {\bibinfo  {journal} {Phys. Rev. Lett.}\ }\textbf
  {\bibinfo {volume} {111}},\ \bibinfo {pages} {236805} (\bibinfo {year}
  {2013})}\BibitemShut {NoStop}%
\bibitem [{\citenamefont {Haegeman}\ \emph {et~al.}(2015)\citenamefont
  {Haegeman}, \citenamefont {Van~Acoleyen}, \citenamefont {Schuch},
  \citenamefont {Cirac},\ and\ \citenamefont {Verstraete}}]{Jutho_2015}%
  \BibitemOpen
  \bibfield  {author} {\bibinfo {author} {\bibfnamefont {J.}~\bibnamefont
  {Haegeman}}, \bibinfo {author} {\bibfnamefont {K.}~\bibnamefont
  {Van~Acoleyen}}, \bibinfo {author} {\bibfnamefont {N.}~\bibnamefont
  {Schuch}}, \bibinfo {author} {\bibfnamefont {J.~I.}\ \bibnamefont {Cirac}},\
  and\ \bibinfo {author} {\bibfnamefont {F.}~\bibnamefont {Verstraete}},\
  }\bibfield  {title} {\bibinfo {title} {Gauging quantum states: From global to
  local symmetries in many-body systems},\ }\href
  {https://doi.org/10.1103/PhysRevX.5.011024} {\bibfield  {journal} {\bibinfo
  {journal} {Phys. Rev. X}\ }\textbf {\bibinfo {volume} {5}},\ \bibinfo {pages}
  {011024} (\bibinfo {year} {2015})}\BibitemShut {NoStop}%
\bibitem [{\citenamefont {Wegner}(1971)}]{Wegner_1971}%
  \BibitemOpen
  \bibfield  {author} {\bibinfo {author} {\bibfnamefont {F.~J.}\ \bibnamefont
  {Wegner}},\ }\bibfield  {title} {\bibinfo {title} {{Duality in Generalized
  Ising Models and Phase Transitions without Local Order Parameters}},\ }\href
  {https://doi.org/10.1063/1.1665530} {\bibfield  {journal} {\bibinfo
  {journal} {Journal of Mathematical Physics}\ }\textbf {\bibinfo {volume}
  {12}},\ \bibinfo {pages} {2259} (\bibinfo {year} {1971})}\BibitemShut
  {NoStop}%
\bibitem [{\citenamefont {Kitaev}(2006)}]{KITAEV2006}%
  \BibitemOpen
  \bibfield  {author} {\bibinfo {author} {\bibfnamefont {A.}~\bibnamefont
  {Kitaev}},\ }\bibfield  {title} {\bibinfo {title} {Anyons in an exactly
  solved model and beyond},\ }\href
  {https://doi.org/https://doi.org/10.1016/j.aop.2005.10.005} {\bibfield
  {journal} {\bibinfo  {journal} {Annals of Physics}\ }\textbf {\bibinfo
  {volume} {321}},\ \bibinfo {pages} {2} (\bibinfo {year} {2006})}\BibitemShut
  {NoStop}%
\bibitem [{\citenamefont {Baskaran}\ \emph {et~al.}(2007)\citenamefont
  {Baskaran}, \citenamefont {Mandal},\ and\ \citenamefont
  {Shankar}}]{Baskaran2007}%
  \BibitemOpen
  \bibfield  {author} {\bibinfo {author} {\bibfnamefont {G.}~\bibnamefont
  {Baskaran}}, \bibinfo {author} {\bibfnamefont {S.}~\bibnamefont {Mandal}},\
  and\ \bibinfo {author} {\bibfnamefont {R.}~\bibnamefont {Shankar}},\
  }\bibfield  {title} {\bibinfo {title} {Exact results for spin dynamics and
  fractionalization in the kitaev model},\ }\href
  {https://doi.org/10.1103/PhysRevLett.98.247201} {\bibfield  {journal}
  {\bibinfo  {journal} {Phys. Rev. Lett.}\ }\textbf {\bibinfo {volume} {98}},\
  \bibinfo {pages} {247201} (\bibinfo {year} {2007})}\BibitemShut {NoStop}%
\bibitem [{\citenamefont {Tupitsyn}\ \emph {et~al.}(2010)\citenamefont
  {Tupitsyn}, \citenamefont {Kitaev}, \citenamefont {Prokof'ev},\ and\
  \citenamefont {Stamp}}]{Tupitsyn_2010}%
  \BibitemOpen
  \bibfield  {author} {\bibinfo {author} {\bibfnamefont {I.~S.}\ \bibnamefont
  {Tupitsyn}}, \bibinfo {author} {\bibfnamefont {A.}~\bibnamefont {Kitaev}},
  \bibinfo {author} {\bibfnamefont {N.~V.}\ \bibnamefont {Prokof'ev}},\ and\
  \bibinfo {author} {\bibfnamefont {P.~C.~E.}\ \bibnamefont {Stamp}},\
  }\bibfield  {title} {\bibinfo {title} {Topological multicritical point in the
  phase diagram of the toric code model and three-dimensional lattice gauge
  higgs model},\ }\href {https://doi.org/10.1103/PhysRevB.82.085114} {\bibfield
   {journal} {\bibinfo  {journal} {Phys. Rev. B}\ }\textbf {\bibinfo {volume}
  {82}},\ \bibinfo {pages} {085114} (\bibinfo {year} {2010})}\BibitemShut
  {NoStop}%
\bibitem [{\citenamefont {Gu}\ and\ \citenamefont {Wen}(2014)}]{Gu_2014}%
  \BibitemOpen
  \bibfield  {author} {\bibinfo {author} {\bibfnamefont {Z.-C.}\ \bibnamefont
  {Gu}}\ and\ \bibinfo {author} {\bibfnamefont {X.-G.}\ \bibnamefont {Wen}},\
  }\bibfield  {title} {\bibinfo {title} {Symmetry-protected topological orders
  for interacting fermions: Fermionic topological nonlinear
  $\ensuremath{\sigma}$ models and a special group supercohomology theory},\
  }\href {https://doi.org/10.1103/PhysRevB.90.115141} {\bibfield  {journal}
  {\bibinfo  {journal} {Phys. Rev. B}\ }\textbf {\bibinfo {volume} {90}},\
  \bibinfo {pages} {115141} (\bibinfo {year} {2014})}\BibitemShut {NoStop}%
\bibitem [{\citenamefont {Patel}\ and\ \citenamefont
  {Trivedi}(2019)}]{Nandini2019}%
  \BibitemOpen
  \bibfield  {author} {\bibinfo {author} {\bibfnamefont {N.~D.}\ \bibnamefont
  {Patel}}\ and\ \bibinfo {author} {\bibfnamefont {N.}~\bibnamefont
  {Trivedi}},\ }\bibfield  {title} {\bibinfo {title} {Magnetic field-induced
  intermediate quantum spin liquid with a spinon fermi surface},\ }\href
  {https://doi.org/10.1073/pnas.1821406116} {\bibfield  {journal} {\bibinfo
  {journal} {Proc. Natl. Acad. Sci. U.S.A.}\ }\textbf {\bibinfo {volume}
  {116}},\ \bibinfo {pages} {12199} (\bibinfo {year} {2019})}\BibitemShut
  {NoStop}%
\bibitem [{\citenamefont {Gohlke}\ \emph {et~al.}(2018)\citenamefont {Gohlke},
  \citenamefont {Moessner},\ and\ \citenamefont {Pollmann}}]{Pollmann2018}%
  \BibitemOpen
  \bibfield  {author} {\bibinfo {author} {\bibfnamefont {M.}~\bibnamefont
  {Gohlke}}, \bibinfo {author} {\bibfnamefont {R.}~\bibnamefont {Moessner}},\
  and\ \bibinfo {author} {\bibfnamefont {F.}~\bibnamefont {Pollmann}},\
  }\bibfield  {title} {\bibinfo {title} {Dynamical and topological properties
  of the kitaev model in a [111] magnetic field},\ }\href
  {https://doi.org/10.1103/PhysRevB.98.014418} {\bibfield  {journal} {\bibinfo
  {journal} {Phys. Rev. B}\ }\textbf {\bibinfo {volume} {98}},\ \bibinfo
  {pages} {014418} (\bibinfo {year} {2018})}\BibitemShut {NoStop}%
\bibitem [{\citenamefont {Zhu}\ \emph {et~al.}(2018)\citenamefont {Zhu},
  \citenamefont {Kimchi}, \citenamefont {Sheng},\ and\ \citenamefont
  {Fu}}]{LiangFu2018}%
  \BibitemOpen
  \bibfield  {author} {\bibinfo {author} {\bibfnamefont {Z.}~\bibnamefont
  {Zhu}}, \bibinfo {author} {\bibfnamefont {I.}~\bibnamefont {Kimchi}},
  \bibinfo {author} {\bibfnamefont {D.~N.}\ \bibnamefont {Sheng}},\ and\
  \bibinfo {author} {\bibfnamefont {L.}~\bibnamefont {Fu}},\ }\bibfield
  {title} {\bibinfo {title} {Robust non-abelian spin liquid and a possible
  intermediate phase in the antiferromagnetic kitaev model with magnetic
  field},\ }\href {https://doi.org/10.1103/PhysRevB.97.241110} {\bibfield
  {journal} {\bibinfo  {journal} {Phys. Rev. B}\ }\textbf {\bibinfo {volume}
  {97}},\ \bibinfo {pages} {241110} (\bibinfo {year} {2018})}\BibitemShut
  {NoStop}%
\bibitem [{\citenamefont {Jiang}\ \emph {et~al.}(2018)\citenamefont {Jiang},
  \citenamefont {Wang}, \citenamefont {Huang},\ and\ \citenamefont
  {Lu}}]{MiangLu2018}%
  \BibitemOpen
  \bibfield  {author} {\bibinfo {author} {\bibfnamefont {H.-C.}\ \bibnamefont
  {Jiang}}, \bibinfo {author} {\bibfnamefont {C.-Y.}\ \bibnamefont {Wang}},
  \bibinfo {author} {\bibfnamefont {B.}~\bibnamefont {Huang}},\ and\ \bibinfo
  {author} {\bibfnamefont {Y.-M.}\ \bibnamefont {Lu}},\ }\bibfield  {title}
  {\bibinfo {title} {Field induced quantum spin liquid with spinon fermi
  surfaces in the kitaev model},\ }\bibfield  {journal} {\bibinfo  {journal}
  {arXiv}\ }\href {https://doi.org/10.48550/ARXIV.1809.08247}
  {10.48550/ARXIV.1809.08247} (\bibinfo {year} {2018})\BibitemShut {NoStop}%
\bibitem [{\citenamefont {Hickey}\ and\ \citenamefont
  {Trebst}(2019)}]{Trebst2019}%
  \BibitemOpen
  \bibfield  {author} {\bibinfo {author} {\bibfnamefont {C.}~\bibnamefont
  {Hickey}}\ and\ \bibinfo {author} {\bibfnamefont {S.}~\bibnamefont
  {Trebst}},\ }\bibfield  {title} {\bibinfo {title} {Emergence of a
  field-driven u (1) spin liquid in the kitaev honeycomb model},\ }\href
  {https://doi.org/10.1038/s41467-019-08459-9} {\bibfield  {journal} {\bibinfo
  {journal} {Nat. Commun.}\ }\textbf {\bibinfo {volume} {10}},\ \bibinfo
  {pages} {1} (\bibinfo {year} {2019})}\BibitemShut {NoStop}%
\bibitem [{\citenamefont {Kaib}\ \emph {et~al.}(2019)\citenamefont {Kaib},
  \citenamefont {Winter},\ and\ \citenamefont {Valent\'{\i}}}]{Valenti2019}%
  \BibitemOpen
  \bibfield  {author} {\bibinfo {author} {\bibfnamefont {D.~A.~S.}\
  \bibnamefont {Kaib}}, \bibinfo {author} {\bibfnamefont {S.~M.}\ \bibnamefont
  {Winter}},\ and\ \bibinfo {author} {\bibfnamefont {R.}~\bibnamefont
  {Valent\'{\i}}},\ }\bibfield  {title} {\bibinfo {title} {Kitaev honeycomb
  models in magnetic fields: Dynamical response and dual models},\ }\href
  {https://doi.org/10.1103/PhysRevB.100.144445} {\bibfield  {journal} {\bibinfo
   {journal} {Phys. Rev. B}\ }\textbf {\bibinfo {volume} {100}},\ \bibinfo
  {pages} {144445} (\bibinfo {year} {2019})}\BibitemShut {NoStop}%
\bibitem [{\citenamefont {S\o{}rensen}\ \emph {et~al.}(2021)\citenamefont
  {S\o{}rensen}, \citenamefont {Catuneanu}, \citenamefont {Gordon},\ and\
  \citenamefont {Kee}}]{Sorensen2021}%
  \BibitemOpen
  \bibfield  {author} {\bibinfo {author} {\bibfnamefont {E.~S.}\ \bibnamefont
  {S\o{}rensen}}, \bibinfo {author} {\bibfnamefont {A.}~\bibnamefont
  {Catuneanu}}, \bibinfo {author} {\bibfnamefont {J.~S.}\ \bibnamefont
  {Gordon}},\ and\ \bibinfo {author} {\bibfnamefont {H.-Y.}\ \bibnamefont
  {Kee}},\ }\bibfield  {title} {\bibinfo {title} {Heart of entanglement:
  Chiral, nematic, and incommensurate phases in the kitaev-gamma ladder in a
  field},\ }\href {https://doi.org/10.1103/PhysRevX.11.011013} {\bibfield
  {journal} {\bibinfo  {journal} {Phys. Rev. X}\ }\textbf {\bibinfo {volume}
  {11}},\ \bibinfo {pages} {011013} (\bibinfo {year} {2021})}\BibitemShut
  {NoStop}%
\bibitem [{\citenamefont {Lahtinen}\ \emph {et~al.}(2012)\citenamefont
  {Lahtinen}, \citenamefont {Ludwig}, \citenamefont {Pachos},\ and\
  \citenamefont {Trebst}}]{Trebst_2012}%
  \BibitemOpen
  \bibfield  {author} {\bibinfo {author} {\bibfnamefont {V.}~\bibnamefont
  {Lahtinen}}, \bibinfo {author} {\bibfnamefont {A.~W.~W.}\ \bibnamefont
  {Ludwig}}, \bibinfo {author} {\bibfnamefont {J.~K.}\ \bibnamefont {Pachos}},\
  and\ \bibinfo {author} {\bibfnamefont {S.}~\bibnamefont {Trebst}},\
  }\bibfield  {title} {\bibinfo {title} {Topological liquid nucleation induced
  by vortex-vortex interactions in kitaev's honeycomb model},\ }\href
  {https://doi.org/10.1103/PhysRevB.86.075115} {\bibfield  {journal} {\bibinfo
  {journal} {Phys. Rev. B}\ }\textbf {\bibinfo {volume} {86}},\ \bibinfo
  {pages} {075115} (\bibinfo {year} {2012})}\BibitemShut {NoStop}%
\bibitem [{\citenamefont {Lahtinen}\ \emph {et~al.}(2014)\citenamefont
  {Lahtinen}, \citenamefont {Ludwig},\ and\ \citenamefont
  {Trebst}}]{Trebst_2014}%
  \BibitemOpen
  \bibfield  {author} {\bibinfo {author} {\bibfnamefont {V.}~\bibnamefont
  {Lahtinen}}, \bibinfo {author} {\bibfnamefont {A.~W.~W.}\ \bibnamefont
  {Ludwig}},\ and\ \bibinfo {author} {\bibfnamefont {S.}~\bibnamefont
  {Trebst}},\ }\bibfield  {title} {\bibinfo {title} {Perturbed vortex lattices
  and the stability of nucleated topological phases},\ }\href
  {https://doi.org/10.1103/PhysRevB.89.085121} {\bibfield  {journal} {\bibinfo
  {journal} {Phys. Rev. B}\ }\textbf {\bibinfo {volume} {89}},\ \bibinfo
  {pages} {085121} (\bibinfo {year} {2014})}\BibitemShut {NoStop}%
\bibitem [{\citenamefont {Zhang}\ \emph {et~al.}(2019)\citenamefont {Zhang},
  \citenamefont {Wang}, \citenamefont {Hal\'asz},\ and\ \citenamefont
  {Batista}}]{Batista2019}%
  \BibitemOpen
  \bibfield  {author} {\bibinfo {author} {\bibfnamefont {S.-S.}\ \bibnamefont
  {Zhang}}, \bibinfo {author} {\bibfnamefont {Z.}~\bibnamefont {Wang}},
  \bibinfo {author} {\bibfnamefont {G.~B.}\ \bibnamefont {Hal\'asz}},\ and\
  \bibinfo {author} {\bibfnamefont {C.~D.}\ \bibnamefont {Batista}},\
  }\bibfield  {title} {\bibinfo {title} {Vison crystals in an extended kitaev
  model on the honeycomb lattice},\ }\href
  {https://doi.org/10.1103/PhysRevLett.123.057201} {\bibfield  {journal}
  {\bibinfo  {journal} {Phys. Rev. Lett.}\ }\textbf {\bibinfo {volume} {123}},\
  \bibinfo {pages} {057201} (\bibinfo {year} {2019})}\BibitemShut {NoStop}%
\bibitem [{\citenamefont {Zhang}\ \emph {et~al.}(2020)\citenamefont {Zhang},
  \citenamefont {Batista},\ and\ \citenamefont
  {Hal\'asz}}]{Zhang_BAtista_2020}%
  \BibitemOpen
  \bibfield  {author} {\bibinfo {author} {\bibfnamefont {S.-S.}\ \bibnamefont
  {Zhang}}, \bibinfo {author} {\bibfnamefont {C.~D.}\ \bibnamefont {Batista}},\
  and\ \bibinfo {author} {\bibfnamefont {G.~B.}\ \bibnamefont {Hal\'asz}},\
  }\bibfield  {title} {\bibinfo {title} {Toward kitaev's sixteenfold way in a
  honeycomb lattice model},\ }\href
  {https://doi.org/10.1103/PhysRevResearch.2.023334} {\bibfield  {journal}
  {\bibinfo  {journal} {Phys. Rev. Res.}\ }\textbf {\bibinfo {volume} {2}},\
  \bibinfo {pages} {023334} (\bibinfo {year} {2020})}\BibitemShut {NoStop}%
\bibitem [{\citenamefont {Biswas}(2013)}]{Rudro2013}%
  \BibitemOpen
  \bibfield  {author} {\bibinfo {author} {\bibfnamefont {R.~R.}\ \bibnamefont
  {Biswas}},\ }\bibfield  {title} {\bibinfo {title} {Majorana fermions in
  vortex lattices},\ }\href {https://doi.org/10.1103/PhysRevLett.111.136401}
  {\bibfield  {journal} {\bibinfo  {journal} {Phys. Rev. Lett.}\ }\textbf
  {\bibinfo {volume} {111}},\ \bibinfo {pages} {136401} (\bibinfo {year}
  {2013})}\BibitemShut {NoStop}%
\bibitem [{\citenamefont {Nasu}\ \emph {et~al.}(2017)\citenamefont {Nasu},
  \citenamefont {Yoshitake},\ and\ \citenamefont {Motome}}]{Nasu2017}%
  \BibitemOpen
  \bibfield  {author} {\bibinfo {author} {\bibfnamefont {J.}~\bibnamefont
  {Nasu}}, \bibinfo {author} {\bibfnamefont {J.}~\bibnamefont {Yoshitake}},\
  and\ \bibinfo {author} {\bibfnamefont {Y.}~\bibnamefont {Motome}},\
  }\bibfield  {title} {\bibinfo {title} {Thermal transport in the kitaev
  model},\ }\href {https://doi.org/10.1103/PhysRevLett.119.127204} {\bibfield
  {journal} {\bibinfo  {journal} {Phys. Rev. Lett.}\ }\textbf {\bibinfo
  {volume} {119}},\ \bibinfo {pages} {127204} (\bibinfo {year}
  {2017})}\BibitemShut {NoStop}%
\bibitem [{\citenamefont {Fuchs}\ \emph {et~al.}(2020)\citenamefont {Fuchs},
  \citenamefont {Patil},\ and\ \citenamefont {Vidal}}]{Vidal_2020}%
  \BibitemOpen
  \bibfield  {author} {\bibinfo {author} {\bibfnamefont {J.-N.}\ \bibnamefont
  {Fuchs}}, \bibinfo {author} {\bibfnamefont {S.}~\bibnamefont {Patil}},\ and\
  \bibinfo {author} {\bibfnamefont {J.}~\bibnamefont {Vidal}},\ }\bibfield
  {title} {\bibinfo {title} {Parity of chern numbers in the kitaev honeycomb
  model and the sixteenfold way},\ }\href
  {https://doi.org/10.1103/PhysRevB.102.115130} {\bibfield  {journal} {\bibinfo
   {journal} {Phys. Rev. B}\ }\textbf {\bibinfo {volume} {102}},\ \bibinfo
  {pages} {115130} (\bibinfo {year} {2020})}\BibitemShut {NoStop}%
\bibitem [{\citenamefont {Koga}\ \emph {et~al.}(2021)\citenamefont {Koga},
  \citenamefont {Murakami},\ and\ \citenamefont {Nasu}}]{Nasu2021}%
  \BibitemOpen
  \bibfield  {author} {\bibinfo {author} {\bibfnamefont {A.}~\bibnamefont
  {Koga}}, \bibinfo {author} {\bibfnamefont {Y.}~\bibnamefont {Murakami}},\
  and\ \bibinfo {author} {\bibfnamefont {J.}~\bibnamefont {Nasu}},\ }\bibfield
  {title} {\bibinfo {title} {Majorana correlations in the kitaev model with
  ordered-flux structures},\ }\href
  {https://doi.org/10.1103/PhysRevB.103.214421} {\bibfield  {journal} {\bibinfo
   {journal} {Phys. Rev. B}\ }\textbf {\bibinfo {volume} {103}},\ \bibinfo
  {pages} {214421} (\bibinfo {year} {2021})}\BibitemShut {NoStop}%
\bibitem [{\citenamefont {Chulliparambil}\ \emph {et~al.}(2021)\citenamefont
  {Chulliparambil}, \citenamefont {Janssen}, \citenamefont {Vojta},
  \citenamefont {Tu},\ and\ \citenamefont {Seifert}}]{Vojta2021}%
  \BibitemOpen
  \bibfield  {author} {\bibinfo {author} {\bibfnamefont {S.}~\bibnamefont
  {Chulliparambil}}, \bibinfo {author} {\bibfnamefont {L.}~\bibnamefont
  {Janssen}}, \bibinfo {author} {\bibfnamefont {M.}~\bibnamefont {Vojta}},
  \bibinfo {author} {\bibfnamefont {H.-H.}\ \bibnamefont {Tu}},\ and\ \bibinfo
  {author} {\bibfnamefont {U.~F.~P.}\ \bibnamefont {Seifert}},\ }\bibfield
  {title} {\bibinfo {title} {Flux crystals, majorana metals, and flat bands in
  exactly solvable spin-orbital liquids},\ }\href
  {https://doi.org/10.1103/PhysRevB.103.075144} {\bibfield  {journal} {\bibinfo
   {journal} {Phys. Rev. B}\ }\textbf {\bibinfo {volume} {103}},\ \bibinfo
  {pages} {075144} (\bibinfo {year} {2021})}\BibitemShut {NoStop}%
\bibitem [{\citenamefont {Hashimoto}\ \emph {et~al.}(2023)\citenamefont
  {Hashimoto}, \citenamefont {Murakami},\ and\ \citenamefont
  {Koga}}]{Koga2023}%
  \BibitemOpen
  \bibfield  {author} {\bibinfo {author} {\bibfnamefont {A.}~\bibnamefont
  {Hashimoto}}, \bibinfo {author} {\bibfnamefont {Y.}~\bibnamefont
  {Murakami}},\ and\ \bibinfo {author} {\bibfnamefont {A.}~\bibnamefont
  {Koga}},\ }\bibfield  {title} {\bibinfo {title} {Majorana gap formation in
  the anisotropic kitaev model with ordered flux configuration},\ }\href
  {https://doi.org/10.1103/PhysRevB.107.174428} {\bibfield  {journal} {\bibinfo
   {journal} {Phys. Rev. B}\ }\textbf {\bibinfo {volume} {107}},\ \bibinfo
  {pages} {174428} (\bibinfo {year} {2023})}\BibitemShut {NoStop}%
\bibitem [{\citenamefont {Alspaugh}\ \emph {et~al.}(2024)\citenamefont
  {Alspaugh}, \citenamefont {Fuchs}, \citenamefont {Ritz-Zwilling},\ and\
  \citenamefont {Vidal}}]{Vidal_2024}%
  \BibitemOpen
  \bibfield  {author} {\bibinfo {author} {\bibfnamefont {D.~J.}\ \bibnamefont
  {Alspaugh}}, \bibinfo {author} {\bibfnamefont {J.-N.}\ \bibnamefont {Fuchs}},
  \bibinfo {author} {\bibfnamefont {A.}~\bibnamefont {Ritz-Zwilling}},\ and\
  \bibinfo {author} {\bibfnamefont {J.}~\bibnamefont {Vidal}},\ }\bibfield
  {title} {\bibinfo {title} {Effective models for dense vortex lattices in the
  kitaev honeycomb model},\ }\href
  {https://doi.org/10.1103/PhysRevB.109.115107} {\bibfield  {journal} {\bibinfo
   {journal} {Phys. Rev. B}\ }\textbf {\bibinfo {volume} {109}},\ \bibinfo
  {pages} {115107} (\bibinfo {year} {2024})}\BibitemShut {NoStop}%
\bibitem [{\citenamefont {Ledwith}\ \emph {et~al.}(2023)\citenamefont
  {Ledwith}, \citenamefont {Vishwanath},\ and\ \citenamefont
  {Parker}}]{Ledwith2023}%
  \BibitemOpen
  \bibfield  {author} {\bibinfo {author} {\bibfnamefont {P.~J.}\ \bibnamefont
  {Ledwith}}, \bibinfo {author} {\bibfnamefont {A.}~\bibnamefont
  {Vishwanath}},\ and\ \bibinfo {author} {\bibfnamefont {D.~E.}\ \bibnamefont
  {Parker}},\ }\bibfield  {title} {\bibinfo {title} {Vortexability: A unifying
  criterion for ideal fractional chern insulators},\ }\href
  {https://doi.org/10.1103/PhysRevB.108.205144} {\bibfield  {journal} {\bibinfo
   {journal} {Phys. Rev. B}\ }\textbf {\bibinfo {volume} {108}},\ \bibinfo
  {pages} {205144} (\bibinfo {year} {2023})}\BibitemShut {NoStop}%
\bibitem [{\citenamefont {Otten}\ \emph {et~al.}(2019)\citenamefont {Otten},
  \citenamefont {Roy},\ and\ \citenamefont {Hassler}}]{Fabian2019}%
  \BibitemOpen
  \bibfield  {author} {\bibinfo {author} {\bibfnamefont {D.}~\bibnamefont
  {Otten}}, \bibinfo {author} {\bibfnamefont {A.}~\bibnamefont {Roy}},\ and\
  \bibinfo {author} {\bibfnamefont {F.}~\bibnamefont {Hassler}},\ }\bibfield
  {title} {\bibinfo {title} {Dynamical structure factor in the non-abelian
  phase of the kitaev honeycomb model in the presence of quenched disorder},\
  }\href {https://doi.org/10.1103/PhysRevB.99.035137} {\bibfield  {journal}
  {\bibinfo  {journal} {Phys. Rev. B}\ }\textbf {\bibinfo {volume} {99}},\
  \bibinfo {pages} {035137} (\bibinfo {year} {2019})}\BibitemShut {NoStop}%
\bibitem [{\citenamefont {Feng}\ \emph {et~al.}(2020)\citenamefont {Feng},
  \citenamefont {Perkins},\ and\ \citenamefont {Burnell}}]{Feng2020}%
  \BibitemOpen
  \bibfield  {author} {\bibinfo {author} {\bibfnamefont {K.}~\bibnamefont
  {Feng}}, \bibinfo {author} {\bibfnamefont {N.~B.}\ \bibnamefont {Perkins}},\
  and\ \bibinfo {author} {\bibfnamefont {F.~J.}\ \bibnamefont {Burnell}},\
  }\bibfield  {title} {\bibinfo {title} {Further insights into the
  thermodynamics of the kitaev honeycomb model},\ }\href
  {https://doi.org/10.1103/PhysRevB.102.224402} {\bibfield  {journal} {\bibinfo
   {journal} {Phys. Rev. B}\ }\textbf {\bibinfo {volume} {102}},\ \bibinfo
  {pages} {224402} (\bibinfo {year} {2020})}\BibitemShut {NoStop}%
\bibitem [{\citenamefont {Metavitsiadis}\ and\ \citenamefont
  {Brenig}(2021)}]{Brenig2021}%
  \BibitemOpen
  \bibfield  {author} {\bibinfo {author} {\bibfnamefont {A.}~\bibnamefont
  {Metavitsiadis}}\ and\ \bibinfo {author} {\bibfnamefont {W.}~\bibnamefont
  {Brenig}},\ }\bibfield  {title} {\bibinfo {title} {Flux mobility
  delocalization in the kitaev spin ladder},\ }\href
  {https://doi.org/10.1103/PhysRevB.103.195102} {\bibfield  {journal} {\bibinfo
   {journal} {Phys. Rev. B}\ }\textbf {\bibinfo {volume} {103}},\ \bibinfo
  {pages} {195102} (\bibinfo {year} {2021})}\BibitemShut {NoStop}%
\bibitem [{\citenamefont {Zhu}\ and\ \citenamefont {Heyl}(2021)}]{Heyl2021}%
  \BibitemOpen
  \bibfield  {author} {\bibinfo {author} {\bibfnamefont {G.-Y.}\ \bibnamefont
  {Zhu}}\ and\ \bibinfo {author} {\bibfnamefont {M.}~\bibnamefont {Heyl}},\
  }\bibfield  {title} {\bibinfo {title} {Subdiffusive dynamics and critical
  quantum correlations in a disorder-free localized kitaev honeycomb model out
  of equilibrium},\ }\href {https://doi.org/10.1103/PhysRevResearch.3.L032069}
  {\bibfield  {journal} {\bibinfo  {journal} {Phys. Rev. Research}\ }\textbf
  {\bibinfo {volume} {3}},\ \bibinfo {pages} {L032069} (\bibinfo {year}
  {2021})}\BibitemShut {NoStop}%
\bibitem [{\citenamefont {Kao}\ \emph {et~al.}(2021)\citenamefont {Kao},
  \citenamefont {Knolle}, \citenamefont {Hal\'asz}, \citenamefont {Moessner},\
  and\ \citenamefont {Perkins}}]{Perkins2021}%
  \BibitemOpen
  \bibfield  {author} {\bibinfo {author} {\bibfnamefont {W.-H.}\ \bibnamefont
  {Kao}}, \bibinfo {author} {\bibfnamefont {J.}~\bibnamefont {Knolle}},
  \bibinfo {author} {\bibfnamefont {G.~B.}\ \bibnamefont {Hal\'asz}}, \bibinfo
  {author} {\bibfnamefont {R.}~\bibnamefont {Moessner}},\ and\ \bibinfo
  {author} {\bibfnamefont {N.~B.}\ \bibnamefont {Perkins}},\ }\bibfield
  {title} {\bibinfo {title} {Vacancy-induced low-energy density of states in
  the kitaev spin liquid},\ }\href {https://doi.org/10.1103/PhysRevX.11.011034}
  {\bibfield  {journal} {\bibinfo  {journal} {Phys. Rev. X}\ }\textbf {\bibinfo
  {volume} {11}},\ \bibinfo {pages} {011034} (\bibinfo {year}
  {2021})}\BibitemShut {NoStop}%
\bibitem [{\citenamefont {Kao}\ and\ \citenamefont
  {Perkins}(2021)}]{Kao2021168506}%
  \BibitemOpen
  \bibfield  {author} {\bibinfo {author} {\bibfnamefont {W.-H.}\ \bibnamefont
  {Kao}}\ and\ \bibinfo {author} {\bibfnamefont {N.~B.}\ \bibnamefont
  {Perkins}},\ }\bibfield  {title} {\bibinfo {title} {Disorder upon disorder:
  Localization effects in the kitaev spin liquid},\ }\href
  {https://doi.org/https://doi.org/10.1016/j.aop.2021.168506} {\bibfield
  {journal} {\bibinfo  {journal} {Annals of Physics}\ }\textbf {\bibinfo
  {volume} {435}},\ \bibinfo {pages} {168506} (\bibinfo {year} {2021})},\
  \bibinfo {note} {special issue on Philip W. Anderson}\BibitemShut {NoStop}%
\bibitem [{\citenamefont {Pereira}\ and\ \citenamefont
  {Egger}(2020)}]{Reinhold2020}%
  \BibitemOpen
  \bibfield  {author} {\bibinfo {author} {\bibfnamefont {R.~G.}\ \bibnamefont
  {Pereira}}\ and\ \bibinfo {author} {\bibfnamefont {R.}~\bibnamefont
  {Egger}},\ }\bibfield  {title} {\bibinfo {title} {Electrical access to ising
  anyons in kitaev spin liquids},\ }\href
  {https://doi.org/10.1103/PhysRevLett.125.227202} {\bibfield  {journal}
  {\bibinfo  {journal} {Phys. Rev. Lett.}\ }\textbf {\bibinfo {volume} {125}},\
  \bibinfo {pages} {227202} (\bibinfo {year} {2020})}\BibitemShut {NoStop}%
\bibitem [{\citenamefont {Udagawa}\ \emph {et~al.}(2021)\citenamefont
  {Udagawa}, \citenamefont {Takayoshi},\ and\ \citenamefont
  {Oka}}]{takashi2021}%
  \BibitemOpen
  \bibfield  {author} {\bibinfo {author} {\bibfnamefont {M.}~\bibnamefont
  {Udagawa}}, \bibinfo {author} {\bibfnamefont {S.}~\bibnamefont {Takayoshi}},\
  and\ \bibinfo {author} {\bibfnamefont {T.}~\bibnamefont {Oka}},\ }\bibfield
  {title} {\bibinfo {title} {Scanning tunneling microscopy as a single majorana
  detector of kitaev's chiral spin liquid},\ }\href
  {https://doi.org/10.1103/PhysRevLett.126.127201} {\bibfield  {journal}
  {\bibinfo  {journal} {Phys. Rev. Lett.}\ }\textbf {\bibinfo {volume} {126}},\
  \bibinfo {pages} {127201} (\bibinfo {year} {2021})}\BibitemShut {NoStop}%
\bibitem [{\citenamefont {Kang}\ and\ \citenamefont {Vafek}(2018)}]{Vafek2018}%
  \BibitemOpen
  \bibfield  {author} {\bibinfo {author} {\bibfnamefont {J.}~\bibnamefont
  {Kang}}\ and\ \bibinfo {author} {\bibfnamefont {O.}~\bibnamefont {Vafek}},\
  }\bibfield  {title} {\bibinfo {title} {Symmetry, maximally localized wannier
  states, and a low-energy model for twisted bilayer graphene narrow bands},\
  }\href {https://doi.org/10.1103/PhysRevX.8.031088} {\bibfield  {journal}
  {\bibinfo  {journal} {Phys. Rev. X}\ }\textbf {\bibinfo {volume} {8}},\
  \bibinfo {pages} {031088} (\bibinfo {year} {2018})}\BibitemShut {NoStop}%
\bibitem [{\citenamefont {{Peotta}}\ and\ \citenamefont
  {{T{\"o}rm{\"a}}}(2015)}]{Peotta_2015}%
  \BibitemOpen
  \bibfield  {author} {\bibinfo {author} {\bibfnamefont {S.}~\bibnamefont
  {{Peotta}}}\ and\ \bibinfo {author} {\bibfnamefont {P.}~\bibnamefont
  {{T{\"o}rm{\"a}}}},\ }\bibfield  {title} {\bibinfo {title} {{Superfluidity in
  topologically nontrivial flat bands}},\ }\href
  {https://doi.org/10.1038/ncomms9944} {\bibfield  {journal} {\bibinfo
  {journal} {Nature Communications}\ }\textbf {\bibinfo {volume} {6}},\
  \bibinfo {eid} {8944} (\bibinfo {year} {2015})}\BibitemShut {NoStop}%
\bibitem [{\citenamefont {Herzog-Arbeitman}\ \emph
  {et~al.}(2022{\natexlab{a}})\citenamefont {Herzog-Arbeitman}, \citenamefont
  {Peri}, \citenamefont {Schindler}, \citenamefont {Huber},\ and\ \citenamefont
  {Bernevig}}]{Herzog_2022}%
  \BibitemOpen
  \bibfield  {author} {\bibinfo {author} {\bibfnamefont {J.}~\bibnamefont
  {Herzog-Arbeitman}}, \bibinfo {author} {\bibfnamefont {V.}~\bibnamefont
  {Peri}}, \bibinfo {author} {\bibfnamefont {F.}~\bibnamefont {Schindler}},
  \bibinfo {author} {\bibfnamefont {S.~D.}\ \bibnamefont {Huber}},\ and\
  \bibinfo {author} {\bibfnamefont {B.~A.}\ \bibnamefont {Bernevig}},\
  }\bibfield  {title} {\bibinfo {title} {Superfluid weight bounds from symmetry
  and quantum geometry in flat bands},\ }\href
  {https://doi.org/10.1103/PhysRevLett.128.087002} {\bibfield  {journal}
  {\bibinfo  {journal} {Phys. Rev. Lett.}\ }\textbf {\bibinfo {volume} {128}},\
  \bibinfo {pages} {087002} (\bibinfo {year} {2022}{\natexlab{a}})}\BibitemShut
  {NoStop}%
\bibitem [{\citenamefont {Zou}\ \emph {et~al.}(2018)\citenamefont {Zou},
  \citenamefont {Po}, \citenamefont {Vishwanath},\ and\ \citenamefont
  {Senthil}}]{Zou_Vishwanath_2018}%
  \BibitemOpen
  \bibfield  {author} {\bibinfo {author} {\bibfnamefont {L.}~\bibnamefont
  {Zou}}, \bibinfo {author} {\bibfnamefont {H.~C.}\ \bibnamefont {Po}},
  \bibinfo {author} {\bibfnamefont {A.}~\bibnamefont {Vishwanath}},\ and\
  \bibinfo {author} {\bibfnamefont {T.}~\bibnamefont {Senthil}},\ }\bibfield
  {title} {\bibinfo {title} {Band structure of twisted bilayer graphene:
  Emergent symmetries, commensurate approximants, and wannier obstructions},\
  }\href {https://doi.org/10.1103/PhysRevB.98.085435} {\bibfield  {journal}
  {\bibinfo  {journal} {Phys. Rev. B}\ }\textbf {\bibinfo {volume} {98}},\
  \bibinfo {pages} {085435} (\bibinfo {year} {2018})}\BibitemShut {NoStop}%
\bibitem [{\citenamefont {Kruchkov}(2022)}]{Kruchkov_2022}%
  \BibitemOpen
  \bibfield  {author} {\bibinfo {author} {\bibfnamefont {A.}~\bibnamefont
  {Kruchkov}},\ }\bibfield  {title} {\bibinfo {title} {Quantum geometry, flat
  chern bands, and wannier orbital quantization},\ }\href
  {https://doi.org/10.1103/PhysRevB.105.L241102} {\bibfield  {journal}
  {\bibinfo  {journal} {Phys. Rev. B}\ }\textbf {\bibinfo {volume} {105}},\
  \bibinfo {pages} {L241102} (\bibinfo {year} {2022})}\BibitemShut {NoStop}%
\bibitem [{\citenamefont {Koshino}\ \emph {et~al.}(2018)\citenamefont
  {Koshino}, \citenamefont {Yuan}, \citenamefont {Koretsune}, \citenamefont
  {Ochi}, \citenamefont {Kuroki},\ and\ \citenamefont {Fu}}]{Koshino_2018}%
  \BibitemOpen
  \bibfield  {author} {\bibinfo {author} {\bibfnamefont {M.}~\bibnamefont
  {Koshino}}, \bibinfo {author} {\bibfnamefont {N.~F.~Q.}\ \bibnamefont
  {Yuan}}, \bibinfo {author} {\bibfnamefont {T.}~\bibnamefont {Koretsune}},
  \bibinfo {author} {\bibfnamefont {M.}~\bibnamefont {Ochi}}, \bibinfo {author}
  {\bibfnamefont {K.}~\bibnamefont {Kuroki}},\ and\ \bibinfo {author}
  {\bibfnamefont {L.}~\bibnamefont {Fu}},\ }\bibfield  {title} {\bibinfo
  {title} {Maximally localized wannier orbitals and the extended hubbard model
  for twisted bilayer graphene},\ }\href
  {https://doi.org/10.1103/PhysRevX.8.031087} {\bibfield  {journal} {\bibinfo
  {journal} {Phys. Rev. X}\ }\textbf {\bibinfo {volume} {8}},\ \bibinfo {pages}
  {031087} (\bibinfo {year} {2018})}\BibitemShut {NoStop}%
\bibitem [{\citenamefont {Schindler}\ \emph {et~al.}(2020)\citenamefont
  {Schindler}, \citenamefont {Bradlyn}, \citenamefont {Fischer},\ and\
  \citenamefont {Neupert}}]{Schindler_Neupert_2020}%
  \BibitemOpen
  \bibfield  {author} {\bibinfo {author} {\bibfnamefont {F.}~\bibnamefont
  {Schindler}}, \bibinfo {author} {\bibfnamefont {B.}~\bibnamefont {Bradlyn}},
  \bibinfo {author} {\bibfnamefont {M.~H.}\ \bibnamefont {Fischer}},\ and\
  \bibinfo {author} {\bibfnamefont {T.}~\bibnamefont {Neupert}},\ }\bibfield
  {title} {\bibinfo {title} {Pairing obstructions in topological
  superconductors},\ }\href {https://doi.org/10.1103/PhysRevLett.124.247001}
  {\bibfield  {journal} {\bibinfo  {journal} {Phys. Rev. Lett.}\ }\textbf
  {\bibinfo {volume} {124}},\ \bibinfo {pages} {247001} (\bibinfo {year}
  {2020})}\BibitemShut {NoStop}%
\bibitem [{Note1()}]{Note1}%
  \BibitemOpen
  \bibinfo {note} {We make an approximation that the single-particle
  dispersion, many-body interaction, and superconducting order parameters are
  short-ranged, restricting to a few nearest neighbors only. (This truncation
  of the Fourier series to a polynomial of few sites gives a finite width of
  the single-particle states in both position and momentum space, and the
  number of nearest neighbors $\protect \mathcal {N}$ to be considered is
  determined within a numerical procedure by fitting to the band structure at
  all ${\protect \bf k}$-points. This yields the so-called compact localized
  orbitals for the flat band in the Wannierization procedure).}\BibitemShut
  {Stop}%
\bibitem [{Note2()}]{Note2}%
  \BibitemOpen
  \bibinfo {note} {Note that in our procedure, it is easy to implement the
  lattice (point-/space-) group symmetry by doing the invariant rotation on the
  Bloch phase spinor ${\protect \bf Z}_{n}({\protect \bf k})$.}\BibitemShut
  {Stop}%
\bibitem [{\citenamefont {Rhim}\ and\ \citenamefont {Yang}(2019)}]{Rhim_2019}%
  \BibitemOpen
  \bibfield  {author} {\bibinfo {author} {\bibfnamefont {J.-W.}\ \bibnamefont
  {Rhim}}\ and\ \bibinfo {author} {\bibfnamefont {B.-J.}\ \bibnamefont
  {Yang}},\ }\bibfield  {title} {\bibinfo {title} {Classification of flat bands
  according to the band-crossing singularity of bloch wave functions},\ }\href
  {https://doi.org/10.1103/PhysRevB.99.045107} {\bibfield  {journal} {\bibinfo
  {journal} {Phys. Rev. B}\ }\textbf {\bibinfo {volume} {99}},\ \bibinfo
  {pages} {045107} (\bibinfo {year} {2019})}\BibitemShut {NoStop}%
\bibitem [{\citenamefont {Zhang}\ and\ \citenamefont
  {Jin}(2020{\natexlab{a}})}]{Zhang_Jin_2020}%
  \BibitemOpen
  \bibfield  {author} {\bibinfo {author} {\bibfnamefont {S.~M.}\ \bibnamefont
  {Zhang}}\ and\ \bibinfo {author} {\bibfnamefont {L.}~\bibnamefont {Jin}},\
  }\bibfield  {title} {\bibinfo {title} {Compact localized states and
  localization dynamics in the dice lattice},\ }\href
  {https://doi.org/10.1103/PhysRevB.102.054301} {\bibfield  {journal} {\bibinfo
   {journal} {Phys. Rev. B}\ }\textbf {\bibinfo {volume} {102}},\ \bibinfo
  {pages} {054301} (\bibinfo {year} {2020}{\natexlab{a}})}\BibitemShut
  {NoStop}%
\bibitem [{\citenamefont {{Onishi}}\ and\ \citenamefont
  {{Fu}}(2024)}]{Liang_2024}%
  \BibitemOpen
  \bibfield  {author} {\bibinfo {author} {\bibfnamefont {Y.}~\bibnamefont
  {{Onishi}}}\ and\ \bibinfo {author} {\bibfnamefont {L.}~\bibnamefont
  {{Fu}}},\ }\bibfield  {title} {\bibinfo {title} {{Fundamental Bound on
  Topological Gap}},\ }\href {https://doi.org/10.1103/PhysRevX.14.011052}
  {\bibfield  {journal} {\bibinfo  {journal} {Physical Review X}\ }\textbf
  {\bibinfo {volume} {14}},\ \bibinfo {eid} {011052} (\bibinfo {year}
  {2024})},\ \Eprint {https://arxiv.org/abs/2306.00078} {arXiv:2306.00078
  [cond-mat.mes-hall]} \BibitemShut {NoStop}%
\bibitem [{\citenamefont {Bouhon}\ \emph {et~al.}(2023)\citenamefont {Bouhon},
  \citenamefont {Timmel},\ and\ \citenamefont {Slager}}]{Bouhon2023}%
  \BibitemOpen
  \bibfield  {author} {\bibinfo {author} {\bibfnamefont {A.}~\bibnamefont
  {Bouhon}}, \bibinfo {author} {\bibfnamefont {A.}~\bibnamefont {Timmel}},\
  and\ \bibinfo {author} {\bibfnamefont {R.-J.}\ \bibnamefont {Slager}},\
  }\href {https://arxiv.org/abs/2303.02180} {\bibinfo {title} {Quantum geometry
  beyond projective single bands}} (\bibinfo {year} {2023}),\ \Eprint
  {https://arxiv.org/abs/2303.02180} {arXiv:2303.02180 [cond-mat.mes-hall]}
  \BibitemShut {NoStop}%
\bibitem [{\citenamefont {Das}(2013)}]{DasWeyl}%
  \BibitemOpen
  \bibfield  {author} {\bibinfo {author} {\bibfnamefont {T.}~\bibnamefont
  {Das}},\ }\bibfield  {title} {\bibinfo {title} {Weyl semimetal and
  superconductor designed in an orbital-selective superlattice},\ }\href
  {https://doi.org/10.1103/PhysRevB.88.035444} {\bibfield  {journal} {\bibinfo
  {journal} {Phys. Rev. B}\ }\textbf {\bibinfo {volume} {88}},\ \bibinfo
  {pages} {035444} (\bibinfo {year} {2013})}\BibitemShut {NoStop}%
\bibitem [{Note3()}]{Note3}%
  \BibitemOpen
  \bibinfo {note} {The higher-energy bands also carry finite Chern numbers,
  resulting in a total Chern number of $C=\pm 2$ for the filled and empty
  bands, consistent with Ref.~\cite {Vidal_2020}. However, for the construction
  of the effective lattice model, we focus exclusively on the two low-energy
  bands with $C=\pm 1$.}\BibitemShut {Stop}%
\bibitem [{\citenamefont {Mostofi}\ \emph {et~al.}(2014)\citenamefont
  {Mostofi}, \citenamefont {Yates}, \citenamefont {Pizzi}, \citenamefont {Lee},
  \citenamefont {Souza}, \citenamefont {Vanderbilt},\ and\ \citenamefont
  {Marzari}}]{Wannier90_MOSTOFI}%
  \BibitemOpen
  \bibfield  {author} {\bibinfo {author} {\bibfnamefont {A.~A.}\ \bibnamefont
  {Mostofi}}, \bibinfo {author} {\bibfnamefont {J.~R.}\ \bibnamefont {Yates}},
  \bibinfo {author} {\bibfnamefont {G.}~\bibnamefont {Pizzi}}, \bibinfo
  {author} {\bibfnamefont {Y.-S.}\ \bibnamefont {Lee}}, \bibinfo {author}
  {\bibfnamefont {I.}~\bibnamefont {Souza}}, \bibinfo {author} {\bibfnamefont
  {D.}~\bibnamefont {Vanderbilt}},\ and\ \bibinfo {author} {\bibfnamefont
  {N.}~\bibnamefont {Marzari}},\ }\bibfield  {title} {\bibinfo {title} {An
  updated version of wannier90: A tool for obtaining maximally-localised
  wannier functions},\ }\href
  {https://doi.org/https://doi.org/10.1016/j.cpc.2014.05.003} {\bibfield
  {journal} {\bibinfo  {journal} {Computer Physics Communications}\ }\textbf
  {\bibinfo {volume} {185}},\ \bibinfo {pages} {2309} (\bibinfo {year}
  {2014})}\BibitemShut {NoStop}%
\bibitem [{\citenamefont {Harada}\ \emph {et~al.}(2023)\citenamefont {Harada},
  \citenamefont {Ono},\ and\ \citenamefont {Nasu}}]{Nasu_2023}%
  \BibitemOpen
  \bibfield  {author} {\bibinfo {author} {\bibfnamefont {C.}~\bibnamefont
  {Harada}}, \bibinfo {author} {\bibfnamefont {A.}~\bibnamefont {Ono}},\ and\
  \bibinfo {author} {\bibfnamefont {J.}~\bibnamefont {Nasu}},\ }\bibfield
  {title} {\bibinfo {title} {Field-driven spatiotemporal manipulation of
  majorana zero modes in a kitaev spin liquid},\ }\href
  {https://doi.org/10.1103/PhysRevB.108.L241118} {\bibfield  {journal}
  {\bibinfo  {journal} {Phys. Rev. B}\ }\textbf {\bibinfo {volume} {108}},\
  \bibinfo {pages} {L241118} (\bibinfo {year} {2023})}\BibitemShut {NoStop}%
\bibitem [{\citenamefont {Thouless}\ \emph {et~al.}(1982)\citenamefont
  {Thouless}, \citenamefont {Kohmoto}, \citenamefont {Nightingale},\ and\
  \citenamefont {den Nijs}}]{Thouless_1982}%
  \BibitemOpen
  \bibfield  {author} {\bibinfo {author} {\bibfnamefont {D.~J.}\ \bibnamefont
  {Thouless}}, \bibinfo {author} {\bibfnamefont {M.}~\bibnamefont {Kohmoto}},
  \bibinfo {author} {\bibfnamefont {M.~P.}\ \bibnamefont {Nightingale}},\ and\
  \bibinfo {author} {\bibfnamefont {M.}~\bibnamefont {den Nijs}},\ }\bibfield
  {title} {\bibinfo {title} {Quantized hall conductance in a two-dimensional
  periodic potential},\ }\href {https://doi.org/10.1103/PhysRevLett.49.405}
  {\bibfield  {journal} {\bibinfo  {journal} {Phys. Rev. Lett.}\ }\textbf
  {\bibinfo {volume} {49}},\ \bibinfo {pages} {405} (\bibinfo {year}
  {1982})}\BibitemShut {NoStop}%
\bibitem [{Note4()}]{Note4}%
  \BibitemOpen
  \bibinfo {note} {The symmetric tensor $\eta _{\mu \nu }$ for the hexagonal
  lattice becomes $\eta _{\mu \nu } = \protect \hat {\protect \mathbf {G}_{\mu
  }}.\protect \hat {\protect \mathbf {G}_{\nu }} = \begin {pmatrix} 1 & -1/2 \\
  -1/2 & 1 \end {pmatrix}. $}\BibitemShut {NoStop}%
\bibitem [{\citenamefont {Kitaev}\ and\ \citenamefont
  {Preskill}(2006)}]{TEE_Kitaev}%
  \BibitemOpen
  \bibfield  {author} {\bibinfo {author} {\bibfnamefont {A.}~\bibnamefont
  {Kitaev}}\ and\ \bibinfo {author} {\bibfnamefont {J.}~\bibnamefont
  {Preskill}},\ }\bibfield  {title} {\bibinfo {title} {Topological entanglement
  entropy},\ }\href {https://doi.org/10.1103/PhysRevLett.96.110404} {\bibfield
  {journal} {\bibinfo  {journal} {Phys. Rev. Lett.}\ }\textbf {\bibinfo
  {volume} {96}},\ \bibinfo {pages} {110404} (\bibinfo {year}
  {2006})}\BibitemShut {NoStop}%
\bibitem [{\citenamefont {Levin}\ and\ \citenamefont {Wen}(2006)}]{Levin_TEE}%
  \BibitemOpen
  \bibfield  {author} {\bibinfo {author} {\bibfnamefont {M.}~\bibnamefont
  {Levin}}\ and\ \bibinfo {author} {\bibfnamefont {X.-G.}\ \bibnamefont
  {Wen}},\ }\bibfield  {title} {\bibinfo {title} {Detecting topological order
  in a ground state wave function},\ }\href
  {https://doi.org/10.1103/PhysRevLett.96.110405} {\bibfield  {journal}
  {\bibinfo  {journal} {Phys. Rev. Lett.}\ }\textbf {\bibinfo {volume} {96}},\
  \bibinfo {pages} {110405} (\bibinfo {year} {2006})}\BibitemShut {NoStop}%
\bibitem [{\citenamefont {Nehra}\ \emph {et~al.}(2020)\citenamefont {Nehra},
  \citenamefont {Bhakuni}, \citenamefont {Ramachandran},\ and\ \citenamefont
  {Sharma}}]{Nehra_2020}%
  \BibitemOpen
  \bibfield  {author} {\bibinfo {author} {\bibfnamefont {R.}~\bibnamefont
  {Nehra}}, \bibinfo {author} {\bibfnamefont {D.~S.}\ \bibnamefont {Bhakuni}},
  \bibinfo {author} {\bibfnamefont {A.}~\bibnamefont {Ramachandran}},\ and\
  \bibinfo {author} {\bibfnamefont {A.}~\bibnamefont {Sharma}},\ }\bibfield
  {title} {\bibinfo {title} {Flat bands and entanglement in the kitaev
  ladder},\ }\href {https://doi.org/10.1103/PhysRevResearch.2.013175}
  {\bibfield  {journal} {\bibinfo  {journal} {Phys. Rev. Res.}\ }\textbf
  {\bibinfo {volume} {2}},\ \bibinfo {pages} {013175} (\bibinfo {year}
  {2020})}\BibitemShut {NoStop}%
\bibitem [{\citenamefont {Kuno}(2020)}]{Kuno_2020}%
  \BibitemOpen
  \bibfield  {author} {\bibinfo {author} {\bibfnamefont {Y.}~\bibnamefont
  {Kuno}},\ }\bibfield  {title} {\bibinfo {title} {Extended flat band,
  entanglement, and topological properties in a creutz ladder},\ }\href
  {https://doi.org/10.1103/PhysRevB.101.184112} {\bibfield  {journal} {\bibinfo
   {journal} {Phys. Rev. B}\ }\textbf {\bibinfo {volume} {101}},\ \bibinfo
  {pages} {184112} (\bibinfo {year} {2020})}\BibitemShut {NoStop}%
\bibitem [{\citenamefont {Brouder}\ \emph {et~al.}(2007)\citenamefont
  {Brouder}, \citenamefont {Panati}, \citenamefont {Calandra}, \citenamefont
  {Mourougane},\ and\ \citenamefont {Marzari}}]{Christian2007}%
  \BibitemOpen
  \bibfield  {author} {\bibinfo {author} {\bibfnamefont {C.}~\bibnamefont
  {Brouder}}, \bibinfo {author} {\bibfnamefont {G.}~\bibnamefont {Panati}},
  \bibinfo {author} {\bibfnamefont {M.}~\bibnamefont {Calandra}}, \bibinfo
  {author} {\bibfnamefont {C.}~\bibnamefont {Mourougane}},\ and\ \bibinfo
  {author} {\bibfnamefont {N.}~\bibnamefont {Marzari}},\ }\bibfield  {title}
  {\bibinfo {title} {Exponential localization of wannier functions in
  insulators},\ }\href {https://doi.org/10.1103/PhysRevLett.98.046402}
  {\bibfield  {journal} {\bibinfo  {journal} {Phys. Rev. Lett.}\ }\textbf
  {\bibinfo {volume} {98}},\ \bibinfo {pages} {046402} (\bibinfo {year}
  {2007})}\BibitemShut {NoStop}%
\bibitem [{\citenamefont {Monaco}\ \emph {et~al.}(2018)\citenamefont {Monaco},
  \citenamefont {Panati}, \citenamefont {Pisante},\ and\ \citenamefont
  {Teufel}}]{Monaco2018}%
  \BibitemOpen
  \bibfield  {author} {\bibinfo {author} {\bibfnamefont {D.}~\bibnamefont
  {Monaco}}, \bibinfo {author} {\bibfnamefont {G.}~\bibnamefont {Panati}},
  \bibinfo {author} {\bibfnamefont {A.}~\bibnamefont {Pisante}},\ and\ \bibinfo
  {author} {\bibfnamefont {S.}~\bibnamefont {Teufel}},\ }\bibfield  {title}
  {\bibinfo {title} {Optimal decay of wannier functions in chern and quantum
  hall insulators},\ }\href {https://doi.org/10.1007/s00220-017-3067-7}
  {\bibfield  {journal} {\bibinfo  {journal} {Communications in Mathematical
  Physics}\ }\textbf {\bibinfo {volume} {359}},\ \bibinfo {pages} {61}
  (\bibinfo {year} {2018})}\BibitemShut {NoStop}%
\bibitem [{\citenamefont {Peotta}\ and\ \citenamefont
  {Törmä}(2015)}]{Peotta2015}%
  \BibitemOpen
  \bibfield  {author} {\bibinfo {author} {\bibfnamefont {S.}~\bibnamefont
  {Peotta}}\ and\ \bibinfo {author} {\bibfnamefont {P.}~\bibnamefont
  {Törmä}},\ }\bibfield  {title} {\bibinfo {title} {Superfluidity in
  topologically nontrivial flat bands},\ }\bibfield  {journal} {\bibinfo
  {journal} {Nature Communications}\ }\textbf {\bibinfo {volume} {6}},\ \href
  {https://doi.org/10.1038/ncomms9944} {10.1038/ncomms9944} (\bibinfo {year}
  {2015})\BibitemShut {NoStop}%
\bibitem [{\citenamefont {Törmä}\ \emph {et~al.}(2022)\citenamefont
  {Törmä}, \citenamefont {Peotta},\ and\ \citenamefont
  {Bernevig}}]{Torma2022}%
  \BibitemOpen
  \bibfield  {author} {\bibinfo {author} {\bibfnamefont {P.}~\bibnamefont
  {Törmä}}, \bibinfo {author} {\bibfnamefont {S.}~\bibnamefont {Peotta}},\
  and\ \bibinfo {author} {\bibfnamefont {B.~A.}\ \bibnamefont {Bernevig}},\
  }\bibfield  {title} {\bibinfo {title} {Superconductivity, superfluidity and
  quantum geometry in twisted multilayer systems},\ }\href
  {https://doi.org/10.1038/s42254-022-00466-y} {\bibfield  {journal} {\bibinfo
  {journal} {Nature Reviews Physics}\ }\textbf {\bibinfo {volume} {4}},\
  \bibinfo {pages} {528} (\bibinfo {year} {2022})}\BibitemShut {NoStop}%
\bibitem [{\citenamefont {Neupert}\ \emph {et~al.}(2011)\citenamefont
  {Neupert}, \citenamefont {Santos}, \citenamefont {Chamon},\ and\
  \citenamefont {Mudry}}]{Neupert_2011}%
  \BibitemOpen
  \bibfield  {author} {\bibinfo {author} {\bibfnamefont {T.}~\bibnamefont
  {Neupert}}, \bibinfo {author} {\bibfnamefont {L.}~\bibnamefont {Santos}},
  \bibinfo {author} {\bibfnamefont {C.}~\bibnamefont {Chamon}},\ and\ \bibinfo
  {author} {\bibfnamefont {C.}~\bibnamefont {Mudry}},\ }\bibfield  {title}
  {\bibinfo {title} {Fractional quantum hall states at zero magnetic field},\
  }\href {https://doi.org/10.1103/PhysRevLett.106.236804} {\bibfield  {journal}
  {\bibinfo  {journal} {Phys. Rev. Lett.}\ }\textbf {\bibinfo {volume} {106}},\
  \bibinfo {pages} {236804} (\bibinfo {year} {2011})}\BibitemShut {NoStop}%
\bibitem [{\citenamefont {Tang}\ \emph {et~al.}(2011)\citenamefont {Tang},
  \citenamefont {Mei},\ and\ \citenamefont {Wen}}]{Tang_2011}%
  \BibitemOpen
  \bibfield  {author} {\bibinfo {author} {\bibfnamefont {E.}~\bibnamefont
  {Tang}}, \bibinfo {author} {\bibfnamefont {J.-W.}\ \bibnamefont {Mei}},\ and\
  \bibinfo {author} {\bibfnamefont {X.-G.}\ \bibnamefont {Wen}},\ }\bibfield
  {title} {\bibinfo {title} {High-temperature fractional quantum hall states},\
  }\href {https://doi.org/10.1103/PhysRevLett.106.236802} {\bibfield  {journal}
  {\bibinfo  {journal} {Phys. Rev. Lett.}\ }\textbf {\bibinfo {volume} {106}},\
  \bibinfo {pages} {236802} (\bibinfo {year} {2011})}\BibitemShut {NoStop}%
\bibitem [{\citenamefont {Sun}\ \emph {et~al.}(2011)\citenamefont {Sun},
  \citenamefont {Gu}, \citenamefont {Katsura},\ and\ \citenamefont
  {Das~Sarma}}]{Sun_2011}%
  \BibitemOpen
  \bibfield  {author} {\bibinfo {author} {\bibfnamefont {K.}~\bibnamefont
  {Sun}}, \bibinfo {author} {\bibfnamefont {Z.}~\bibnamefont {Gu}}, \bibinfo
  {author} {\bibfnamefont {H.}~\bibnamefont {Katsura}},\ and\ \bibinfo {author}
  {\bibfnamefont {S.}~\bibnamefont {Das~Sarma}},\ }\bibfield  {title} {\bibinfo
  {title} {Nearly flatbands with nontrivial topology},\ }\href
  {https://doi.org/10.1103/PhysRevLett.106.236803} {\bibfield  {journal}
  {\bibinfo  {journal} {Phys. Rev. Lett.}\ }\textbf {\bibinfo {volume} {106}},\
  \bibinfo {pages} {236803} (\bibinfo {year} {2011})}\BibitemShut {NoStop}%
\bibitem [{\citenamefont {Neupert}\ \emph {et~al.}(2015)\citenamefont
  {Neupert}, \citenamefont {Chamon}, \citenamefont {Iadecola}, \citenamefont
  {Santos},\ and\ \citenamefont {Mudry}}]{FCI_Neupert_2015}%
  \BibitemOpen
  \bibfield  {author} {\bibinfo {author} {\bibfnamefont {T.}~\bibnamefont
  {Neupert}}, \bibinfo {author} {\bibfnamefont {C.}~\bibnamefont {Chamon}},
  \bibinfo {author} {\bibfnamefont {T.}~\bibnamefont {Iadecola}}, \bibinfo
  {author} {\bibfnamefont {L.~H.}\ \bibnamefont {Santos}},\ and\ \bibinfo
  {author} {\bibfnamefont {C.}~\bibnamefont {Mudry}},\ }\bibfield  {title}
  {\bibinfo {title} {Fractional (chern and topological) insulators},\ }\href
  {https://doi.org/10.1088/0031-8949/2015/T164/014005} {\bibfield  {journal}
  {\bibinfo  {journal} {Physica Scripta}\ }\textbf {\bibinfo {volume} {2015}},\
  \bibinfo {pages} {014005} (\bibinfo {year} {2015})}\BibitemShut {NoStop}%
\bibitem [{\citenamefont {Regnault}\ and\ \citenamefont
  {Bernevig}(2011)}]{Regnault_2011}%
  \BibitemOpen
  \bibfield  {author} {\bibinfo {author} {\bibfnamefont {N.}~\bibnamefont
  {Regnault}}\ and\ \bibinfo {author} {\bibfnamefont {B.~A.}\ \bibnamefont
  {Bernevig}},\ }\bibfield  {title} {\bibinfo {title} {Fractional chern
  insulator},\ }\href {https://doi.org/10.1103/PhysRevX.1.021014} {\bibfield
  {journal} {\bibinfo  {journal} {Phys. Rev. X}\ }\textbf {\bibinfo {volume}
  {1}},\ \bibinfo {pages} {021014} (\bibinfo {year} {2011})}\BibitemShut
  {NoStop}%
\bibitem [{\citenamefont {Liu}\ \emph {et~al.}(2012)\citenamefont {Liu},
  \citenamefont {Bergholtz}, \citenamefont {Fan},\ and\ \citenamefont
  {L\"auchli}}]{FCI_Liu_2012}%
  \BibitemOpen
  \bibfield  {author} {\bibinfo {author} {\bibfnamefont {Z.}~\bibnamefont
  {Liu}}, \bibinfo {author} {\bibfnamefont {E.~J.}\ \bibnamefont {Bergholtz}},
  \bibinfo {author} {\bibfnamefont {H.}~\bibnamefont {Fan}},\ and\ \bibinfo
  {author} {\bibfnamefont {A.~M.}\ \bibnamefont {L\"auchli}},\ }\bibfield
  {title} {\bibinfo {title} {Fractional chern insulators in topological flat
  bands with higher chern number},\ }\href
  {https://doi.org/10.1103/PhysRevLett.109.186805} {\bibfield  {journal}
  {\bibinfo  {journal} {Phys. Rev. Lett.}\ }\textbf {\bibinfo {volume} {109}},\
  \bibinfo {pages} {186805} (\bibinfo {year} {2012})}\BibitemShut {NoStop}%
\bibitem [{\citenamefont {Parameswaran}\ \emph {et~al.}(2013)\citenamefont
  {Parameswaran}, \citenamefont {Roy},\ and\ \citenamefont
  {Sondhi}}]{PARAMESWARAN_2013}%
  \BibitemOpen
  \bibfield  {author} {\bibinfo {author} {\bibfnamefont {S.~A.}\ \bibnamefont
  {Parameswaran}}, \bibinfo {author} {\bibfnamefont {R.}~\bibnamefont {Roy}},\
  and\ \bibinfo {author} {\bibfnamefont {S.~L.}\ \bibnamefont {Sondhi}},\
  }\bibfield  {title} {\bibinfo {title} {Fractional quantum hall physics in
  topological flat bands},\ }\href
  {https://doi.org/https://doi.org/10.1016/j.crhy.2013.04.003} {\bibfield
  {journal} {\bibinfo  {journal} {Comptes Rendus Physique}\ }\textbf {\bibinfo
  {volume} {14}},\ \bibinfo {pages} {816} (\bibinfo {year} {2013})},\ \bibinfo
  {note} {topological insulators / Isolants topologiques}\BibitemShut {NoStop}%
\bibitem [{\citenamefont {Barkeshli}\ \emph {et~al.}(2015)\citenamefont
  {Barkeshli}, \citenamefont {Yao},\ and\ \citenamefont
  {Laumann}}]{FCI_Laumann_2015}%
  \BibitemOpen
  \bibfield  {author} {\bibinfo {author} {\bibfnamefont {M.}~\bibnamefont
  {Barkeshli}}, \bibinfo {author} {\bibfnamefont {N.~Y.}\ \bibnamefont {Yao}},\
  and\ \bibinfo {author} {\bibfnamefont {C.~R.}\ \bibnamefont {Laumann}},\
  }\bibfield  {title} {\bibinfo {title} {Continuous preparation of a fractional
  chern insulator},\ }\href {https://doi.org/10.1103/PhysRevLett.115.026802}
  {\bibfield  {journal} {\bibinfo  {journal} {Phys. Rev. Lett.}\ }\textbf
  {\bibinfo {volume} {115}},\ \bibinfo {pages} {026802} (\bibinfo {year}
  {2015})}\BibitemShut {NoStop}%
\bibitem [{\citenamefont {M\"oller}\ and\ \citenamefont
  {Cooper}(2015)}]{FCI_Cooper_2015}%
  \BibitemOpen
  \bibfield  {author} {\bibinfo {author} {\bibfnamefont {G.}~\bibnamefont
  {M\"oller}}\ and\ \bibinfo {author} {\bibfnamefont {N.~R.}\ \bibnamefont
  {Cooper}},\ }\bibfield  {title} {\bibinfo {title} {Fractional chern
  insulators in harper-hofstadter bands with higher chern number},\ }\href
  {https://doi.org/10.1103/PhysRevLett.115.126401} {\bibfield  {journal}
  {\bibinfo  {journal} {Phys. Rev. Lett.}\ }\textbf {\bibinfo {volume} {115}},\
  \bibinfo {pages} {126401} (\bibinfo {year} {2015})}\BibitemShut {NoStop}%
\bibitem [{\citenamefont {Behrmann}\ \emph {et~al.}(2016)\citenamefont
  {Behrmann}, \citenamefont {Liu},\ and\ \citenamefont
  {Bergholtz}}]{FCI_Behrmann_2016}%
  \BibitemOpen
  \bibfield  {author} {\bibinfo {author} {\bibfnamefont {J.}~\bibnamefont
  {Behrmann}}, \bibinfo {author} {\bibfnamefont {Z.}~\bibnamefont {Liu}},\ and\
  \bibinfo {author} {\bibfnamefont {E.~J.}\ \bibnamefont {Bergholtz}},\
  }\bibfield  {title} {\bibinfo {title} {Model fractional chern insulators},\
  }\href {https://doi.org/10.1103/PhysRevLett.116.216802} {\bibfield  {journal}
  {\bibinfo  {journal} {Phys. Rev. Lett.}\ }\textbf {\bibinfo {volume} {116}},\
  \bibinfo {pages} {216802} (\bibinfo {year} {2016})}\BibitemShut {NoStop}%
\bibitem [{\citenamefont {Lee}\ \emph {et~al.}(2017)\citenamefont {Lee},
  \citenamefont {Claassen},\ and\ \citenamefont {Thomale}}]{FCI_Thomale_2017}%
  \BibitemOpen
  \bibfield  {author} {\bibinfo {author} {\bibfnamefont {C.~H.}\ \bibnamefont
  {Lee}}, \bibinfo {author} {\bibfnamefont {M.}~\bibnamefont {Claassen}},\ and\
  \bibinfo {author} {\bibfnamefont {R.}~\bibnamefont {Thomale}},\ }\bibfield
  {title} {\bibinfo {title} {Band structure engineering of ideal fractional
  chern insulators},\ }\href {https://doi.org/10.1103/PhysRevB.96.165150}
  {\bibfield  {journal} {\bibinfo  {journal} {Phys. Rev. B}\ }\textbf {\bibinfo
  {volume} {96}},\ \bibinfo {pages} {165150} (\bibinfo {year}
  {2017})}\BibitemShut {NoStop}%
\bibitem [{\citenamefont {Liu}\ and\ \citenamefont
  {Bergholtz}(2024)}]{FCI_Zhao_2024}%
  \BibitemOpen
  \bibfield  {author} {\bibinfo {author} {\bibfnamefont {Z.}~\bibnamefont
  {Liu}}\ and\ \bibinfo {author} {\bibfnamefont {E.~J.}\ \bibnamefont
  {Bergholtz}},\ }\bibfield  {title} {\bibinfo {title} {Recent developments in
  fractional chern insulators},\ }in\ \href
  {https://doi.org/https://doi.org/10.1016/B978-0-323-90800-9.00136-0} {\emph
  {\bibinfo {booktitle} {Encyclopedia of Condensed Matter Physics (Second
  Edition)}}},\ \bibinfo {editor} {edited by\ \bibinfo {editor} {\bibfnamefont
  {T.}~\bibnamefont {Chakraborty}}}\ (\bibinfo  {publisher} {Academic Press},\
  \bibinfo {address} {Oxford},\ \bibinfo {year} {2024})\ \bibinfo {edition}
  {second edition}\ ed.,\ pp.\ \bibinfo {pages} {515--538}\BibitemShut
  {NoStop}%
\bibitem [{\citenamefont {Gupta}\ and\ \citenamefont {Das}(2017)}]{Gupta_2017}%
  \BibitemOpen
  \bibfield  {author} {\bibinfo {author} {\bibfnamefont {G.~K.}\ \bibnamefont
  {Gupta}}\ and\ \bibinfo {author} {\bibfnamefont {T.}~\bibnamefont {Das}},\
  }\bibfield  {title} {\bibinfo {title} {Quantum spin hall density wave
  insulator of correlated fermions},\ }\href
  {https://doi.org/10.1103/PhysRevB.95.161109} {\bibfield  {journal} {\bibinfo
  {journal} {Phys. Rev. B}\ }\textbf {\bibinfo {volume} {95}},\ \bibinfo
  {pages} {161109} (\bibinfo {year} {2017})}\BibitemShut {NoStop}%
\bibitem [{\citenamefont {Roy}(2014)}]{Roy_2014}%
  \BibitemOpen
  \bibfield  {author} {\bibinfo {author} {\bibfnamefont {R.}~\bibnamefont
  {Roy}},\ }\bibfield  {title} {\bibinfo {title} {Band geometry of fractional
  topological insulators},\ }\href {https://doi.org/10.1103/PhysRevB.90.165139}
  {\bibfield  {journal} {\bibinfo  {journal} {Phys. Rev. B}\ }\textbf {\bibinfo
  {volume} {90}},\ \bibinfo {pages} {165139} (\bibinfo {year}
  {2014})}\BibitemShut {NoStop}%
\bibitem [{\citenamefont {Simon}\ and\ \citenamefont
  {Rudner}(2020)}]{Simon_2020}%
  \BibitemOpen
  \bibfield  {author} {\bibinfo {author} {\bibfnamefont {S.~H.}\ \bibnamefont
  {Simon}}\ and\ \bibinfo {author} {\bibfnamefont {M.~S.}\ \bibnamefont
  {Rudner}},\ }\bibfield  {title} {\bibinfo {title} {Contrasting lattice
  geometry dependent versus independent quantities: Ramifications for berry
  curvature, energy gaps, and dynamics},\ }\href
  {https://doi.org/10.1103/PhysRevB.102.165148} {\bibfield  {journal} {\bibinfo
   {journal} {Phys. Rev. B}\ }\textbf {\bibinfo {volume} {102}},\ \bibinfo
  {pages} {165148} (\bibinfo {year} {2020})}\BibitemShut {NoStop}%
\bibitem [{\citenamefont {Elitzur}(1975)}]{Elitzur_1975}%
  \BibitemOpen
  \bibfield  {author} {\bibinfo {author} {\bibfnamefont {S.}~\bibnamefont
  {Elitzur}},\ }\bibfield  {title} {\bibinfo {title} {Impossibility of
  spontaneously breaking local symmetries},\ }\href
  {https://doi.org/10.1103/PhysRevD.12.3978} {\bibfield  {journal} {\bibinfo
  {journal} {Phys. Rev. D}\ }\textbf {\bibinfo {volume} {12}},\ \bibinfo
  {pages} {3978} (\bibinfo {year} {1975})}\BibitemShut {NoStop}%
\bibitem [{\citenamefont {Wen}(2002)}]{wen_psg_2002}%
  \BibitemOpen
  \bibfield  {author} {\bibinfo {author} {\bibfnamefont {X.-G.}\ \bibnamefont
  {Wen}},\ }\bibfield  {title} {\bibinfo {title} {Quantum orders and symmetric
  spin liquids},\ }\href {https://doi.org/10.1103/PhysRevB.65.165113}
  {\bibfield  {journal} {\bibinfo  {journal} {Phys. Rev. B}\ }\textbf {\bibinfo
  {volume} {65}},\ \bibinfo {pages} {165113} (\bibinfo {year}
  {2002})}\BibitemShut {NoStop}%
\bibitem [{\citenamefont {Wang}\ and\ \citenamefont
  {Vishwanath}(2006)}]{Wang_Vishwanath_2006}%
  \BibitemOpen
  \bibfield  {author} {\bibinfo {author} {\bibfnamefont {F.}~\bibnamefont
  {Wang}}\ and\ \bibinfo {author} {\bibfnamefont {A.}~\bibnamefont
  {Vishwanath}},\ }\bibfield  {title} {\bibinfo {title} {Spin-liquid states on
  the triangular and kagom\'e lattices: A projective-symmetry-group analysis of
  schwinger boson states},\ }\href {https://doi.org/10.1103/PhysRevB.74.174423}
  {\bibfield  {journal} {\bibinfo  {journal} {Phys. Rev. B}\ }\textbf {\bibinfo
  {volume} {74}},\ \bibinfo {pages} {174423} (\bibinfo {year}
  {2006})}\BibitemShut {NoStop}%
\bibitem [{\citenamefont {{Song}}\ \emph {et~al.}(2019)\citenamefont {{Song}},
  \citenamefont {{Wang}}, \citenamefont {{Vishwanath}},\ and\ \citenamefont
  {{He}}}]{song_vishwanath_2020}%
  \BibitemOpen
  \bibfield  {author} {\bibinfo {author} {\bibfnamefont {X.-Y.}\ \bibnamefont
  {{Song}}}, \bibinfo {author} {\bibfnamefont {C.}~\bibnamefont {{Wang}}},
  \bibinfo {author} {\bibfnamefont {A.}~\bibnamefont {{Vishwanath}}},\ and\
  \bibinfo {author} {\bibfnamefont {Y.-C.}\ \bibnamefont {{He}}},\ }\bibfield
  {title} {\bibinfo {title} {{Unifying description of competing orders in
  two-dimensional quantum magnets}},\ }\href
  {https://doi.org/10.1038/s41467-019-11727-3} {\bibfield  {journal} {\bibinfo
  {journal} {Nature Communications}\ }\textbf {\bibinfo {volume} {10}},\
  \bibinfo {eid} {4254} (\bibinfo {year} {2019})},\ \Eprint
  {https://arxiv.org/abs/1811.11186} {arXiv:1811.11186 [cond-mat.str-el]}
  \BibitemShut {NoStop}%
\bibitem [{\citenamefont {Jang}\ \emph {et~al.}(2021)\citenamefont {Jang},
  \citenamefont {Kato},\ and\ \citenamefont {Motome}}]{Motome2021}%
  \BibitemOpen
  \bibfield  {author} {\bibinfo {author} {\bibfnamefont {S.-H.}\ \bibnamefont
  {Jang}}, \bibinfo {author} {\bibfnamefont {Y.}~\bibnamefont {Kato}},\ and\
  \bibinfo {author} {\bibfnamefont {Y.}~\bibnamefont {Motome}},\ }\bibfield
  {title} {\bibinfo {title} {Vortex creation and control in the kitaev spin
  liquid by local bond modulations},\ }\href
  {https://doi.org/10.1103/PhysRevB.104.085142} {\bibfield  {journal} {\bibinfo
   {journal} {Phys. Rev. B}\ }\textbf {\bibinfo {volume} {104}},\ \bibinfo
  {pages} {085142} (\bibinfo {year} {2021})}\BibitemShut {NoStop}%
\bibitem [{\citenamefont {Wang}\ \emph {et~al.}(2022)\citenamefont {Wang},
  \citenamefont {Laav}, \citenamefont {Volotsenko}, \citenamefont {Frydman},\
  and\ \citenamefont {Kalisky}}]{Kalisky_2022VisualSNet}%
  \BibitemOpen
  \bibfield  {author} {\bibinfo {author} {\bibfnamefont {X.}~\bibnamefont
  {Wang}}, \bibinfo {author} {\bibfnamefont {M.}~\bibnamefont {Laav}}, \bibinfo
  {author} {\bibfnamefont {I.}~\bibnamefont {Volotsenko}}, \bibinfo {author}
  {\bibfnamefont {A.}~\bibnamefont {Frydman}},\ and\ \bibinfo {author}
  {\bibfnamefont {B.}~\bibnamefont {Kalisky}},\ }\bibfield  {title} {\bibinfo
  {title} {Visualizing current in superconducting networks},\ }\href
  {https://doi.org/10.1103/PhysRevApplied.17.024073} {\bibfield  {journal}
  {\bibinfo  {journal} {Phys. Rev. Appl.}\ }\textbf {\bibinfo {volume} {17}},\
  \bibinfo {pages} {024073} (\bibinfo {year} {2022})}\BibitemShut {NoStop}%
\bibitem [{\citenamefont {Senthil}\ \emph {et~al.}(2004)\citenamefont
  {Senthil}, \citenamefont {Vishwanath}, \citenamefont {Balents}, \citenamefont
  {Sachdev},\ and\ \citenamefont {Fisher}}]{Senthil_Deconf_2004}%
  \BibitemOpen
  \bibfield  {author} {\bibinfo {author} {\bibfnamefont {T.}~\bibnamefont
  {Senthil}}, \bibinfo {author} {\bibfnamefont {A.}~\bibnamefont {Vishwanath}},
  \bibinfo {author} {\bibfnamefont {L.}~\bibnamefont {Balents}}, \bibinfo
  {author} {\bibfnamefont {S.}~\bibnamefont {Sachdev}},\ and\ \bibinfo {author}
  {\bibfnamefont {M.~P.~A.}\ \bibnamefont {Fisher}},\ }\bibfield  {title}
  {\bibinfo {title} {Deconfined quantum critical points},\ }\href
  {https://doi.org/10.1126/science.1091806} {\bibfield  {journal} {\bibinfo
  {journal} {Science}\ }\textbf {\bibinfo {volume} {303}},\ \bibinfo {pages}
  {1490} (\bibinfo {year} {2004})}\BibitemShut {NoStop}%
\bibitem [{\citenamefont {{Senthil}}(2023)}]{Senthil_Deconf_Review}%
  \BibitemOpen
  \bibfield  {author} {\bibinfo {author} {\bibfnamefont {T.}~\bibnamefont
  {{Senthil}}},\ }\bibfield  {title} {\bibinfo {title} {{Deconfined quantum
  critical points: a review}},\ }\href
  {https://doi.org/10.48550/arXiv.2306.12638} {\bibfield  {journal} {\bibinfo
  {journal} {arXiv e-prints}\ ,\ \bibinfo {eid} {arXiv:2306.12638}} (\bibinfo
  {year} {2023})},\ \Eprint {https://arxiv.org/abs/2306.12638}
  {arXiv:2306.12638 [cond-mat.str-el]} \BibitemShut {NoStop}%
\bibitem [{\citenamefont {{Kohmoto}}(1985)}]{TI_Kohmoto_1985}%
  \BibitemOpen
  \bibfield  {author} {\bibinfo {author} {\bibfnamefont {M.}~\bibnamefont
  {{Kohmoto}}},\ }\bibfield  {title} {\bibinfo {title} {{Topological invariant
  and the quantization of the Hall conductance}},\ }\href
  {https://doi.org/10.1016/0003-4916(85)90148-4} {\bibfield  {journal}
  {\bibinfo  {journal} {Annals of Physics}\ }\textbf {\bibinfo {volume}
  {160}},\ \bibinfo {pages} {343} (\bibinfo {year} {1985})}\BibitemShut
  {NoStop}%
\bibitem [{\citenamefont {Thonhauser}\ and\ \citenamefont
  {Vanderbilt}(2006)}]{TI_Thonhauser_2006}%
  \BibitemOpen
  \bibfield  {author} {\bibinfo {author} {\bibfnamefont {T.}~\bibnamefont
  {Thonhauser}}\ and\ \bibinfo {author} {\bibfnamefont {D.}~\bibnamefont
  {Vanderbilt}},\ }\bibfield  {title} {\bibinfo {title}
  {Insulator/chern-insulator transition in the haldane model},\ }\href
  {https://doi.org/10.1103/PhysRevB.74.235111} {\bibfield  {journal} {\bibinfo
  {journal} {Phys. Rev. B}\ }\textbf {\bibinfo {volume} {74}},\ \bibinfo
  {pages} {235111} (\bibinfo {year} {2006})}\BibitemShut {NoStop}%
\bibitem [{\citenamefont {Soluyanov}\ and\ \citenamefont
  {Vanderbilt}(2011)}]{Soluyanov2011}%
  \BibitemOpen
  \bibfield  {author} {\bibinfo {author} {\bibfnamefont {A.~A.}\ \bibnamefont
  {Soluyanov}}\ and\ \bibinfo {author} {\bibfnamefont {D.}~\bibnamefont
  {Vanderbilt}},\ }\bibfield  {title} {\bibinfo {title} {Wannier representation
  of ${\mathbb{z}}_{2}$ topological insulators},\ }\href
  {https://doi.org/10.1103/PhysRevB.83.035108} {\bibfield  {journal} {\bibinfo
  {journal} {Phys. Rev. B}\ }\textbf {\bibinfo {volume} {83}},\ \bibinfo
  {pages} {035108} (\bibinfo {year} {2011})}\BibitemShut {NoStop}%
\bibitem [{\citenamefont {Yu}\ \emph {et~al.}(2011)\citenamefont {Yu},
  \citenamefont {Qi}, \citenamefont {Bernevig}, \citenamefont {Fang},\ and\
  \citenamefont {Dai}}]{TI_Rui_2011}%
  \BibitemOpen
  \bibfield  {author} {\bibinfo {author} {\bibfnamefont {R.}~\bibnamefont
  {Yu}}, \bibinfo {author} {\bibfnamefont {X.~L.}\ \bibnamefont {Qi}}, \bibinfo
  {author} {\bibfnamefont {A.}~\bibnamefont {Bernevig}}, \bibinfo {author}
  {\bibfnamefont {Z.}~\bibnamefont {Fang}},\ and\ \bibinfo {author}
  {\bibfnamefont {X.}~\bibnamefont {Dai}},\ }\bibfield  {title} {\bibinfo
  {title} {Equivalent expression of ${\mathbb{z}}_{2}$ topological invariant
  for band insulators using the non-abelian berry connection},\ }\href
  {https://doi.org/10.1103/PhysRevB.84.075119} {\bibfield  {journal} {\bibinfo
  {journal} {Phys. Rev. B}\ }\textbf {\bibinfo {volume} {84}},\ \bibinfo
  {pages} {075119} (\bibinfo {year} {2011})}\BibitemShut {NoStop}%
\bibitem [{\citenamefont {Qi}(2011)}]{TI_Qi_2011}%
  \BibitemOpen
  \bibfield  {author} {\bibinfo {author} {\bibfnamefont {X.-L.}\ \bibnamefont
  {Qi}},\ }\bibfield  {title} {\bibinfo {title} {Generic wave-function
  description of fractional quantum anomalous hall states and fractional
  topological insulators},\ }\href
  {https://doi.org/10.1103/PhysRevLett.107.126803} {\bibfield  {journal}
  {\bibinfo  {journal} {Phys. Rev. Lett.}\ }\textbf {\bibinfo {volume} {107}},\
  \bibinfo {pages} {126803} (\bibinfo {year} {2011})}\BibitemShut {NoStop}%
\bibitem [{\citenamefont {Gunawardana}\ \emph {et~al.}(2024)\citenamefont
  {Gunawardana}, \citenamefont {Turner},\ and\ \citenamefont
  {Barnett}}]{TI_Gunawardana_2024}%
  \BibitemOpen
  \bibfield  {author} {\bibinfo {author} {\bibfnamefont {T.~M.}\ \bibnamefont
  {Gunawardana}}, \bibinfo {author} {\bibfnamefont {A.~M.}\ \bibnamefont
  {Turner}},\ and\ \bibinfo {author} {\bibfnamefont {R.}~\bibnamefont
  {Barnett}},\ }\bibfield  {title} {\bibinfo {title} {Optimally localized
  single-band wannier functions for two-dimensional chern insulators},\ }\href
  {https://doi.org/10.1103/PhysRevResearch.6.023046} {\bibfield  {journal}
  {\bibinfo  {journal} {Phys. Rev. Res.}\ }\textbf {\bibinfo {volume} {6}},\
  \bibinfo {pages} {023046} (\bibinfo {year} {2024})}\BibitemShut {NoStop}%
\bibitem [{\citenamefont {Xie}\ \emph {et~al.}(2024)\citenamefont {Xie},
  \citenamefont {Fang}, \citenamefont {Chen}, \citenamefont {Cano},\ and\
  \citenamefont {Si}}]{TI_Xie_2024}%
  \BibitemOpen
  \bibfield  {author} {\bibinfo {author} {\bibfnamefont {F.}~\bibnamefont
  {Xie}}, \bibinfo {author} {\bibfnamefont {Y.}~\bibnamefont {Fang}}, \bibinfo
  {author} {\bibfnamefont {L.}~\bibnamefont {Chen}}, \bibinfo {author}
  {\bibfnamefont {J.}~\bibnamefont {Cano}},\ and\ \bibinfo {author}
  {\bibfnamefont {Q.}~\bibnamefont {Si}},\ }\bibfield  {title} {\bibinfo
  {title} {Chern bands' optimally localized wannier functions and fractional
  chern insulators},\ }\href {https://arxiv.org/abs/2407.08920} {\bibfield
  {journal} {\bibinfo  {journal} {arXiv}\ } (\bibinfo {year} {2024})},\ \Eprint
  {https://arxiv.org/abs/2407.08920} {arXiv:2407.08920 [cond-mat.mes-hall]}
  \BibitemShut {NoStop}%
\bibitem [{\citenamefont {Wang}\ \emph {et~al.}(2024)\citenamefont {Wang},
  \citenamefont {Shi}, \citenamefont {Liu},\ and\ \citenamefont
  {Wang}}]{wang2024_Landau}%
  \BibitemOpen
  \bibfield  {author} {\bibinfo {author} {\bibfnamefont {H.}~\bibnamefont
  {Wang}}, \bibinfo {author} {\bibfnamefont {R.}~\bibnamefont {Shi}}, \bibinfo
  {author} {\bibfnamefont {Z.}~\bibnamefont {Liu}},\ and\ \bibinfo {author}
  {\bibfnamefont {J.}~\bibnamefont {Wang}},\ }\href
  {https://arxiv.org/abs/2411.13071} {\bibinfo {title} {Orbital description of
  landau levels}} (\bibinfo {year} {2024}),\ \Eprint
  {https://arxiv.org/abs/2411.13071} {arXiv:2411.13071 [cond-mat.mes-hall]}
  \BibitemShut {NoStop}%
\bibitem [{\citenamefont {Strinati}(1978)}]{GP_Strinati_1978}%
  \BibitemOpen
  \bibfield  {author} {\bibinfo {author} {\bibfnamefont {G.}~\bibnamefont
  {Strinati}},\ }\bibfield  {title} {\bibinfo {title} {Multipole wave functions
  for photoelectrons in crystals. iii. the role of singular points in the band
  structure and the tails of the wannier functions},\ }\href
  {https://doi.org/10.1103/PhysRevB.18.4104} {\bibfield  {journal} {\bibinfo
  {journal} {Phys. Rev. B}\ }\textbf {\bibinfo {volume} {18}},\ \bibinfo
  {pages} {4104} (\bibinfo {year} {1978})}\BibitemShut {NoStop}%
\bibitem [{\citenamefont {Bergman}\ \emph {et~al.}(2008)\citenamefont
  {Bergman}, \citenamefont {Wu},\ and\ \citenamefont
  {Balents}}]{GP_Bergman_2008}%
  \BibitemOpen
  \bibfield  {author} {\bibinfo {author} {\bibfnamefont {D.~L.}\ \bibnamefont
  {Bergman}}, \bibinfo {author} {\bibfnamefont {C.}~\bibnamefont {Wu}},\ and\
  \bibinfo {author} {\bibfnamefont {L.}~\bibnamefont {Balents}},\ }\bibfield
  {title} {\bibinfo {title} {Band touching from real-space topology in
  frustrated hopping models},\ }\href
  {https://doi.org/10.1103/PhysRevB.78.125104} {\bibfield  {journal} {\bibinfo
  {journal} {Phys. Rev. B}\ }\textbf {\bibinfo {volume} {78}},\ \bibinfo
  {pages} {125104} (\bibinfo {year} {2008})}\BibitemShut {NoStop}%
\bibitem [{\citenamefont {Herzog-Arbeitman}\ \emph
  {et~al.}(2022{\natexlab{b}})\citenamefont {Herzog-Arbeitman}, \citenamefont
  {Peri}, \citenamefont {Schindler}, \citenamefont {Huber},\ and\ \citenamefont
  {Bernevig}}]{Herzog2022}%
  \BibitemOpen
  \bibfield  {author} {\bibinfo {author} {\bibfnamefont {J.}~\bibnamefont
  {Herzog-Arbeitman}}, \bibinfo {author} {\bibfnamefont {V.}~\bibnamefont
  {Peri}}, \bibinfo {author} {\bibfnamefont {F.}~\bibnamefont {Schindler}},
  \bibinfo {author} {\bibfnamefont {S.~D.}\ \bibnamefont {Huber}},\ and\
  \bibinfo {author} {\bibfnamefont {B.~A.}\ \bibnamefont {Bernevig}},\
  }\bibfield  {title} {\bibinfo {title} {Superfluid weight bounds from symmetry
  and quantum geometry in flat bands},\ }\href
  {https://doi.org/10.1103/PhysRevLett.128.087002} {\bibfield  {journal}
  {\bibinfo  {journal} {Phys. Rev. Lett.}\ }\textbf {\bibinfo {volume} {128}},\
  \bibinfo {pages} {087002} (\bibinfo {year} {2022}{\natexlab{b}})}\BibitemShut
  {NoStop}%
\bibitem [{\citenamefont {Flach}\ \emph {et~al.}(2014)\citenamefont {Flach},
  \citenamefont {Leykam}, \citenamefont {Bodyfelt}, \citenamefont {Matthies},\
  and\ \citenamefont {Desyatnikov}}]{FB_Flach_2014}%
  \BibitemOpen
  \bibfield  {author} {\bibinfo {author} {\bibfnamefont {S.}~\bibnamefont
  {Flach}}, \bibinfo {author} {\bibfnamefont {D.}~\bibnamefont {Leykam}},
  \bibinfo {author} {\bibfnamefont {J.~D.}\ \bibnamefont {Bodyfelt}}, \bibinfo
  {author} {\bibfnamefont {P.}~\bibnamefont {Matthies}},\ and\ \bibinfo
  {author} {\bibfnamefont {A.~S.}\ \bibnamefont {Desyatnikov}},\ }\bibfield
  {title} {\bibinfo {title} {Detangling flat bands into fano lattices},\ }\href
  {https://doi.org/10.1209/0295-5075/105/30001} {\bibfield  {journal} {\bibinfo
   {journal} {Europhysics Letters}\ }\textbf {\bibinfo {volume} {105}},\
  \bibinfo {pages} {30001} (\bibinfo {year} {2014})}\BibitemShut {NoStop}%
\bibitem [{\citenamefont {Dubail}\ and\ \citenamefont
  {Read}(2015)}]{FB_Dubail_2015}%
  \BibitemOpen
  \bibfield  {author} {\bibinfo {author} {\bibfnamefont {J.}~\bibnamefont
  {Dubail}}\ and\ \bibinfo {author} {\bibfnamefont {N.}~\bibnamefont {Read}},\
  }\bibfield  {title} {\bibinfo {title} {Tensor network trial states for chiral
  topological phases in two dimensions and a no-go theorem in any dimension},\
  }\href {https://doi.org/10.1103/PhysRevB.92.205307} {\bibfield  {journal}
  {\bibinfo  {journal} {Phys. Rev. B}\ }\textbf {\bibinfo {volume} {92}},\
  \bibinfo {pages} {205307} (\bibinfo {year} {2015})}\BibitemShut {NoStop}%
\bibitem [{\citenamefont {Morales-Inostroza}\ and\ \citenamefont
  {Vicencio}(2016)}]{FB_Morales_2016}%
  \BibitemOpen
  \bibfield  {author} {\bibinfo {author} {\bibfnamefont {L.}~\bibnamefont
  {Morales-Inostroza}}\ and\ \bibinfo {author} {\bibfnamefont {R.~A.}\
  \bibnamefont {Vicencio}},\ }\bibfield  {title} {\bibinfo {title} {Simple
  method to construct flat-band lattices},\ }\href
  {https://doi.org/10.1103/PhysRevA.94.043831} {\bibfield  {journal} {\bibinfo
  {journal} {Phys. Rev. A}\ }\textbf {\bibinfo {volume} {94}},\ \bibinfo
  {pages} {043831} (\bibinfo {year} {2016})}\BibitemShut {NoStop}%
\bibitem [{\citenamefont {Maimaiti}\ \emph {et~al.}(2017)\citenamefont
  {Maimaiti}, \citenamefont {Andreanov}, \citenamefont {Park}, \citenamefont
  {Gendelman},\ and\ \citenamefont {Flach}}]{FB_Maimaiti_2017}%
  \BibitemOpen
  \bibfield  {author} {\bibinfo {author} {\bibfnamefont {W.}~\bibnamefont
  {Maimaiti}}, \bibinfo {author} {\bibfnamefont {A.}~\bibnamefont {Andreanov}},
  \bibinfo {author} {\bibfnamefont {H.~C.}\ \bibnamefont {Park}}, \bibinfo
  {author} {\bibfnamefont {O.}~\bibnamefont {Gendelman}},\ and\ \bibinfo
  {author} {\bibfnamefont {S.}~\bibnamefont {Flach}},\ }\bibfield  {title}
  {\bibinfo {title} {Compact localized states and flat-band generators in one
  dimension},\ }\href {https://doi.org/10.1103/PhysRevB.95.115135} {\bibfield
  {journal} {\bibinfo  {journal} {Phys. Rev. B}\ }\textbf {\bibinfo {volume}
  {95}},\ \bibinfo {pages} {115135} (\bibinfo {year} {2017})}\BibitemShut
  {NoStop}%
\bibitem [{\citenamefont {Read}(2017)}]{FB_Read_2017}%
  \BibitemOpen
  \bibfield  {author} {\bibinfo {author} {\bibfnamefont {N.}~\bibnamefont
  {Read}},\ }\bibfield  {title} {\bibinfo {title} {Compactly supported wannier
  functions and algebraic $k$-theory},\ }\href
  {https://doi.org/10.1103/PhysRevB.95.115309} {\bibfield  {journal} {\bibinfo
  {journal} {Phys. Rev. B}\ }\textbf {\bibinfo {volume} {95}},\ \bibinfo
  {pages} {115309} (\bibinfo {year} {2017})}\BibitemShut {NoStop}%
\bibitem [{\citenamefont {Zhang}\ and\ \citenamefont
  {Jin}(2020{\natexlab{b}})}]{FB_Zhang_2020}%
  \BibitemOpen
  \bibfield  {author} {\bibinfo {author} {\bibfnamefont {S.~M.}\ \bibnamefont
  {Zhang}}\ and\ \bibinfo {author} {\bibfnamefont {L.}~\bibnamefont {Jin}},\
  }\bibfield  {title} {\bibinfo {title} {Compact localized states and
  localization dynamics in the dice lattice},\ }\href
  {https://doi.org/10.1103/PhysRevB.102.054301} {\bibfield  {journal} {\bibinfo
   {journal} {Phys. Rev. B}\ }\textbf {\bibinfo {volume} {102}},\ \bibinfo
  {pages} {054301} (\bibinfo {year} {2020}{\natexlab{b}})}\BibitemShut
  {NoStop}%
\bibitem [{\citenamefont {Bradlyn}\ \emph {et~al.}(2017)\citenamefont
  {Bradlyn}, \citenamefont {Elcoro}, \citenamefont {Cano}, \citenamefont
  {Vergniory}, \citenamefont {Wang}, \citenamefont {Felser}, \citenamefont
  {Aroyo},\ and\ \citenamefont {Bernevig}}]{AOI_Bradlyn_2017}%
  \BibitemOpen
  \bibfield  {author} {\bibinfo {author} {\bibfnamefont {B.}~\bibnamefont
  {Bradlyn}}, \bibinfo {author} {\bibfnamefont {L.}~\bibnamefont {Elcoro}},
  \bibinfo {author} {\bibfnamefont {J.}~\bibnamefont {Cano}}, \bibinfo {author}
  {\bibfnamefont {M.~G.}\ \bibnamefont {Vergniory}}, \bibinfo {author}
  {\bibfnamefont {Z.}~\bibnamefont {Wang}}, \bibinfo {author} {\bibfnamefont
  {C.}~\bibnamefont {Felser}}, \bibinfo {author} {\bibfnamefont {M.~I.}\
  \bibnamefont {Aroyo}},\ and\ \bibinfo {author} {\bibfnamefont {B.~A.}\
  \bibnamefont {Bernevig}},\ }\bibfield  {title} {\bibinfo {title} {Topological
  quantum chemistry},\ }\href {https://doi.org/10.1038/nature23268} {\bibfield
  {journal} {\bibinfo  {journal} {Nature}\ }\textbf {\bibinfo {volume} {547}},\
  \bibinfo {pages} {298–305} (\bibinfo {year} {2017})}\BibitemShut {NoStop}%
\bibitem [{\citenamefont {Po}\ \emph {et~al.}(2017)\citenamefont {Po},
  \citenamefont {Vishwanath},\ and\ \citenamefont {Watanabe}}]{AOI_Po_2017}%
  \BibitemOpen
  \bibfield  {author} {\bibinfo {author} {\bibfnamefont {H.~C.}\ \bibnamefont
  {Po}}, \bibinfo {author} {\bibfnamefont {A.}~\bibnamefont {Vishwanath}},\
  and\ \bibinfo {author} {\bibfnamefont {H.}~\bibnamefont {Watanabe}},\
  }\bibfield  {title} {\bibinfo {title} {Symmetry-based indicators of band
  topology in the 230 space groups},\ }\href
  {http://dx.doi.org/10.1038/s41467-017-00133-2} {\bibfield  {journal}
  {\bibinfo  {journal} {Nature Communications}\ }\textbf {\bibinfo {volume}
  {8}} (\bibinfo {year} {2017})}\BibitemShut {NoStop}%
\bibitem [{\citenamefont {Schindler}\ and\ \citenamefont
  {Bernevig}(2021)}]{AOI_Schnidler_2021}%
  \BibitemOpen
  \bibfield  {author} {\bibinfo {author} {\bibfnamefont {F.}~\bibnamefont
  {Schindler}}\ and\ \bibinfo {author} {\bibfnamefont {B.~A.}\ \bibnamefont
  {Bernevig}},\ }\bibfield  {title} {\bibinfo {title} {Noncompact atomic
  insulators},\ }\href {https://doi.org/10.1103/PhysRevB.104.L201114}
  {\bibfield  {journal} {\bibinfo  {journal} {Phys. Rev. B}\ }\textbf {\bibinfo
  {volume} {104}},\ \bibinfo {pages} {L201114} (\bibinfo {year}
  {2021})}\BibitemShut {NoStop}%
\bibitem [{\citenamefont {Chen}\ \emph {et~al.}(2023)\citenamefont {Chen},
  \citenamefont {Lin},\ and\ \citenamefont {Kao}}]{AOI_Chen_2023}%
  \BibitemOpen
  \bibfield  {author} {\bibinfo {author} {\bibfnamefont {Y.-C.}\ \bibnamefont
  {Chen}}, \bibinfo {author} {\bibfnamefont {Y.-P.}\ \bibnamefont {Lin}},\ and\
  \bibinfo {author} {\bibfnamefont {Y.-J.}\ \bibnamefont {Kao}},\ }\bibfield
  {title} {\bibinfo {title} {Stably protected gapless edge states without
  wannier obstruction},\ }\href {https://doi.org/10.1103/PhysRevB.107.075126}
  {\bibfield  {journal} {\bibinfo  {journal} {Phys. Rev. B}\ }\textbf {\bibinfo
  {volume} {107}},\ \bibinfo {pages} {075126} (\bibinfo {year}
  {2023})}\BibitemShut {NoStop}%
\end{thebibliography}%

\end{document}